\documentstyle[prl,aps,epsfig,multicol]{revtex}
\newcommand{\beq}{\begin{equation}}
\newcommand{\eeq}{\end{equation}}

\begin{document}

\title{Atom Interference using microfabricated structures}
\author{B. Dubetsky and P. R. Berman}

\address{Physics Department, University of Michigan, Ann Arbor, Michigan 48109-1120,\\
USA\\
in ''{\it Atom Interferometry,'' }P. R. Berman (ed.) (Academic, Chestnut\\
Hill, 1997).}

\date{June 3, 1996}
\maketitle

\begin{center}
\large
\bf{Table of Contents}\\
\normalsize
\end{center}%
%

$
\begin{array}{ll}
\text{I. Introduction} & 1 \\ 
\text{II. Qualitative Consideration} & 4 \\ 
\hspace{0.5in}\text{A. Fresnel Difraction} & 4 \\ 
\hspace{0.5in}\text{B. Fraunhofer Difraction} & 4 \\ 
\hspace{0.5in}\text{C. Talbot-Lau Regime} & 4 \\ 
\hspace{0.5in}\text{D. Classical Scattering} & 5 \\ 
\text{III. Talbot-Effect} & 6 \\ 
\hspace{0.5in}\text{A. Talbot-Effect as a Recoil Effect} & 6 \\ 
\hspace{0.5in}\text{B. Calculation of the Atomic Density Profile} & 7 \\ 
\hspace{0.5in}\text{C. Higher-Order Gratings using Talbot Technique} & 8 \\ 
\text{IV. Shadow Effect with Microfabricated Structures} & 9 \\ 
\hspace{0.5in}\text{A Dephasing-Rephasing Processes in Two Spatially
Separated MS} & 10 \\ 
\hspace{0.5in}\text{B. Particles Distribution Profile} & 12 \\ 
\hspace{0.5in}\text{C. Main Features} & 15 \\ 
\text{V. Talbot-Lau Effect Using Microfabricated Structures} & 16 \\ 
\hspace{0.5in}\text{A. Grating formation} & 17 \\ 
\hspace{0.5in}\text{B. Higher-Order Gratings Induced by MS using Talbot-Lau
Technique} & 20 \\ 
\hspace{0.5in}\text{C. Comparison of the Talbot and Talbot-Lau effects} & 23
\\ 
\hspace{0.5in}\text{D. Additional Examples Including a Quantum Talbot-Lau
Effect} & 24 \\ 
\text{VI. Talbot and Talbot-Lau Effects in a thermal Atomic beam} & 25 \\ 
\hspace{0.5in}\text{A. Atomic Density Profile for a Thermal Beam} & 26 \\ 
\text{VII. Conclusion} & 29 \\ 
\text{Appendix} & 30 \\ 
\,\text{References} & 32
\end{array}
$
\begin{multicols}{2}

\section{Introduction}

In 1973, Altschuler and Frantz patented an idea for creating an atom
interferometer (Altschuler and Frantz, $1973)$. The beam splitter in their
apparatus was a standing-wave optical field. Their ideas were rekindled by
Dubetsky et al. $\left( 1984\right) $, who presented detailed calculations
of atomic scattering by standing-wave fields in the context of atom
interferometry. It was not until recently, however, that experimentalists
were successful in constructing the first atom interferometers. Double-slit
interference (Carnal and Mlynek, 1991; Shimizu et al., 1992), Fraunhofer
diffraction by microfabricated structures (MS) (Keith et al., 1991; Ekstrom
et al., 1995) or by resonant standing wave fields (SW) (Rasel et al., 1995;
Giltner et al., 1995), and Fresnel diffraction by one (Chapman et al., 1995)
or two (Clauser and Li, 1994) MS have all been observed using atomic beams
as matter waves.

Two types of atom-optical elements have been used as beam splitters in these
experiments, standing wave fields or microfabricated structures. A SW beam
splitter allows one to operate with relatively dense atomic beams, having
densities up to $10^{10}cm^{-3}$ and flow densities up to $%
10^{15}cm^{-2}s^{-1}$. Moreover, by varying the atom field detuning, one can
use SW beamsplitters as either amplitude or phase gratings. Additional
degrees of freedom are provided by the polarization of the field which can
act selectively on targeted magnetic state sublevels. A theory of atom
interference in standing wave fields has been developed by Altschuler and
Frantz $\left( 1973\right) ,$ Dubetsky et al. $\left( 1984\right) $,
Chebotayev et al. $\left( 1985\right) $, Bord\'{e} $\left( 1989\right) $,
Friedberg and Hartmann $\left( 1993,1993a\right) $, Dubetsky and Berman $%
\left( 1994\right) $ and Janicke and Wilkens $\left( 1994\right) $. In
contrast to SW beam splitters, MS usually scatter atoms in a state
independent manner; as a consequence, most experiments involving MS use
atoms in their ground (or, possibly, metastable) states. Microfabricated
structures provide 100\% modulation of the incident atomic beam. They offer
the additional advantage that their period and duty cycle (ratio of slit
opening to period) can be chosen arbitrarily within the limits of current
lithographic technology. A theory of atom interference using MS has been
developed by Turchette et al. $\left( 1992\right) ,$ Clauser and Reinsch $%
\left( 1992\right) $ and Carnal et al. $\left( 1995\right) .$

\begin{figure}
\begin{minipage}{0.99\linewidth}
\begin{center}
\epsfxsize=.95\linewidth \epsfbox{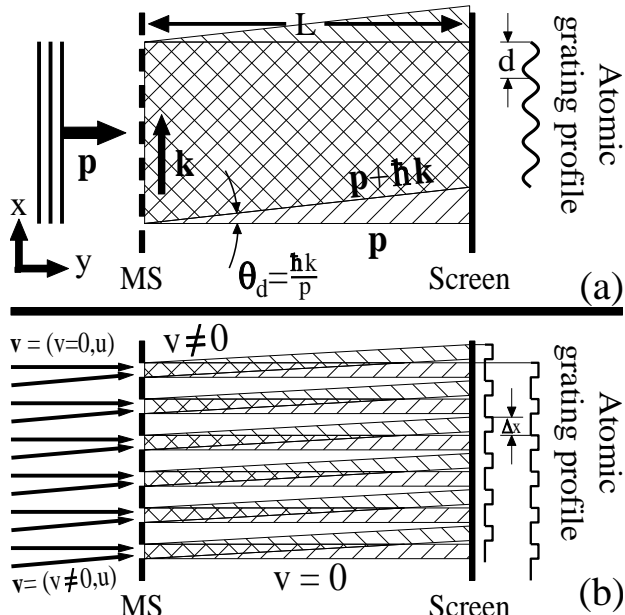}
\end{center}
\end{minipage}
\begin{minipage}{0.99\linewidth} \caption{Matter-wave interference 
$\left( a\right) $ and shadow effect $%
\left( b\right) $ for a collimated atomic beam incident on a microfabricated
structure $\left( MS\right) .$
$\left( a\right) $ When an
atom having given center of mass momentum ${\bf p\parallel \hat{y}}$
scatters from a MS having wave vector ${\bf k=}2\protect\pi {\bf \hat{x}/}d,$
where $d$ is the period of the MS, the output wave function consists of a
set of momenta states ${\bf p+}n\hbar {\bf k}$ (two of them $\left| {\bf p}%
\right\rangle $ and $\left| {\bf p+}\hbar {\bf k}\right\rangle $ are shown).
A superposition of these states' wave functions leads to an interference
pattern on the screen, having the same period $d$ as the MS's.
$\left( b
\right) $ When a collimated beam of particles
moving along classical trajectories scatters from MS, the beam profile
imprinted by MS is copied on the screen. A beam consisting of two velocity
subgoups, having zero $\left( v=0\right) $ and nonzero $\left( v\neq
0\right) $ velocity projections on the $x-$axis is shown. Gratings
associated with these two subgroups are shifted on the screen from on
another by $\Delta x=vt=\frac{v}{u}L,$ where $t$ is the time of flight
between the MS and the screen. 
\label{intrf1}}
\end{minipage}
\end{figure}

Both the splitting of an atomic beam into two or more beams coherent with
respect to one another and the recombining of the scattered beams are
physical processes that are essential to the operation of an atom
interferometer. We consider scattering of atoms by an ideal MS, having an
infinite number of slits, period $d$, duty cycle $f$, and 100\% transmission
through the slits. Each MS is normal to the $y$-axis, and the slits are
oriented in the $z$-direction, so that the axis of the MS is in the $x$%
-direction (see Fig. \ref{intrf1}).
After scattering from a MS, each in-coming atomic state $\psi $ having $x$%
-component of center-of-mass momentum $p$ splits into a set of out-going
states $\psi _{n}$ having $x$-components of momenta $p+n\hbar k\ $($k=2\pi
/d,$ $n$ is an integer) which evolve as 
\begin{equation}
\psi _{n}\propto \exp \left[ \frac{i}{\hbar }\left( p+n\hbar k\right) x%
\right] .  \label{intr1}
\end{equation}
Interference of two components, such as $\psi _{0}$ and $\psi _{1},$ on a
screen (see Fig. \ref{intrf1}a) leads to an atomic density grating 
\begin{equation}
\rho \propto 
\mathop{\rm Re}%
[\psi _{0}\psi _{1}^{\ast }]=\cos \left[ kx\right]  \label{intr2}
\end{equation}
having the same period $d$ as the MS.

Observation of this grating in the experiments listed above often has been
considered as direct evidence for matter-wave interference. Nevertheless,
one can easily see that such a conclusion is not necessarily justified. For
particles moving along classical trajectories (Fig. \ref{intrf1}b), and for
a beam whose angular divergence is sufficiently small to satisfy 
\begin{equation}
\theta _{b}\ll \frac{d}{L},  \label{intr3}
\end{equation}
where $L$ is a distance on the order of the distance between the MS and the
screen, a shadow of the MS can be seen on the screen at distances $L$ where
all matter-wave effects are completely negligible. This example is the
simplest manifestation of the classical shadow effect (Chebotayev et al.,
1985; Dubetsky and Berman 1994).

To distinguish quantum matter-wave interference from the classical shadow
effect, one needs to observe additional features of the phenomena or to
choose a scheme where one of the effects is excluded. Young's double-slit
experiment (Carnal and Mlynek, 1991; Shimizu et al., 1992), as well as
interference produced by a phase grating created using light that is
far-detuned from atomic transition frequencies (Rasel et al., 1995), cannot
be explained in terms of atoms moving on classical trajectories. A
matter-wave interpretation is also necessary if one observes a shift in the
fringe pattern resulting from an index change in one of the arms of an
interferometer (Ekstrom et al., 1995). We determine below those particular
conditions for which pure quantum interference can be obtained using MS.

\begin{figure}
\begin{minipage}{0.99\linewidth}
\begin{center}
\epsfxsize=.95\linewidth \epsfysize=0.71\epsfxsize \epsfbox{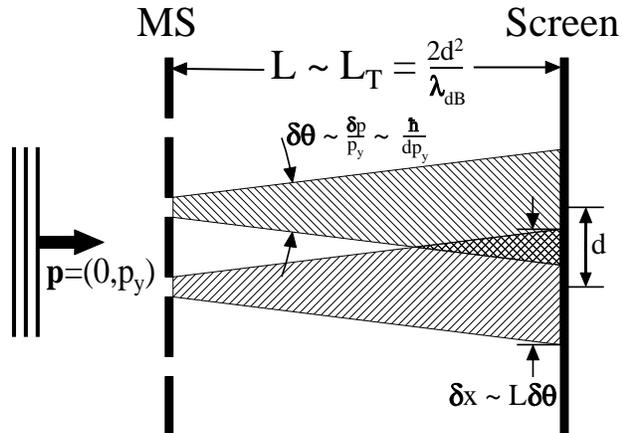}
\end{center}
\end{minipage}
\begin{minipage}{0.99\linewidth} \caption{The incident atomic wave 
function after scattering from MS
transforms into a superposition of the divergent waves emitted from the each
slit, having angular divergence $\protect\delta \protect\theta $ inverse to
the MS's period $d$. (for the given duty cycle). One estimates Talbot
distance as a distance between MS and screen large enough to provide
overlapping of neighboring divergent waves and their interference. 
\label{intrf2}}
\end{minipage}
\end{figure}

Let us estimate a typical distance for which quantum interference effects
have to be included. Consider an incident beam which has zero angular
divergence. The atoms are assumed to be in a pure state having momentum $%
{\bf p=}\left( 0,p_{y},0\right) $ (see Fig. \ref{intrf2}). 
Localization of the atoms inside each slit leads to an uncertainty in the
x-component of atomic momentum $\delta p\sim \frac{\hbar }{d}$, where it is
assumed that the slit width $fd$ is comparable with the MS period $d.$ A
beam passing through the slits acquires an angular divergence $\delta \theta
\sim \frac{\delta p}{p_{y}}$, and atoms passing through a given slit are
deposited on the screen with a spot size $\delta x\sim L\delta \theta \sim L%
\frac{\hbar }{dp_{y}}.$ Interference occurs when spots produced by
neighboring slits overlap, i. e. when $\delta x\sim d.$ One finds,
therefore, that the characteristic distance for which matter-wave
interference plays an essential role is given by 
\begin{equation}
L\sim L_{T},  \label{intr4}
\end{equation}
where 
\begin{equation}
L_{T}=2d^{2}/\lambda _{dB}  \label{intr5}
\end{equation}
is the so-called Talbot distance and $\lambda _{dB}=\frac{h}{p_{y}}$ is the
atomic de Broglie wavelength.

The manner in which this distance appears in the theory of the optical or
atomic Talbot-effect is well known in the context of the Fresnel-Kirchhoff
theory of diffraction [see, for example, (Patorski, 1989; Winthrop and
Worthington, 1965) for optical Talbot-effect theory and (Chapman et al.,
1995; Turchette et al., 1992; Clauser and Reinsch, 1992; Carnal et al.,
1995) for atomic Talbot-effect theory]. One can obtain Eq. (\ref{intr4}, \ref
{intr5}) using another description. When the number of slits in the MS and
the area of the incident atomic beam are infinite, constructive interference
occurs only for those directions in which $p$ is changed by an integral
multiple of the recoil momentum $\hbar k.$ In the atomic ''rest-frame'' (a
frame moving along the $y-$axis with velocity 
\begin{equation}
u=p_{y}/M,  \label{intr6}
\end{equation}
where $M$ is the atomic mass) an out-going state with momentum $p+n\hbar k$ (%
\ref{intr1}) acquires a phase $\phi _{n}=\epsilon _{p+n\hbar k}t/\hbar ,$
where $\epsilon _{p}=\frac{p^{2}}{2M}$ is the kinetic energy associated with
atomic motion along the $x$-axis and $t$ is the time after scattering from
the MS. Comparing this phase with the phase $\phi _{0}\ $that an atom would
acquire in the absence of the MS, one sees that a dephasing $\delta \phi
=\phi _{n}-\phi _{0}$ occurs for the different out-going state amplitudes as
a result of diffraction. The relative dephasing is 
\begin{equation}
\delta \phi =n\phi _{D}+n^{2}\phi _{t},  \label{intr7}
\end{equation}
where 
\begin{equation}
\phi _{D}=kvt,  \eqnum{7a}  \label{intr7a}
\end{equation}
\begin{equation}
\phi _{t}=\omega _{k}t,  \eqnum{7b}  \label{intr7b}
\end{equation}
$v=\frac{p}{M}$ is the $x$ component of atomic velocity, and 
\begin{equation}
\omega _{k}=\frac{\hbar k^{2}}{2M}  \label{intr8}
\end{equation}
is a recoil frequency, related to the energy that an atom, having initial
momentum $p=0,$ acquires as a result of recoil during scattering. The two
contributions to the dephasing (\ref{intr7}) have different origins. The
phase $\phi _{D}\ $is a Doppler phase that does not disappear in the
classical limit $\hbar \rightarrow 0;$ consequently, it must be classical in
nature. It is evident from Fig. \ref{intrf1}b that, for atoms incident with
a nonzero value of $p$, the shadow moves along the $x-$axis together with
atoms. The nodes in the shadow are displaced by a distance $\Delta x=vt$
along the $x-$axis, which corresponds to a phase shift of $2\pi \frac{\Delta
x}{d}$ for the atomic grating, a phase shift that coincides with $\phi _{D}.$
The phase $\phi _{D}$ is analogous to the phase a moving dipole driven by an
optical field would acquire in its rest frame as a result of the Doppler
frequency shift.

The phase $\phi _t$ in Eq. (\ref{intr7}) is a quantum addition to the
dephasing, resulting from recoil. This contribution is responsible for
atomic scattering and for matter-wave interference. Quantum effects have to
be included when this phase is of order unity, that is, for times of order 
\begin{equation}
t_T=2\pi /\omega _k.  \label{intr9}
\end{equation}
One finds that length associated with this time in laboratory frame
coincides with the Talbot distance 
\begin{equation}
L_T=ut_T.  \label{intr10}
\end{equation}
Thus, we are led to the same conclusion that we reached above by considering
the scattering from two adjacent slits; for $\omega _kt_T\gtrsim 1$ or $%
L\gtrsim L_T$, one must use a quantized description of the atomic
center-of-mass motion.

In this chapter we consider the Talbot-effect and other interference
phenomena as a consequence of the recoil-effect. In the context of the
nonlinear interaction of optical fields with an atomic vapor, the recoil
effect was considered by Kol'chenko et al. $\left( 1968\right) $ and
observed by Hall et al. $\left( 1976\right) $. Quantum structure resulting
from the scattering of atoms by a resonant standing wave (resonant
Kapitza-Dirac effect), which can be attributed to atomic recoil, was
discussed theoretically by Kazantsev et al. $\left( 1980\right) $ and
observed by Moskovitz et al. $\left( 1983\right) $. Splitting of optical
Ramsey fringes (Baklanov et al., 1976) associated with the resonant
Kapitza-Dirac effect was discussed theoretically by Dubetsky and Semibalamut 
$\left( 1978\right) $ and observed by Barger et al. $\left( 1979\right) $.
Matter-wave interference resulting from resonant Kapitsa-Dirac scattering in
a standing wave field has been studied theoretically by Altschuler and
Frantz $\left( 1973\right) ,$ Dubetsky et al. $\left( 1984\right) $ and
observed by Rasel et al. $\left( 1995\right) $. The theory of atom
interference presented here, based on an interpretation of scattering of
atoms by MS in terms of the recoil effect, is a natural extension of the
work involving standing-wave fields.

This chapter is organized as follows: in the next Section we discuss
conditions necessary for the observation of matter-wave interference in
different regimes. Rigorous proof of the equivalence of theories based on
Fresnel-Kirchhoff integrals and on the recoil effect is given in Section
III, as is a discusion of the atomic gratings that can be produced as a
consequence of the Talbot effect. The classical shadow effect is analyzed in
Section IV. Section V is devoted to a theory of the Talbot-Lau effect. The
Talbot and Talbot-Lau effects for a thermal beam are considered in the
Section VI, in the limit where the characteristic length scale in the
problem is larger than the Talbot length $L_T$. A discussion of the results
is given in Section VII.

\section{Qualitative considerations}

The scattering of atoms by gratings can be separated roughly into three
categories: classical scattering, Fresnel diffraction, and Fraunhofer
diffraction. The limit of Bragg scattering (Martin et al., 1988), in which $%
\omega _k\ell ^{\prime }/u\geq 1$, where $\ell ^{\prime }$ is the grating
thickness, is not discussed in this chapter.

\subsection{Fresnel Diffraction}

The Fresnel diffraction limit occurs when $\omega _kt\sim 1$ or $L\sim L_T$.
Owing to the angular divergence of the incident beam, it is possible that
the diffraction pattern at $L\sim L_T$ will be washed out. To ensure that
this does not occur, it is necessary that the spread of Doppler phases, $%
kut_T\theta _b=kL_T\theta _b,$ be smaller than unity. This requirement
corresponds to inequality (\ref{intr3}) when $L\sim L_T$. Using Eq. (\ref
{intr5}), the condition on $\theta _b$ can be restated as 
\begin{equation}
\theta _b\ll \theta _d,  \label{intr11}
\end{equation}
where 
\begin{equation}
\theta _d=\frac{\hbar k}{p_y}=\frac{\lambda _{dB}}d  \label{intr12}
\end{equation}
is the angle associated with a single atomic recoil at the MS ($\delta
p=\hbar k)$. The Talbot effect refers to the self-imaging of a grating in
the Fresnel diffraction limit. For self-imaging to occur, the displacement
of the scattered atomic beam $L_T\theta _d$ must be much smaller than the
beam diameter $D,$ which translates into the condition 
\begin{equation}
D\gg d.  \label{intr13}
\end{equation}
In this limit, one can consider the beam diameter to be infinite to first
approximation; finite beam effects (or, equivalently, gratings with finite
slit number) are discussed by Clauser and Reinsch $\left( 1992\right) .$

Conditions (\ref{intr4}, \ref{intr11}, \ref{intr13}) are sufficient to
observe the Talbot effect. In this case, the contribution to the wave
function's phase resulting from atomic recoil is of order unity, the
scattered beams overlap almost entirely with one another on the screen, and
the atomic gratings are not washed out after averaging over atomic
velocities $v$ in the incident beam. Matter-wave interference is a critical
component of the Talbot effect.

\subsection{Fraunhofer Diffraction}

Although the Talbot effect illustrates a matter-wave interference
phenomenon, it does not result in an atom interferometer having two arms
that are nonoverlapping. We refer to the Fraunhofer diffraction limit as one
in which the various diffraction orders are nonoverlapping at a distance $L$
from a single grating. The grating then serves as a beam splitter that
physically separates the incident beam into two or more beams. To physically
separate the various diffraction orders over a distance $L$, one must
require that 
\begin{equation}
L\theta _d\gg D.  \label{intr14}
\end{equation}
Using the fact that $\theta _d=\lambda _{dB}/d$ and setting $t=\frac Lu,$
one can recast this inequality as 
\begin{equation}
\omega _kt\gg D/d\text{ ; }\hspace{0.250in}L/L_T\gg D/d,  \label{intr15}
\end{equation}
which requires the quantum phase $\phi _t$ (\ref{intr7b}) to be larger than $%
D/d>>1$. Consequently, quantum effects play an essential role in an atom
interferometer having nonoverlapping beam paths. Note that the Fraunhofer
limit cannot be reached for a beam having infinite diameter. We do not
consider matter-wave interference in the Fraunhofer limit in this work.

\subsection{Talbot-Lau Regime.}

To achieve spatial separation of the beams in the Fraunhofer limit and to
observe the Talbot effect, the angular divergence $\theta _b$ of the
incident beam must be less than $\theta _d$. For typical values $d\sim 200$ $%
nm$, $M\sim 20$ $A.u.,\,\,u\sim 10^5$ $cm/s,$ the deflection angle $\theta
_d\sim 10^{-4}$ $rad.$ Atomic beams having $\theta _b\ll \theta _d$ have
been used to observe atomic scattering by standing waves (Moskovitz et. al.,
1983), to build a two-arm atom interferometer (Keith et al., 1991; Rasel et
al., 1995; Giltner et al., 1995), and to observe the Talbot-effect (Chapman
et al., 1995). Such strong collimation results in a decrease in the atomic
flux and a corresponding decrease in signal strength that may be a limiting
factor in certain applications of matter-wave interference, such as atom
lithography (Timp et al., 1992). Alternatively, one can observe matter-wave
interference in beams having larger angular divergence using the atomic
Talbot-Lau effect [see for example (Patorski, 1989)]. In the atomic
Talbot-Lau effect two or more MS are used. Doppler dephasing following the
first MS washes out the normal Talbot effect, but subsequent scattering by a
second MS can result in a Doppler rephasing that ultimately leads to a
Talbot-like interference pattern. The dephasing-rephasing process is
analogous to that occurring in the production of photon echoes (Dubetsky et
al., 1984). The atomic Talbot-Lau effect has been observed recently by
Clauser and Li $\left( 1994\right) $ using a K beam scattered by MS. The
incident beam is not separated into nonoverlapping beams in the Talbot-Lau
effect, but the origin of the interference pattern can still be traced to
matter-wave interference since it is related to Fresnel diffraction.

\subsection{Classical Scattering}

It is worthwhile at this point to return to the classical shadow effect. The
shadow effect in a collimated beam is obvious; the atomic grating produced
by the MS simply propagates in space over a distance in which diffraction
can be ignored ($L/L_{T}\ll 1$). If one has an ensemble of atoms incident on
a MS from different angles, each velocity subgroup creates its own grating.
Just after passing through the MS, all the gratings are the same, having the
profile of the MS. Downstream from the MS, the atomic gratings corresponding
to different velocity classes move in different directions (two of them are
shown in Fig.\ref{intrf1}b). The grating of the ensemble as a whole is
washed out at a distance $\tilde{\ell}$ from the MS, provided that 
\begin{equation}
\tilde{\ell}\theta _{b}\geq d.  \label{intr16}
\end{equation}
For 
\begin{equation}
L\theta _{b}\gg d,  \label{intr17}
\end{equation}
the shadow of the atomic grating is washed out well before reaching the
screen or any subsequent MS in the experimental setup$.$

There are two ways to view the washing out of the grating. As discussed
above, the washing out is a result of the different classical trajectories
of the atoms. An alternative view allows one to relate this phenomenon to
that encountered in optical coherent transients. Imagine that the incident
beam consists of a number of velocity subgroups, having $v_x\equiv $ $v\sim
u\theta _b$. When the atoms are scattered by a MS having a transmission
function $\chi _1(x)$, the atomic density immediately following the grating
is proportional to $\chi _1(x)$. Downstream from the MS, the atomic grating
in the $x$-direction simply propagates with velocity $v$, leading to a
density distribution that varies as $\chi _1(x-vt)$, where $t=y/u$ and $y$
is the distance from the grating. If this density distribution is expanded
in a Fourier series, one finds terms in the sum that vary as $\cos
[nk(x-vt)] $, where $n$ is an integer. In this picture, particles in
velocity subgroup $v$ acquire a Doppler phase (\ref{intr7a}) of order 
\begin{equation}
\phi _D\sim kvt\sim \theta _bL/d\gg 1,  \label{intr18}
\end{equation}
as they propagate a distance of order $L$ from the MS$.$ The decay of the
macroscopic grating is analogous to the free-induction-decay of the
macroscopic polarization of a Doppler-broadened atomic vapor following
excitation by an optical pulse. In this approach, one can draw on many
processes that are well known in the theory of optical coherent transients.

Although the atomic grating is washed out following the interaction with the
MS, it is possible to restore the original macroscopic atomic grating by
placing a second MS between the first MS and the screen. This effect has
been observed recently by Batelaan et al. $\left( 1996\right) $ using a
metastable Ar beam. In the classical trajectory picture, the restoration of
the grating corresponds to a Moir\'{e} pattern. In the Doppler dephasing
picture, the restoration is analogous to the dephasing-rephasing process
that occurs for a photon echo. The first MS starts a dephasing process for
the different velocity subgroups and the second MS results in a rephasing
process (see below, section IV. A). At a particular focal plane, where the
rephasing is complete, a macroscopic grating appears. For 
\begin{equation}
L\ll L_T  \label{intr19}
\end{equation}
effects relating to quantization of the atomic center-of-mass motion play no
role\cite{1}.

If $L\sim L_T$, one has to include recoil effects. Gratings appearing in
this regime are usually associated with the Talbot-Lau effect, i. e. with
the interference of light for the optical case or quantum interference of
matter waves in the case of the atomic Talbot-Lau effect. It is shown in
Sec. V, however, that the position of the focal planes and gratings' periods
are often the same as in the classical case. From this point of view, the
Talbot-Lau effect is a quantum generalization of the shadow effect.

In summary, one can conclude that interference is qualitatively different
for collimated beams and beams with large angular divergence. For collimated
beams ($\theta _{b}L\ll d)$ one has two interesting regimes, 
\begin{equation}
L\sim L_{T}\text{\hspace{0.25in} }[\omega _{k}\text{ }t\sim 1]
\label{intr20}
\end{equation}
and\ 
\begin{equation}
L/L_{T}\gg D/d\text{\hspace{0.25in}}\left[ \omega _{k}t\gg D/d\right] , 
\eqnum{20a}  \label{intr20a}
\end{equation}
corresponding to Fresnel (Talbot effect) and Fraunhofer diffraction
(nonoverlapping scattered beams), respectively. For beams having angular
divergence $\theta _{b}L\gg d$, the Fresnel and Fraunhofer diffraction
patterns would wash out following a single MS. Restoration of the atomic
gratings in this limit can be achieved using two or more MS. Distances $%
L\sim L_{T}$ correspond to the atomic Talbot-Lau effect and distances 
\begin{equation}
L\ll L_{T}\text{\hspace{0.25in}}\omega _{k}t\ll 1  \label{intr21}
\end{equation}
correspond to the classical shadow effect.

\section{Talbot effect}

\subsection{Talbot effect as a recoil effect.}

In this section we show that the Talbot effect is a consequence of the
recoil an atom undergoes when it passes through a microfabricated structure
(MS). We assume that the MS is located in the plane $y=0,$ normal to the
direction of propagation of the atomic beam. The MS consists of an infinite
number of slits oriented in the $z-$direction; as such, only the $x-$%
dependence of the atomic wave function changes when atoms pass through the
slits. Atomic motion in the $y-$direction can be considered as classical in
nature provided that $\lambda _{dB}/d\ll 1$, but motion along the $x$-axis
must be quantized. In this section, it is assumed that the incident beam is
strongly collimated, $\theta _b\ll d/L$, where $d$ is the period of the MS
and $L$ is the distance from the MS to the screen. As such, we can neglect
any spread in the transverse velocities in the initial beam and consider all
atoms to be incident with transverse momentum $p=0$.

After passing through the MS, the wave function for an atom is given by 
\begin{equation}
\psi (x)=\eta (x),  \label{talbot1}
\end{equation}
where $\eta (x)$ is the amplitude transmission function associated with the
MS$.$ In the momentum representation, $\psi (x)$ can be written as a
superposition of states having momenta $p=m\hbar k,$ where $m$ is an integer
and $k=2\pi /d$. Explicitly, one finds that the Fourier transform of $\psi
(x)$ is given by 
\begin{equation}
\tilde{\psi}(p)=\sqrt{2\pi \hbar }\sum_m\eta _m\delta (p-m\hbar k),
\label{talbot2}
\end{equation}
where 
\begin{equation}
\eta _m=\int \frac{dx}de^{-imkx}\eta (x)  \label{talbot3}
\end{equation}
is a Fourier coefficient. Unless indicated otherwise, all sums run from $%
-\infty $ to $+\infty $. The terms with $m\neq 0$ in Eq. (\ref{talbot2}) can
be associated with atomic scattering at angles $m\hbar k/p_y$, where $p_y$
is the longitudinal momentum in the atomic beam. It is assumed that $p_y$ is
constant for all atoms in the beam - this restriction is relaxed in Sec. VI.
In analogy with electron scattering from a standing wave field
[Kapitza-Dirac effect (Kapitza and Dirac, 1933)] or atomic scattering from a
resonant standing wave field [resonant Kapitza-Dirac effect (Kazantsev et
al., 1980)], the scattering from the MS can be interpreted as arising from
the recoil the atoms undergo when they acquire $m\hbar k$ of momenta by
scattering from the MS.

The classical motion of the atoms in the $y$-direction associates a distance 
$y$ with a time 
\begin{equation}
t=y/u,  \label{talbot6}
\end{equation}
where $u=p_{y}/M$ and $M$ is the atomic mass. For a given $u$, the momentum
space wave function evolves as 
\begin{eqnarray}
\tilde{\psi}(p,t) &=&e^{-i\epsilon _{p}t/\hbar }\tilde{\psi}(p)  \nonumber \\
&=&\sqrt{2\pi \hbar }\sum_{m}\eta _{m}\exp [-im^{2}\omega _{k}t]\delta
(p-m\hbar k),  \label{talbot5}
\end{eqnarray}
where $\epsilon _{p}=p^{2}/2M$ is the kinetic energy of an atom having
center of mass momentum $p$, and $\omega _{k}=\hbar k^{2}/2M$ is a recoil
frequency. In the coordinate representation 
\begin{equation}
\psi (x,t)=\int_{-\infty }^{\infty }\frac{dp}{\sqrt{2\pi \hbar }}%
e^{ipx/\hbar }\tilde{\psi}(p,t)  \label{talbot7}
\end{equation}
one finds 
\begin{equation}
\psi (x,t)=\sum_{m}\eta _{m}\exp [imkx-im^{2}\phi _{t}],  \label{talbot8}
\end{equation}
where the Talbot phase is defined by 
\begin{equation}
\phi _{t}=\omega _{k}t.  \label{talbot36}
\end{equation}

Superposition of the different terms in Eq. (\ref{talbot8}) leads to a
spatial modulation of the atomic density 
\begin{equation}
f(x,t)=\left| \psi (x,t)\right| ^2.  \label{talbot9}
\end{equation}
The interference terms in Eq. (\ref{talbot9}) are a direct manifestation of
matter-wave interference. One can see that, as a function of the time of
flight $t=$ $y/u,$ the wave function (\ref{talbot8}) undergoes oscillations
on a time scale $\omega _k^{-1}$. As a consequence, the atomic spatial
distribution (\ref{talbot9}) contains quantum beats at frequencies $%
(m^2-n^2)\omega _k$, for integral $m,n$. Such quantum beats have been
predicted by Chebotayev et al. (1985) and observed by Chapman et al. $\left(
1995\right) $.

It follows from Eq.(\ref{talbot8}) that the atomic wave function coincides
with the amplitude transmission function of the MS when 
\begin{equation}
t=t_T\equiv 2\pi /\omega _k.  \label{talbot10}
\end{equation}
At this time, atoms are found in the focal plane at 
\begin{equation}
y=L_T=ut_T=2d^2/\lambda _{dB},  \label{talbot11}
\end{equation}
and a self-image of the MS is produced. In general one finds that the atomic
wave function is a periodic function of the Talbot phase $\phi _t$ having
period $2\pi ,$ a periodic function of the time $t$ having period $2\pi
/\omega _k,$ and a periodic function of the distance $y$ having period $L_T$.

The self-imaging of a periodic structure is well known in classical optics
as the Talbot effect. To describe this effect, one usually starts from the
Fresnel-Kirchhoff equation 
\begin{equation}
\psi (x)=\frac 1{\sqrt{i\lambda _{dB}y}}\int_{-\infty }^\infty dx^{\prime
}\eta (x^{\prime })\exp \left[ ik_{dB}(x-x^{\prime })^2/2y\right] ,
\label{talbot12}
\end{equation}
which is written here in the parabolic approximation. To establish an
equivalence between Eqs.(\ref{talbot12}) and (\ref{talbot8}), one can
substitute the Fourier expansion of the function $\eta (x^{\prime })$ in Eq.(%
\ref{talbot12}), carry out the integration, and express $y$ in terms of $t.$

It is possible to derive a useful symmetry property for $f(x,t)$ when the
transmission function $\eta (x)$ is real, as it is for the MS. For real $%
\eta (x)$, there is pure amplitude modulation of the atomic wave function
and $\eta _m=\eta _{-m}^{*}.$ It then follows from Eqs. (\ref{talbot8}) and (%
\ref{talbot9}) that the atomic spatial distribution is invariant under
inversion with respect to the plane $y=L_T/2,$ i. e.

\begin{equation}
\left. f\left( x,t\right) \right| _{\phi _{t}}=\left. f\left( x,t\right)
\right| _{2\pi -\phi _{t}}  \label{talbot38}
\end{equation}
\begin{equation}
f\left( x,t\right) =f\left( x,t_{T}-t\right)   \eqnum{34a}
\label{talbot38a}
\end{equation}
\begin{equation}
\left. f\left( x,t\right) \right| _{y}=\left. f\left( x,t\right) \right|
_{L_{T}-y}.  \eqnum{34b}  \label{talbot38b}
\end{equation}
Thus, one need calculate $f\left( x,t\right) $ in the range $0\leq y\leq
L_{T}/2$ to obtain the distribution for all $y$.

\subsection{Calculation of the atomic density profile}

We have seen that, for a sufficiently collimated atomic beam, self-imaging
of a MS occurs at integral multiples of the Talbot length. To analyze the
diffraction pattern for arbitrary $y$, it is convenient to use Eqs. (\ref
{talbot3}) and (\ref{talbot8}) to reexpress the atomic wave-function as the
convolution (Winthrop and Worthington, 1965) 
\begin{equation}
\psi (x,\phi _t)=(1/d)\int_{x-d}^x\eta (x^{\prime })Z(x-x^{\prime },\phi
_t)dx^{\prime },  \label{talbot13}
\end{equation}
where 
\begin{equation}
Z(x,\phi _t)=\sum_m\exp \left[ -im^2\phi _t+imkx\right] .  \label{talbot14}
\end{equation}
In the following discussion, we calculate $\psi (x,\phi _t=2\pi y/L_T)$ at
fractions of the Talbot length, that is, for 
\begin{equation}
y=L_T/n,  \label{talbot20}
\end{equation}
or, equivalently, for 
\begin{equation}
\phi _t=2\pi /n,  \label{talbot21}
\end{equation}
where $n$ is a positive integer.

When $n=2$, one can show that the diffraction pattern is a self-image of the
MS, shifted by half a period. For $n=2$, $\phi _t=\pi ,$ and 
\begin{equation}
Z(x,\pi )=\left( 1-e^{ikx}\right) \sum_qe^{2iqkx}.  \label{talbot16}
\end{equation}
Using the equality 
\begin{equation}
\sum_qe^{iq\alpha }\equiv 2\pi \sum_s\delta (\alpha -2\pi s),
\label{talbot17}
\end{equation}
one finds 
\begin{equation}
Z(x,\pi )=d\sum_{odd\ s}\delta \left( x-s\frac d2\right) .  \label{talbot18}
\end{equation}
For the integration range in Eq. (\ref{talbot13}), only the $s=1$ term
contributes when Eq. (\ref{talbot18}) is substituted into Eq. (\ref{talbot13}%
), leading to 
\begin{equation}
\psi (x,\pi )=\eta \left( x-\frac d2\right) ,  \label{talbot19}
\end{equation}
i. e. at half-integral multiples of the Talbot length, there are self images
of the MS shifted by half a period.

For arbitrary $n$, it is convenient to write 
\begin{equation}
m=nq+r,  \label{talbot22}
\end{equation}
where $0\leq r\leq n-1$ and $q$ and $r$ are integers. It then follows that 
\begin{equation}
\exp \left( -im^2\phi _t\right) =\exp (-2\pi im^2/n)\equiv \exp \left(
-i2\pi r^2/n\right) .  \label{talbot23}
\end{equation}
and 
\begin{equation}
Z(x,2\pi /n)=d\sum_sa_s\delta \left( x-s\frac dn\right) ,  \label{talbot24}
\end{equation}
where

\begin{equation}
a_{s}(n)=\frac{1}{n}\sum_{r=0}^{n-1}\exp \left[ 2\pi ir(s-r)/n\right] . 
\eqnum{45a}  \label{talbot24a}
\end{equation}
The atomic wave function (\ref{talbot13}) is then given by 
\begin{equation}
\psi (x,2\pi /n)=\sum_{s=0}^{n-1}a_{s}(n)\eta \left( x-s\frac{d}{n}\right) .
\label{talbot25}
\end{equation}

The meaning of this equation is clear. At distances $y=$ $L_{T}/n$ , the
wave function consists of $n$ self-images of the amplitude transmission
function $\eta (x)$, having different amplitudes $a_{s}(n)$ (some of which
might vanish) and spaced from one another by the distance $d/n$. For
appropriately chosen $\eta (x)$ and $n$ (see below), the atomic density (\ref
{talbot9}) associated with the wave-function (\ref{talbot25}) is a periodic
function of $x$ having period 
\begin{equation}
d_{g}=d/n,\ \text{or }d_{g}=2d/n.  \label{talbot26}
\end{equation}
Thus, the Talbot-effect can be used to generate spatial modulation of an
atomic beam having a period which is a fraction of the period of the MS. We
refer to such profiles as ''higher order atomic gratings.''

To simplify the expression for the coefficients $a_{s}(n),$ one can use an
alternative approach for evaluating $Z\left( x,\phi _{t}\right) $ (Winthrop
and Worthington, 1965). The sum in Eq. (\ref{talbot14}) can be written in
the form 
\begin{eqnarray}
Z\left( x,\phi _{t}\right)  &=&\frac{1}{2\pi }\sum_{q}\int dmdf  \nonumber \\
&&\times \exp \left[ -if(q-m)-im^{2}\phi _{t}+imkx\right] .  \label{talbot27}
\end{eqnarray}
Carrying out the integration over $m,$ summation over $q$ [using Eq. (\ref
{talbot17})] and integration over $f,$ one arrives at 
\begin{equation}
Z(x,\phi _{t})=\sqrt{\frac{\pi }{i\phi _{t}}}\sum_{r}z_{r}(x,\phi _{t}),
\label{talbot28}
\end{equation}
where 
\begin{equation}
z_{r}(x,\phi _{t})=\exp \left[ i\left( kx+2\pi r\right) ^{2}/4\phi _{t}%
\right] .  \eqnum{49a}  \label{talbot28a}
\end{equation}
At distances $y=L_{T}/n$ one finds
\end{multicols}
\begin{equation}
z_{r}(x,2\pi /n)=\exp \left[ in\left( kx\right) ^{2}/8\pi \right] \left\{ 
\begin{array}{cc}
\exp \left[ inkx\left( q+\frac{1}{2}\right) +in\frac{\pi }{2}\right] , & 
\text{for }r=2q+1, \\ 
\exp \left( inkxq\right) , & \text{for }r=2q,
\end{array}
\right.   \label{talbot29}
\end{equation}
\begin{multicols}{2}
where $q$ is integer. Substituting this expression into Eq. (\ref{talbot28})
and summing over $r$, one arrives again at Eq. (\ref{talbot24}), but with an
alternative expression for $a_{s}(n)$: 
\begin{equation}
a_{s}(n)=\frac{1}{\sqrt{2in}}\left[ 1+\left( -1\right) ^{s}e^{in\pi /2}%
\right] e^{i\pi s^{2}/2n}.  \label{talbot30}
\end{equation}

\subsection{Higher-order gratings using the Talbot effect}

One can conclude from Eq. (\ref{talbot25}) that, owing to matter-wave
interference, the transmission function $\eta (x)$ imprinted on the atomic
wave function by the MS can be copied $n$ times in the plane $y=L_T/n,$ with
each copy separated by $d/n$. This effect occurs for arbitrary transmission
functions and can be used to generate higher order atomic gratings.

Since the $a_s(n)$ appearing in Eq. (\ref{talbot25}) are not necessarily
equal, the wave function (\ref{talbot25}) is periodic with period $d$, but
not necessarily with period $d_g<d.$ Moreover, it is possible for the
different copies corresponding to different $s$ to overlap. We refer to a
pure, higher order atomic grating as one in which the different grating
images do not overlap and for which $d_g<d$. When the width $fd$ of the
slits in the MS is smaller than spacing $d/n,$ different terms in the
wave-function (\ref{talbot25}) do not overlap with one another and one finds
for the atomic density (\ref{talbot9}) 
\begin{equation}
f(x,t)=\sum_{s=0}^{m-1}\left| a_s\left( n\right) \right| ^2\left| \eta
\left( x-s\frac dn\right) \right| ^2.  \label{talbot31}
\end{equation}

Pure higher order atomic gratings are produced only if the nonvanishing $%
\left| a_s(n)\right| $ are equal. From Eq. (\ref{talbot30}) one sees that 
\begin{equation}
\left| a_s(n)\right| =\sqrt{\frac 2n}\left\{ 
\begin{array}{l}
\left| \cos \left( n\pi /4\right) \right| ,\text{ for even }s \\ 
\left| \sin \left( n\pi /4\right) \right| ,\text{ for odd }s
\end{array}
\right. .  \label{talbot32}
\end{equation}
Three different situations can be distinguished 
\begin{equation}
\begin{array}{l}
1.\ n=2m+1 \\ 
2.\ n=2(2m+1) \\ 
3.\ n=4m
\end{array}
\label{talbot33}
\end{equation}
for integers $m\geq 0$.

In the first case $\left| a_s(n)\right| =\frac 1{\sqrt{n}}$, independent of $%
s$. At distances $y=L_T/3,\ y=L_T/5,\ \ldots $, pure, higher order atomic
gratings having periods $d_g=d/3,\ d/5\ \ldots $ are produced.

In the second case $\left| a_s(n)\right| =0$ for $s$ even and $\left|
a_s(n)\right| =\sqrt{\frac 2n}$, independent of $s$, for $s$ odd. In the
plane $y=L_T/2$ the atomic grating is shifted by a half-period $d/2$ from
the initial grating (as found above); in the plane $y=L_T/6$ only terms
located at $x=\frac d6,\ \frac d2,\ \frac{5d}6$ in Eq. (\ref{talbot31})
contribute to the sum (for $0\leq x<d$), corresponding to an atomic grating
having period $d_g=d/3,$ that is shifted by a distance $d/6$ from the
initial grating$.$ In general in the focal plane $y=L_T/[2\left( 2m+1\right)
],$ one finds an atomic grating having period $d_g=d/\left( 2m+1\right) $
which is shifted by a distance $d/[2\left( 2m+1\right) ]$ from the initial
grating$.$
\begin{figure}
\begin{minipage}{0.99\linewidth}
\begin{center}
\epsfxsize=.95\linewidth \epsfysize=\epsfxsize \epsfbox{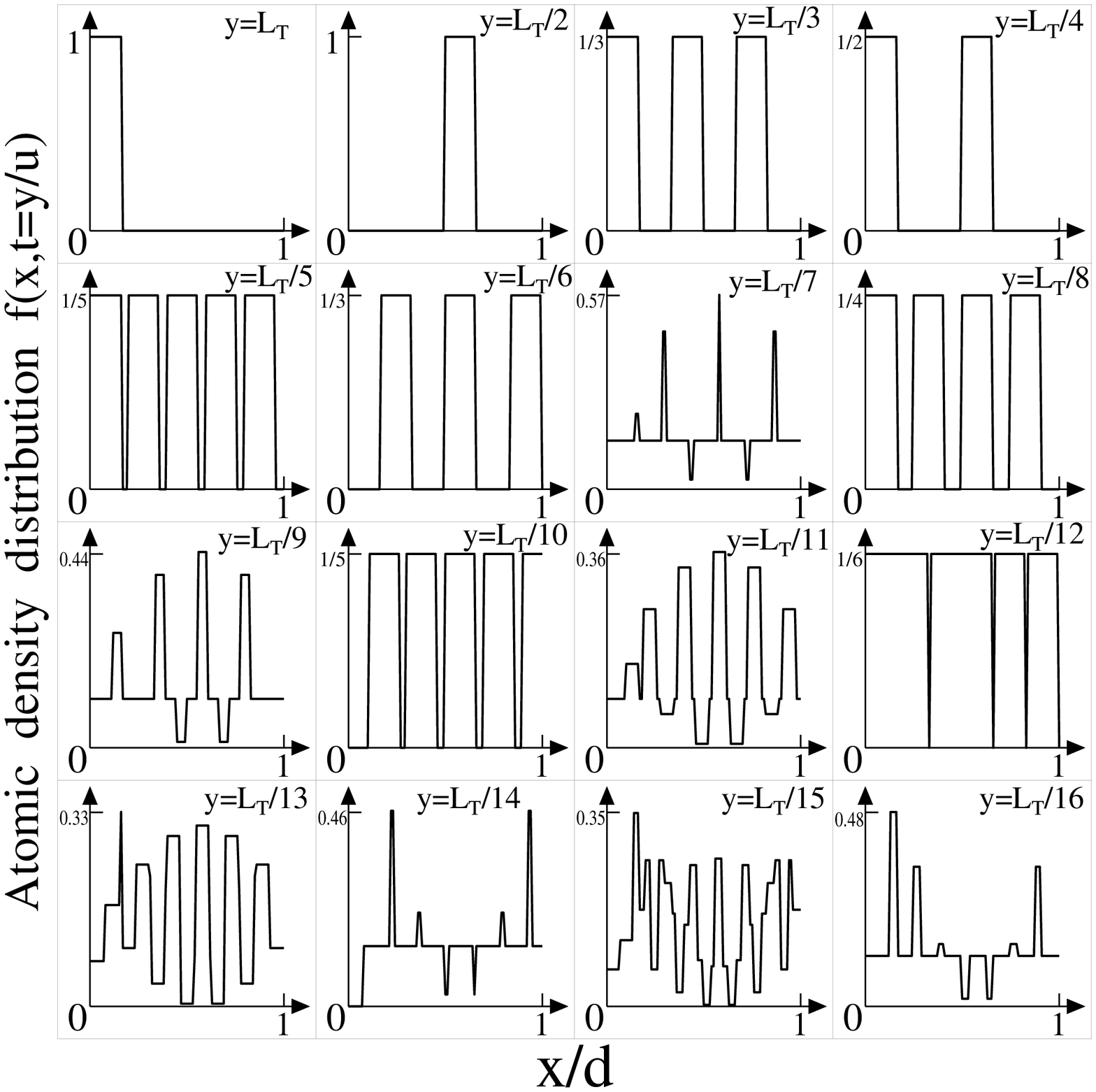}
\end{center}
\end{minipage}
\begin{minipage}{0.99\linewidth} \caption{One period of the atomic 
density spatial modulation in the planes
located at the distances $y=L_{T}/m$ ($m=1,$ $\ldots $ $16$). The case of
the atomic beam modulation with microfabricated structure having
transmission step-function (\ref{talbot35}) with relative width $f=0.16$ is
shown$.$ While expected period $d_{g}$ of the beam self-image greater than
slits' width $\protect\sigma =df,$ one obtains higher-order spatial
gratings.  
\label{talbotf1}}
\end{minipage}
\end{figure}

In the third case $\left| a_s(n)\right| =\sqrt{\frac 2n}$, independent of $s$%
, for $s$ even and $\left| a_s(n)\right| =0$ for $s$ odd. In the planes $%
y=L_T/4,\ y=L_T/8\ \ldots $, one finds atomic gratings having periods $%
d_g=d/2,\ d/4\ \ldots $.

The atomic density profile can no longer be written in the form (\ref
{talbot31}) when 
\begin{equation}
fd>d_g.  \label{talbot34}
\end{equation}
In this limit, different components in the wave-function (\ref{talbot25})
overlap and can interfere with one another in forming the atomic density.
Even though the atomic distribution function can still contain narrow peaks
having a size of order of $d_g,$ the amplitudes of the peaks are not equal,
and the period of the overall diffraction pattern reverts to the period $d$
of the initial grating.

These different regimes are illustrated in Fig. \ref{talbotf1} plotted for
an amplitude transmission function defined in the interval $0\leq x<d$ as 
\begin{equation}
\eta (x)=\left\{ 
\begin{array}{c}
1,\text{ for }0\leq x\leq df \\ 
0,\text{ for }df<x<d
\end{array}
\right. .  \label{talbot35}
\end{equation}
The Talbot effect enables one to create pure, higher-order atomic gratings
having periods that are limited only by the slit widths in the MS. 

\section{Shadow effect with microfabricated structures}

In the previous section, it was assumed that the angular divergence $\theta
_b$ of the incident beam was less than $d/L_T$. If this inequality is not
satisfied, the diffraction patterns associated with different velocity
subgroups in the incident atomic beam result in a washing out of the overall
diffraction pattern. For typical beam parameters, this condition restricts $%
\theta _b$ to be less than $10^{-5}-10^{-4}$ rad$.$ The restriction on $%
\theta _b$ is a limiting factor on the maximum flux of the atomic beam. It
is possible to avoid this restriction and increase the atomic flux if
echo-like techniques are used.

Using echo techniques that are analogous to those encountered in the study
of coherent transients, one can observe matter-wave interference in beams
having a large angular divergence (Dubetsky et al., 1984). It turns out,
however, that the dephasing and rephasing of the atomic gratings which
occurs in such schemes does not depend in any critical manner on
quantization of the atomic center-of-mass motion. In other words, the
dephasing-rephasing mechanism is the same whether or not $L\sim L_T$ (Talbot
effect) or $L\ll L_T$ (classical limit). As such, it makes sense to consider
the limit of classical scattering first, since the analysis is easier and a
simple geometric interpretation can be given to the results (Dubetsky and
Berman, 1994). Thus, we consider the limit $L\ll L_T$ in this section and
defer a discussion of the case $L\sim $ $L_T$ until Sec. V.

In this and the following section we consider the interaction of an atomic
beam with two MS, separated by a distance $L$. The angular divergence of the
incident beam is sufficiently large to satisfy the inequality 
\begin{equation}
\theta _b\gg d/L.  \label{shdslt48}
\end{equation}
The first MS produces a sum of atomic gratings, one for each velocity
subgroup in the initial atomic beam. Immediately following the MS, these
gratings overlap and mirror the transmission function of the grating, but
downstream from the MS, they dephase relative to one another. As a result,
the macroscopic atomic grating is washed out at a distance

\begin{equation}
\tilde{\ell}\sim d/\theta _b\ll L  \label{shdslt49}
\end{equation}
from the MS.

\begin{figure}
\begin{minipage}{0.99\linewidth}
\begin{center}
\epsfxsize=.95\linewidth \epsfysize=\epsfxsize \epsfbox{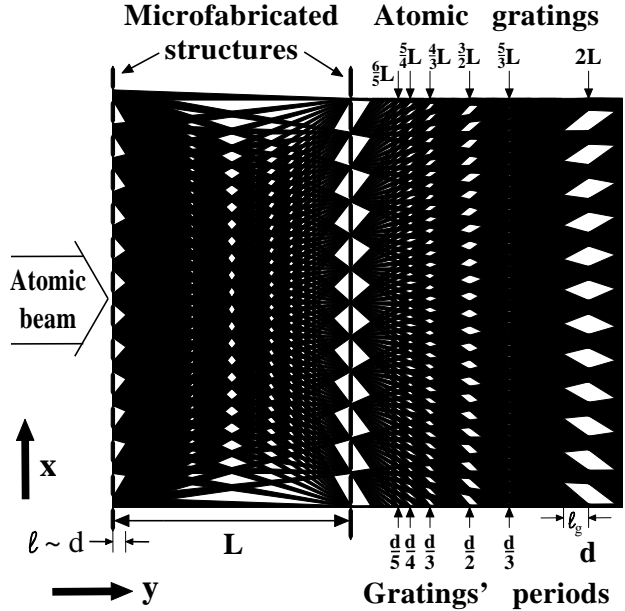}
\end{center}
\end{minipage}
\begin{minipage}{0.99\linewidth} \caption{Shadow effect, geometric 
simulation. Atomic beam having large
angular divergence $\left( \protect\theta _{b}=120^{\circ }\right) $ passes
through two separated MSs having the same periods $d_{1}=d_{2}=d$, duty
cycles $f_{1}=0.5$ and $f_{2}=0.05$ and separated on the distance $L.$
Grating carved by first MS is washed out on the distance $\ell \sim d.$
Other gratings between MSs ($0<y<L$) are fictive, they appeared because only
trajectories for atoms passed through both MSs are shown. Gratings after the
second MS ($y>L$) are real, they appear owing to echo effect. 
\label{sltshf12}}
\end{minipage}
\end{figure}

Although the macroscopic grating produced by the first MS washes out in a
distance of order $\tilde{\ell}\ll L$, it is possible for the second MS to
lead to a restoration of the atomic gratings. For particles moving on
classical trajectories, we refer to this process as a {\em shadow effect}
since it can be interpreted completely by the ''shadow'' of the incident
beam formed by the two MS (Chebotayev et al., 1985) (see Fig. \ref{sltshf12}%
). 
For a beam having large angular divergence $\left( \theta _{b}\sim 1\right) $%
, the initial grating produced by the first MS washes out in a distance
comparable with the MS's period, in accordance with Eq. (\ref{shdslt49}).
After passing through the second MS, however, macroscopic gratings reappear
in specific focal planes. A grating having the same period as the MS is
focused in the focal plane $y=2L$, while higher-order gratings having
periods $d/n$ (for integer $n$) are focused at other locations (to be
determined below). The shadow effect can also be demonstrated easily using
incoherent light (Chebotayev, 1986).

Although the shadow effect occurs for classical particles, it can be
interpreted in terms of a dephasing and rephasing of atomic gratings. The
relevant phases are the Doppler phases associated with various Fourier
components of the atomic density, as discussed in the Introduction. In such
a picture, the final image on the screen depends on a cancellation of
Doppler phases in the spatial regions $0\rightarrow L$ and $L\rightarrow 2L$%
, for example. In other words, the signal is sensitive to the relative
Doppler phases in two spatial regions and is a measure of this relative
phase. Insofar as interferometers are measures of relative phase, the
echo-like rephasing of the atomic gratings can be viewed as a manifestation
of atom interferometry. On the other hand, this rephasing is {\em not}
related to the wave nature of matter. A shadow effect interferometer of this
type was used by Batelaan et al. $\left( 1996\right) $ to measure the
displacement of atomic gratings produced by rotation and by gravity.

The same type of Doppler dephasing and rephasing that occurs using MS can
also occur when atoms interact with two or more nearly resonant standing
wave fields (Baklanov et al., 1976; Barger et. al., 1979; Dubetsky, 1976;
Chebotayev, 1978; Chebotayev et al., 1978a, LeGou\"{e}t and Berman, 1979;
Mossberg et al., 1979; Dubetsky and Semibalamut, 1982; Bord\'{e}, 1989;
Dubetsky and Berman, 1994). When standing wave optical fields are used for
modulation of the atomic spatial distribution, the atomic gratings often are
monitored by applying a probe pulse in the focal planes that transfers the
phase associated with an atomic state population to one associated with an
atomic coherence. Atom interferometers of this type have been used for
precision measurements of gravitational (Kasevich and Chu, 1991) and
inertial (Riehle et al., 1991) phenomena [for a review, see (M\"{u}ller et
al., 1995)]. In these cases, external fields give rise to a displacement of
the atomic gratings.

\subsection{Dephasing-rephasing processes using two spatially separated MS}

Before calculating the particles' distribution function, we derive some
general properties of grating formation. In this subsection, it is
convenient to make a Fourier decomposition of the atomic density profile in
the $x$-direction. The propagation of each of the Fourier components is then
treated separately.

Consider the case when the two MS $\left( 1\text{ and }2\right) $ have
periods $d_1$ and $d_2$ and are separated from one another by a distance $L.$
A MS forms a periodic spatial distribution (shadow) which is the same for
all atomic velocity subgroups just after passing through the MS. The profile
created by the first MS contains a sum of harmonics in the $x-$direction
having spatial periods 
\begin{equation}
d_{m_1}=d/\left| m_1\right| ,  \label{shdslt50}
\end{equation}
where $m_1$ is an integer. Immediately following the MS, the $m_1$th spatial
harmonic varies as $cos(m_1k_1x)$, where $k_1=2\pi /d_1$ is the wave-number
associated with the first MS. As the atoms move downstream from the first
MS, the $m_1$th harmonic acquires a Doppler phase (\ref{intr7a}) given by\ 
\begin{equation}
\phi _{m_1}(t)=m_1k_1vt,  \label{shdslt51}
\end{equation}
where $v$ is the x-component of atomic velocity and $t=y/u$ as before. For a
time 
\begin{equation}
t_d\sim 1/(k_1v)  \label{shdslt52}
\end{equation}
the Doppler phases becomes large, $\phi _{m_1}(t)\gtrsim 1,$ and the
macroscopic grating washes out on averaging over $v.$ Since $v\sim u\theta
_b,$ the time $t_d$ (\ref{shdslt52}) corresponds to the distance $\tilde{\ell%
}$ (\ref{shdslt49}).

\begin{figure}
\begin{minipage}{0.99\linewidth}
\begin{center}
\epsfxsize=.95\linewidth \epsfysize=.93\epsfxsize \epsfbox{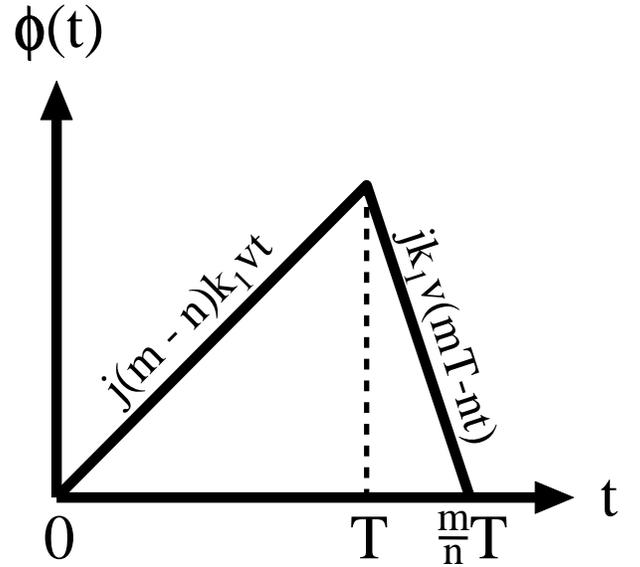}
\end{center}
\end{minipage}
\begin{minipage}{0.99\linewidth} \caption{A dephasing-rephasing process responsible for the atomic grating
focusing at the focal plane $y=\frac{m}{n}L$ produced by two MSs having
periods $d_{1}$ and $d_{2}$ and associated wave numbers $k_{1}$ and $k_{2}$,
such as $jk_{1}=\ell k_{2},$ where $j$ and $\ell $ are integrals. When
atomic beam passes through first MS a number of harmonics having wave
numbers $m_{1}k_{1}$ (harmonics $m_{1}k_{1}$) are induced. Mask, consisting
of harmonics $m_{2}k_{2},$ is superimposed on the atomic spatial profile
when beam passes through second MS. Dephasing of the relevant harmonic $%
j(m-n)k_{1}$ occurs between MSs. Second MS starts rephasing process for
harmonic $j(m-n)k_{1}-\ell mk_{2}=-jnk_{1}$ which leads to the phase
cancellation at the echo-point $t=\frac{m}{n}T,$ where macroscopic atomic
grating is focused. Other harmonics contributing to this grating must have
wave numbers $q$-times larger, where $q$ is integral. Their phase diagrams
differ from plotted only by scaling the phase in $q-$times. 
\label{sltshf13}}
\end{minipage}
\end{figure}

The atoms pass through the second MS at time $T=L/u$ ($y=L$). Downstream
from the second MS, each spatial harmonic acquires an additional phase
(i.e., a phase in addition to $\phi _{m_{1}}(t)$, which, itself, continues
to increase following the second MS) 
\begin{equation}
\phi _{m_{2}}(t)=m_{2}k_{2}v(t-T),  \label{shdslt53}
\end{equation}
where $m_{2}$ is another integer. Since the mask created by the second MS is
superimposed on the shadow from the first MS, the resulting shadow consists
of harmonics having wave numbers 
\begin{equation}
k_{h}=\left| m_{1}k_{1}+m_{2}k_{2}\right|  \label{shdslt57}
\end{equation}
and velocity-dependent Doppler phases 
\begin{equation}
\phi \left( t\right) =\phi _{m_{1}}\left( t\right) +\phi _{m_{2}}\left(
t\right) .  \label{shdslt58}
\end{equation}
The two phases in the rhs of this equation can cancel one another at the
so-called echo time $t_{e}$ defined by 
\begin{equation}
\phi \left( t_{e}\right) =0,  \label{shdslt54}
\end{equation}
corresponding to a focal plane $y_{e}=ut_{e}$. At such times, one produces
an harmonic in the atomic density that is independent of $v$; as a
consequence this grating survives any averaging over the velocity
distribution in the incident beam. From Eq. (\ref{shdslt54}) one sees
gratings are focused when 
\begin{equation}
t_{e}/T=y_{e}/L=\frac{1.}{1+(m_{1}/m_{2})(k_{1}/k_{2})}  \label{shdslt55}
\end{equation}
The dephasing-rephasing process is illustrated in Fig. \ref{sltshf13}. 
[To satisfy Eq. (\ref{shdslt54}), the integers $m_{1}$ and $m_{2}$ must have
opposite signs (for the sake of illustration, we take $m_{1}$ as negative in
the examples given below)].

We assume that the ratio of the MS's periods is rational, 
\begin{equation}
d_1/d_2=k_2/k_1=j/\ell ,  \label{shdslt56}
\end{equation}
where $j$ and $\ell $ are the {\em smallest} positive integers which can be
used to satisfy this equation. In this case an infinite number of harmonics
associated with pairs $\left( m_1,m_2\right) ,$ having the same ratio $\frac{%
m_1}{m_2},$ contribute at a given focal plane. It follows from Eqs. (\ref
{shdslt55}, \ref{shdslt56}) that for 
\begin{equation}
\frac{m_1}{m_2}=-(j/\ell )\left( 1-\frac nm\right) ,  \label{shdslt12}
\end{equation}
where $m$ and $n$ are positive integers having no common factors with $m>n,$
gratings in the atomic spatial distribution appear at echo times 
\begin{equation}
t_e=\frac mnT  \label{shdslt59}
\end{equation}
or at focal planes located at 
\begin{equation}
y_e=\frac mnL.  \label{shdslt11}
\end{equation}
From Eq. (\ref{shdslt57}), one finds that harmonics having 
\begin{equation}
k_h=\frac{\left| m_1\right| k_1}{\frac mn-1}  \label{shdslt100}
\end{equation}
are focused in this plane.

For example, consider the limiting case in which $d_1=d_2\equiv d$ ($j=\ell
=1$), analogous to the situation studied by Dubetsky and Berman $\left(
1994\right) $. In the plane $y=2L$ ($m/n=2$), all harmonics having $%
m_2/m_1=-2$ (i.e. $\{m_1,m_2\}=\{-1,2\};\{-2,4\};\{-3,6\};$etc) are focused.
As a result [see Eq. (\ref{shdslt57})], harmonics having $k_h=k,2k,3k,$etc,
are focused in the plane $y=2L$. The period of this atomic grating $d_g$
corresponds to the smallest value of $k_h,$ namely $d_g=2\pi /k=d$.
Similarly, in the plane $y=3L/2$ ($m/n=3/2$), all harmonics having $%
m_2/m_1=-3$ (i.e. $\{m_1,m_2\}=\{-1,3\};\{-2,6\};\{-3,9\};$etc) are focused.
As a result [see Eq. (\ref{shdslt57})], harmonics having $k_h=2k,4k,6k,$etc,
are focused in the plane $y=(3/2)L$. The period of this atomic grating is $%
d_g=2\pi /2k=d/2$. For $m=(n+1)$, one finds that atomic gratings having
period $d_g=d/n$ are focused in the plane $y=[(n+1)/n]L$.

To treat the case of arbitrary, rational $d_1/d_2=$ $j/\ell $, we set 
\begin{equation}
m_1=-\bar{j}(m-n)q,\ m_2=\bar{\ell}mq,  \label{shdslt12p}
\end{equation}
where $q$ is an integer, 
\begin{equation}
\bar{j}=j/\mu ,\hspace{1.0in}\bar{\ell}=\ell /\mu ,  \label{shdslt101}
\end{equation}
and $\mu $ is the largest common factor of $j(m-n)$ and $\ell m$. Harmonics
having wave numbers (\ref{shdslt100}) 
\begin{equation}
k_h=n\bar{j}k_1\left| q\right| ,  \label{shdslt60}
\end{equation}
are focused in the plane $y_e=\frac mnL$. The minimum possible wave number 
\begin{equation}
k_g=n\bar{j}k_1  \label{shdslt62}
\end{equation}
determines the period of the focused grating 
\begin{equation}
d_g=\frac{2\pi }{k_g}=\frac{d_1}{\bar{j}n}.  \label{shdslt61}
\end{equation}

One concludes that it is possible to create a higher order atomic grating,
having a period that is $\bar{j}n-$times smaller than that of the first MS,
by passing an atomic beam having a large angular divergence through two MS.
Although both the Talbot and shadow effects lead to higher order atomic
gratings, there is a qualitative difference between the two cases. In the
Talbot effect, the structure of the MS is copied $n$ times in the image
plane $y=L_T/n$, giving rise to a profile having period $d_g=d_1/n$ or $%
2d_1/n$, {\em provided that} $d_g>f_1d_1$. The minimum period is determined
by the slit width. In contrast, the period of the atomic grating produced by
the shadow effect in the plane $y_e=\frac mnL$ is given by Eq. (\ref
{shdslt61}) and is not limited by the slit widths of the MS (although the
contrast is determined by the slit widths). The period of the atomic grating
is equal to $d_1/\bar{j}$ in the focal plane $y_e=2L$ and is compressed by a
factor $n$ in the plane $y_e=\frac mnL$. This compression lies at the heart
of the shadow effect's application to atomic lithography (Dubetsky and
Berman, $1994$). Atomic gratings having periods smaller than those of both
the MS and even smaller than the slit widths of the MS can be obtained. In
this respect the shadow effect has yet an additional advantage over the
Talbot-effect, where higher order grating production is not accompanied by
compression.

Before proceeding to calculate the atomic density distribution, we should
like to estimate the depth of focus of the various gratings. The distances
between focal planes are comparable with the distance $L$ between the MS$.$
One can estimate the depth of focus $\tilde{\ell}_g$ from the requirement
that the phase (\ref{shdslt58}) be smaller than unity in the region of the
focal plane. For $m_i$ given by (\ref{shdslt12p}) one finds 
\begin{equation}
\phi \left( t\right) =qk_gv\delta y/u,  \label{shdslt63}
\end{equation}
where $\delta y=(y-\frac mnL)=(y-y_e)$ in the neighborhood of the focal
plane at $y=\frac mnL.$ Setting $q=1,\,\,\phi (t)\sim 1$ and $\delta y=%
\tilde{\ell}_g$, and using Eq. (\ref{shdslt62}) and the fact that $\theta
_b\sim v/u$, one obtains 
\begin{equation}
\tilde{\ell}_g\sim \frac{\tilde{\ell}}{n\bar{j}}\leq \tilde{\ell},
\label{shdslt64}
\end{equation}
where $\tilde{\ell}$ is given by (\ref{shdslt49}) with $d=d_1$. Since $L\gg 
\tilde{\ell}$ has been assumed, it is possible to separate the various
gratings. The sharpening of depth of focus of the higher order gratings
predicted by Eq. (\ref{shdslt64}) is in qualitative agreement with the
results shown in Fig. \ref{sltshf12}.

\subsection{Particles' distribution profile}

We now turn our attention to a calculation of the atomic density profile. In
contrast to the Talbot effect, it is not possible to find self-imaging using
the shadow effect, since the scattering coefficients for the different
spatial harmonics are not the same. Let the transmission functions for the
two MS be denoted by $\chi _s(x)$ ($s=1$ or $2$) [$\chi _s(x)$ is a
transmission function for atomic density, while $\eta _s(x)$ is a
transmission function for atomic state amplitudes - for MS having
transmission of either $1$ or $0$, these functions are identical]. Atoms are
scattered by the MS in the planes $y=0$ and $y=L$, or, equivalently, at
times $T_1=0$ and $T_2=T=L/u$. Calculations are carried out using $t=y/u$ as
a variable. In some sense, this corresponds to working in the atomic rest
frame. As a result of scattering, the atomic density is modified as 
\begin{equation}
f(x,v,T_s^{+})=\chi _s(x)f(x,v,T_s^{-}),  \label{shdslt1}
\end{equation}
where $T_s^{\pm }$ are times just after or before a scattering event.
Following the scattering event, the distribution evolves as 
\begin{equation}
f(x,v,t)=f[x-v(t-T_s),v,T_s^{+}].  \label{shdslt2}
\end{equation}

We assume that, for $t<0,$ the atoms are distributed homogeneously in the
transverse direction, i.e. 
\begin{equation}
\left. f(x,v,t)\right| _{t<0}=1.  \label{shdslt3}
\end{equation}
The assumption of a homogeneous velocity distribution is consistent with a
beam having angular divergence (\ref{shdslt48}), since in this limit, the
transverse velocity distribution is approximately constant over the range $%
d_s/L$. The spatial distribution of the atomic density for $t>T$ is given by 
\begin{equation}
f(x,t)=\left\langle \chi _1(x-vt)\chi _2[x-v(t-T)]\right\rangle ,
\label{shdslt5}
\end{equation}
where $\left\langle \ldots \right\rangle $ represents an average over
velocities. Expanding $\chi _s\left( x\right) $ in a Fourier series 
\begin{equation}
\chi _s(x)=\sum_m\chi _m^{(s)}e^{imk_sx},  \label{shdslt6}
\end{equation}
where $k_s=2\pi /d_s$ is the ''wave-number'' of structure $s$, one finds 
\begin{equation}
f(x,t)=\sum_{m_1,\ m_2}\chi _{m_1}^{(1)}\chi _{m_2}^{(2)}\exp
[i(k_1m_1+k_2m_2)x]\left\langle \exp \left[ -i\phi (t)\right] \right\rangle ,
\label{shdslt7}
\end{equation}
where $\phi \left( t\right) $, as defined by Eq. (\ref{shdslt58}), is also a
function of $m$ and $n$. Owing to condition (\ref{shdslt48}), for $t\sim
T\sim L/u$ and $v\sim u\theta _b$, the phase factor in the brackets of Eq. (%
\ref{shdslt7}) is large, of order $L\theta _b/d_s\gg 1.$ On averaging over
velocities, one finds a nonvanishing contribution only at the particular
focal planes or echo-times given by Eq. (\ref{shdslt54}).

Retaining contributions from only those $m_i$ given by (\ref{shdslt12p})
corresponding to the various focal planes, one finds from Eq. (\ref{shdslt7}%
) that the atomic density in the focal planes is given by 
\begin{equation}
f(x,t_e)=\sum_q\chi _{\bar{j}(n-m)q}^{(1)}\chi _{\bar{\ell}mq}^{(2)}\exp
(iqk_gx),  \label{shdslt13}
\end{equation}
where $k_g$ is the wave number of the focused atomic grating given in Eq. (%
\ref{shdslt62})$.$ Thus, owing to the shadow effect, at the echo-time (\ref
{shdslt59}) an atomic grating is focused having period 
\begin{equation}
d_g=2\pi /k_g=d_1/(\bar{j}n).  \label{shdslt14}
\end{equation}
Note that the period of this grating is $n\bar{j}$ times smaller than the
period of the first microfabricated structure.

To reexpress the density function at the focal planes in terms of the
transmission functions, we write the Fourier harmonic amplitude $\chi
_s^{(j)}$ as 
\begin{equation}
\chi _s^{(j)}=\int_o^{d_j}\frac{dx}{d_j}\chi _j(x)\exp (-ik_jsx),
\label{shdslt15}
\end{equation}
from which one finds the atomic density at the focal planes given by
\end{multicols}
\begin{equation}
f(x,t_e)=\int_0^{d_1}\int_0^{d_2}\frac{dx_1dx_2}{d_1d_2}\sum_q\exp \left\{ iq%
\left[ k_gx+\bar{j}\left( m-n\right) k_1x_1-m\bar{\ell}k_2x_2\right]
\right\} \chi _1\left( x_1\right) \chi _2\left( x_2\right) .
\label{shdslt38}
\end{equation}
The sum over $q$ leads to an infinite set of $\delta -$functions 
\begin{equation}
\sum_qe^{iq\alpha }=2\pi \sum_{s^{\prime }}\delta (\alpha -2\pi s^{\prime })
\label{shdslt16}
\end{equation}
which allows one to carry out the integration over $x_1$, for example. Only
the values 
\begin{equation}
x_1=\frac{d_1}{\bar{j}(m-n)}\left[ s^{\prime }+m\bar{\ell}\frac{x_2}{d_2}-%
\frac x{d_g}\right]  \label{shdslt39}
\end{equation}
in the range $\left[ 0,d_1\right) $ contribute, resulting in the inequality 
\begin{equation}
\frac x{d_g}-m\bar{\ell}\frac{x_2}{d_2}\leq s^{\prime }<\frac x{d_g}-m\bar{%
\ell}\frac{x_2}{d_2}+\bar{j}(m-n)  \label{shdslt40}
\end{equation}
for integer $s^{\prime }.$ One finds that a finite number of terms
contribute, having 
\begin{equation}
s^{\prime }=s-\left[ m\bar{\ell}\frac{x_2}{d_2}-\frac x{d_g}\right] _I,
\label{shdslt41}
\end{equation}
where $s=0,\ 1,\ \ldots \ \bar{j}(m-n)-1$. In Eq. (\ref{shdslt41}) and for
the remainder of this chapter, a notation is adopted, in which we set $A=%
\left[ A\right] _I+\left\{ A\right\} _F,$ where $\left[ A\right] _I$ and $%
\left\{ A\right\} _F$ are the integral and fractional parts of $A$,
respectively. Using Eqs. (\ref{shdslt39}) and (\ref{shdslt41}), we obtain 
\begin{equation}
f(x,t_e)=\frac 1{\bar{j}(m-n)}\sum_{s=0}^{\bar{j}(m-n)-1}\int_0^{d_2}\frac{%
dx_2}{d_2}\chi _2\left( x_2\right) \chi _1\left[ \frac{d_1}{\bar{j}(m-n)}%
\left( s+\left\{ m\bar{\ell}\frac{x_2}{d_2}-\frac x{d_g}\right\} _F\right) %
\right] .  \label{shdslt42}
\end{equation}

Consider in detail the case when the microfabricated structures have duty
cycles (ratio of slit openings to periods) $f_{j}$ and 
\begin{equation}
\chi _{j}(x)=\left\{ 
\begin{array}{l}
1,\ \text{for }\left\{ \frac{x}{d_{j}}\right\} _{F}<f_{j} \\ 
0,\ \text{for }\left\{ \frac{x}{d_{j}}\right\} _{F}>f_{j}
\end{array}
\right. .  \label{shdslt21p}
\end{equation}
Introducing dimensionless variables 
\begin{equation}
w=x/d_{g},\ z=x_{2}/(d_{2}f_{2})  \label{shdslt22}
\end{equation}
and taking into account that the argument of function $\chi _{1}$ in Eq. (%
\ref{shdslt42}) is positive, one obtains 
\begin{equation}
f(x,t_{e})=\frac{f_{2}}{\bar{j}\left( m-n\right) }\sum_{s=0}^{\left[ \beta 
\right] _{I}}h_{s}(w),  \label{shdslt23}
\end{equation}
where 
\begin{equation}
h_{s}(w)=\int_{0}^{1}dz\theta \left[ \beta -\left( s+\left\{ \alpha
z-w\right\} _{F}\right) \right] ,  \eqnum{96a}  \label{shdslt23a}
\end{equation}
\begin{equation}
\alpha =m\bar{\ell}f_{2},\ \beta =\bar{j}\left( m-n\right) f_{1}, 
\eqnum{96b}  \label{shdslt23b}
\end{equation}
and $\theta (x)=\left\{ 
\begin{array}{c}
1,\ for\,\,x>0 \\ 
0,\ for\,\,x<0
\end{array}
\right. $ is the Heaviside step function. It is
sufficient to consider only the range 
\begin{equation}
0\leq w\leq 1.  \label{shdslt47}
\end{equation}
For $0\leq s\leq \lbrack \beta ]_{I}-1$, the integrand in Eq. (\ref
{shdslt23a}) is equal to unity. Therefore, 
\begin{equation}
f(x,t_{e})=\frac{f_{2}}{\bar{j}\left( m-n\right) }\left[ [\beta
]_{I}+h_{[\beta ]_{I}}(w)\right] ,  \label{shdslt24}
\end{equation}
and one needs to evaluate the expression (\ref{shdslt23a}) only for $%
s=[\beta ]_{I},$%
\begin{equation}
h_{\left[ \beta \right] _{I}}(w)=\int_{0}^{1}dz\theta \left( \left\{ \beta
\right\} _{F}-\left\{ \alpha z-w\right\} _{F}\right) .  \label{shdslt25}
\end{equation}
The first term in the Eq. (\ref{shdslt24}) brackets is independent of $w=%
\frac{x}{d_{g}};$ consequently, it is only the second term which corresponds
to the atomic gratings.

A method for evaluating the integral (\ref{shdslt25}) is given in the
Appendix. Using this method one finds 
\begin{equation}
h_{\left[ \beta \right] _I}\left( w\right) =\left[ \left\{ \beta \right\} _F%
\left[ \alpha \right] _I+S\left( w\right) \right] /\alpha ,
\label{shdslt32p}
\end{equation}
where the function $S\left( w\right) $ is given by

\begin{equation}
S\left( w\right) =\left\{ 
\begin{array}{ll}
\left\{ \alpha \right\} _{F}-w, & \text{for }0\leq w\leq 1-\left\{ \beta
\right\} _{F} \\ 
\left\{ \beta \right\} _{F}+\left\{ \alpha \right\} _{F}-1, & \text{for }%
1-\left\{ \beta \right\} _{F}\leq w\leq \left\{ \alpha \right\} _{F} \\ 
\left\{ \beta \right\} _{F}+w-1, & \text{for }\left\{ \alpha \right\}
_{F}\leq w\leq 1+\left\{ \alpha \right\} _{F}-\left\{ \beta \right\} _{F} \\ 
\left\{ \alpha \right\} _{F}, & \text{for }1+\left\{ \alpha \right\}
_{F}-\left\{ \beta \right\} _{F}\leq w\leq 1
\end{array}
\right\} ,\text{ if }\left\{ \beta \right\} _{F}\geq \max \left( \left\{
\alpha \right\} _{F},1-\left\{ \alpha \right\} _{F}\right) ,
\label{shdslt32}
\end{equation}
\begin{equation}
S\left( w\right) =\left\{ 
\begin{array}{ll}
\left\{ \alpha \right\} _{F}-w, & \text{for }0\leq w\leq \left\{ \alpha
\right\} _{F} \\ 
0, & \text{for }\left\{ \alpha \right\} _{F}\leq w\leq 1-\left\{ \beta
\right\} _{F} \\ 
\left\{ \beta \right\} _{F}+w-1, & \text{for }1-\left\{ \beta \right\}
_{F}\leq w\leq 1+\left\{ \alpha \right\} _{F}-\left\{ \beta \right\} _{F} \\ 
\left\{ \alpha \right\} _{F}, & \text{for }1+\left\{ \alpha \right\}
_{F}-\left\{ \beta \right\} _{F}\leq w\leq 1
\end{array}
\right\} ,\text{ if }\left\{ a\right\} _{F}\leq \left\{ \beta \right\}
_{F}\leq 1-\left\{ a\right\} _{F},  \eqnum{101a}  \label{shdslt32a}
\end{equation}
\begin{equation}
S\left( w\right) =\left\{ 
\begin{array}{ll}
\left\{ \beta \right\} _{F}, & \text{for }0\leq w\leq \left\{ \alpha
\right\} _{F}-\left\{ \beta \right\} _{F} \\ 
\left\{ \alpha \right\} _{F}-w, & \text{for }\left\{ \alpha \right\}
_{F}-\left\{ \beta \right\} _{F}\leq w\leq 1-\left\{ \beta \right\} _{F} \\ 
\left\{ \beta \right\} _{F}+\left\{ \alpha \right\} _{F}-1, & \text{for }%
1-\left\{ \beta \right\} _{F}\leq w\leq \left\{ \alpha \right\} _{F} \\ 
\left\{ \beta \right\} _{F}+w-1, & \text{for }\left\{ \alpha \right\}
_{F}\leq w\leq 1
\end{array}
\right\} ,\text{ if }1-\left\{ a\right\} _{F}\leq \left\{ \beta \right\}
_{F}\leq \left\{ a\right\} _{F},  \eqnum{101b}  \label{shdslt32b}
\end{equation}
\begin{equation}
S\left( w\right) =\left\{ 
\begin{array}{ll}
\left\{ \beta \right\} _{F}, & \text{for }0\leq w\leq \left\{ \alpha
\right\} _{F}-\left\{ \beta \right\} _{F} \\ 
\left\{ \alpha \right\} _{F}-w, & \text{for }\left\{ \alpha \right\}
_{F}-\left\{ \beta \right\} _{F}\leq w\leq \left\{ \alpha \right\} _{F} \\ 
0, & \text{for }\left\{ \alpha \right\} _{F}\leq w\leq 1-\left\{ \beta
\right\} _{F} \\ 
\left\{ \beta \right\} _{F}+w-1, & \text{for }1-\left\{ \beta \right\}
_{F}\leq w\leq 1
\end{array}
\right\} ,\text{ if }\left\{ \beta \right\} _{F}\leq \min \left( \left\{
\alpha \right\} _{F},1-\left\{ \alpha \right\} _{F}\right) .  \eqnum{101c}
\label{shdslt32c}
\end{equation}
\begin{multicols}{2}
Substituting Eq. (\ref{shdslt32p}) in the rhs of Eq. (\ref{shdslt24}) one
finds that the atomic beam profile at a given focal plane is equal to 
\begin{eqnarray}
f(x,t_{e}) &=&\left[ \alpha \lbrack \beta ]_{I}+\left\{ \beta \right\}
_{F}[\alpha ]_{I}\right.   \nonumber \\
&&+\left. S\left( w=x/d_{g}\right) \right] /\left[ m(m-n)\bar{\ell}\bar{j}%
\right] .  \label{shdslt31}
\end{eqnarray}

\subsection{Main features}

All dependencies in (\ref{shdslt32}) coincide if 
\begin{equation}
\left\{ \alpha \right\} _F=\left\{ \beta \right\} _F=\ ^1/_2,
\label{shdslt33}
\end{equation}
when 
\begin{equation}
S(w)=\left| w-\frac 12\right| .  \label{shdslt33p}
\end{equation}
In this case the grating amplitude 
\begin{equation}
A=f(x,t_e)_{\max }-f(x,t_e)_{\min },  \label{shdslt34}
\end{equation}
for given $m$,$\ell ,j,n$, achieves a maximum value $A=(2m\bar{\ell}\bar{j}%
(m-n))^{-1}$. To maximize this quantity for a given grating period, one has
to choose $\ell =j=1$ and $m=n+1$ ($\bar{\ell}=\bar{j}=1$), which
corresponds to the focal planes $y=2L\ (n=1),$ $y=\ ^3/_2L\ (n=2),$ $y=\
^4/_3L\ (n=3)\ldots ,$ where gratings having periods $d_g=d_1/n$ are
focused. To satisfy condition (\ref{shdslt33}), one can choose 
\begin{equation}
f_1=1/2,\ f_2=1/[2(n+1)],  \label{shdslt35}
\end{equation}
for which 
\begin{equation}
A=1/[2(n+1)].  \label{shdslt36}
\end{equation}
The constant background term of $f(x,t_e)$ [first two terms in the numerator
of Eq. (\ref{shdslt31})] vanishes since 
\begin{equation}
\alpha =\beta =\ ^1/_2.  \label{shdslt37}
\end{equation}
\begin{figure}
\begin{minipage}{0.99\linewidth}
\begin{center}
\epsfxsize=.95\linewidth \epsfysize=\epsfxsize \epsfbox{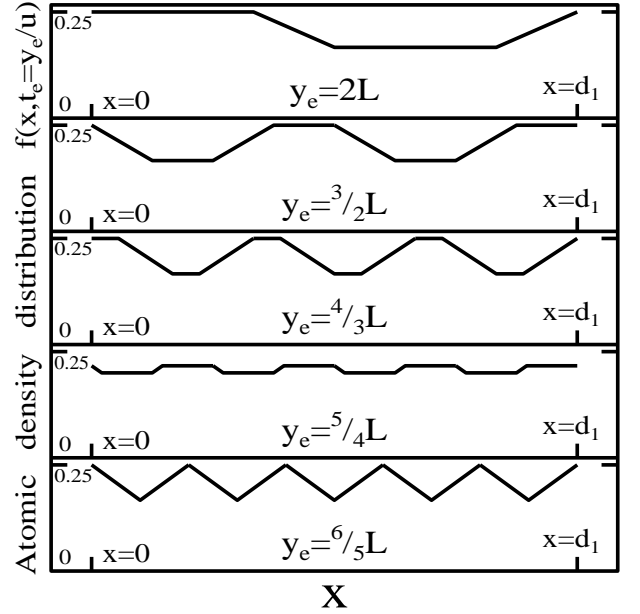}
\end{center}
\end{minipage}
\begin{minipage}{0.99\linewidth} \caption{Gratings' profiles produced 
by two MSs having equal periods $\left(
d_{1}=d_{2}\right) $ and duty cycles $f_{1}=\frac{1}{2},$ $f_{2}=\frac{5}{12}
$ at the different focal planes. Chosen case is optimum for $5-$order
grating focused at the plane $y_{e}=\frac{6}{5}L.$ This grating has
amplitude $A=\frac{1}{12}$ and twice large background term. The amplitude of
this grating is not less than amplitudes of the lower-order gratings focused
at the other planes for given duty cycles. 
\label{sltshf16}}
\end{minipage}
\end{figure}

To achieve this maximum signal, one must use slits in the second
microfabricated structure whose width is smaller than the atomic grating
period ($f_2d_2=d_1/2(n+1)\leq d_g=d_1/n$). Using the shadow-effect
technique, one can also observe atomic gratings having the same amplitude (%
\ref{shdslt36}) whose period is smaller than the slit width, but some
background term appears for these gratings. Indeed, if $f_2=\frac{2q+1}{%
2(n+1)},\ell =j=1,m=n+1$ (leading to $\left\{ \alpha \right\} _F=\,1/2,$ $%
\beta =1/2$)$,$ for positive integers $q\leq n$, then Eq. (\ref{shdslt33})
still holds and provides the grating amplitude (\ref{shdslt36}), but the
background term is $q$ times larger than the grating amplitude.

For illustration we plot in Fig. \ref{sltshf16} the grating profiles at
different focal planes $y_{e}=\frac{n+1}{n}L\,\,\left( n=1,\ldots 5\right) $
for MS having the same periods $d_{1}=d_{2}$, and duty cycles $f_{1}=\frac{1%
}{2},\,\,f_{2}=\frac{5}{12},$ such that the parameters $\alpha $ and $\beta $
are given by $\alpha =\frac{5}{12}(n+1),$ $\beta =\frac{1}{2}.$ 
This case corresponds to the $5th$ order grating at the plane $y_{e}=\frac{6%
}{5}L$ having maximum amplitude. When the amplitude of the $5th$ order
grating is optimized and $f_{2}\simeq 0.5,$ the amplitudes of the gratings
that are focused in the planes $y_{e}=\frac{n+1}{n}L\,\,\left( n=1,\ldots
4\right) $ are less than or equal to the amplitude of the $5th$ order
grating. This feature is seen in Fig. \ref{sltshf16}.

\begin{figure}
\begin{minipage}{0.99\linewidth}
\begin{center}
\epsfxsize=.95\linewidth \epsfysize=.81\epsfxsize \epsfbox{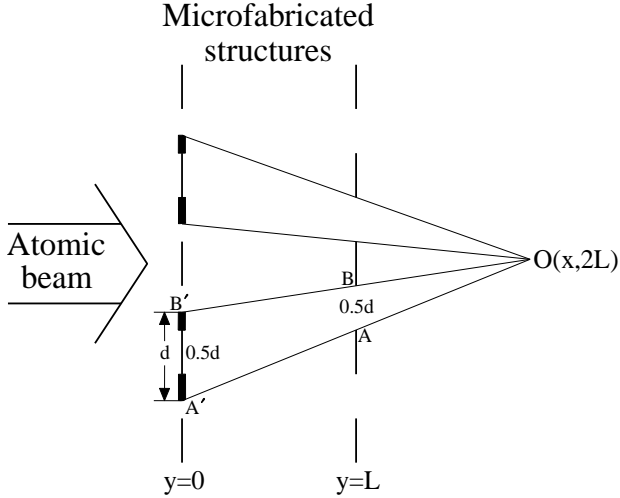}
\end{center}
\end{minipage}
\begin{minipage}{0.99\linewidth} \caption{To explain an absence of 
shadow-effect at the focal plane $y=2L$
for half-open MSs having the same periods. An illumination of an arbitrary
point $O$ at this focal plane from given slit $AB$ of the second MS is
determined by the number of particles moving into the point $O$ inside angle 
$AOB$, i. e. it is proportional to the length of the bold part of $A^{\prime
}B^{\prime },$ given by $s=\left| A^{\prime }B^{\prime }\right| -\frac{d}{2}%
. $ Since $\left| A^{\prime }B^{\prime }\right| =2\left| AB\right| =d,\,\,s$
is always equal to $\frac{d}{2}$ independently on the point $O\,\,x-$%
coordinate. Consequently, any variation of the particles' distribution at
the focal plane $y=2L$ disappears. 
\label{sltshf14}}
\end{minipage}
\end{figure}

The geometric simulation introduced above allows one to obtain the positions
of the atomic gratings. It can also be used to provide some quantitative
results. For example, one finds from Eqs. (\ref{shdslt32}, \ref{shdslt31})
that the shadow effect disappears at the focal plane $y=2L$ if both MS have
the same period $\left( d_{1}=d_{2}=d\right) $ and duty cycles, $%
f_{1}=f_{2}=0.5.$ In this case, $m=2,\,\,n=j=\ell =1,$ and, from Eq. (\ref
{shdslt23b}), one finds that $\alpha =1,\,\,\beta =0.5.$ From Eq. (\ref
{shdslt32a}) one finds that $S\left( w\right) \equiv 0$; there is no atomic
grating. The reason for the absence of the grating under these conditions is
evident from Fig. \ref{sltshf14}. 

\begin{figure}
\begin{minipage}{0.99\linewidth}
\begin{center}
\epsfxsize=.95\linewidth \epsfysize=.96\epsfxsize \epsfbox{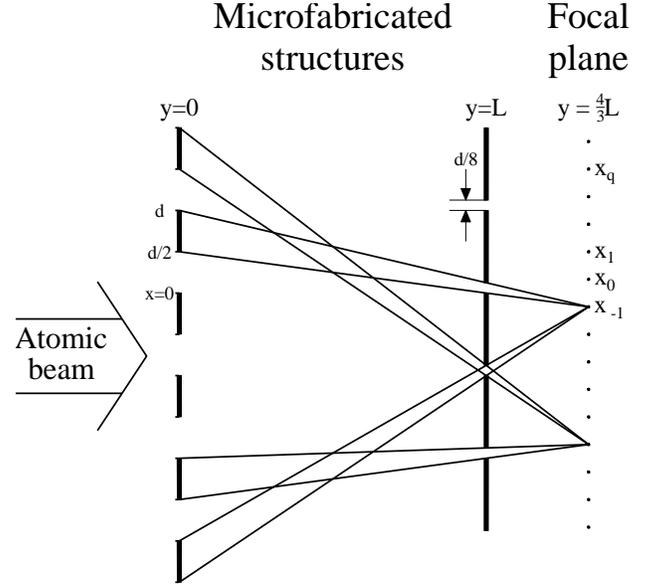}
\end{center}
\end{minipage}
\begin{minipage}{0.99\linewidth} \caption{To prove that two MS having 
equal periods $\left(
d_{1}=d_{2}=d\right) $ and duty cycles $f_{1}=\frac{1}{2}$ and\thinspace
\thinspace $f_{2}=\frac{1}{2\left( n+1\right) }$ (case $n=3$ is shown)
produce a background-free atomic grating at the focal plane $y=\frac{n+1}{n}%
L,$ one can notice that points $x_{q}=d_{g}\left( q+\frac{1}{2}\right) $ are
not achievable for particles ($d_{g}=\frac{d}{n}$ is a grating period). It
follows from the fact that an angle, built from any point $x_{q}$ and
arbitrary slit of the second MS, meets the closed part of the first MS. 
\label{sltshf15}}
\end{minipage}
\end{figure}

The geometric picture can also be used to explain the absence of background
terms at the focal planes $y=\frac{n+1}{n}L$ produced by MS having duty
cycles (\ref{shdslt35}) and equal periods $\left( d_{1}=d_{2}=d\right) $.
One can see from Eq. (\ref{shdslt31}, \ref{shdslt33p}) that the background
disappears because there are no particles at the points $x_{q}=\frac{d}{n}%
\left( q+\frac{1}{2}\right) ,$ where $q$ is integral. A geometric
interpretation of this result is presented in Fig. \ref{sltshf15}. 

\section{Talbot-Lau effect}

When the spatial separation of the MS is increased to the point where 
\begin{equation}
L\sim L_T=2d^2/\lambda _{dB},  \label{tlbtvl63}
\end{equation}
it is no longer possible to neglect quantization of the atomic
center-of-mass in calculating the transverse motion of the atoms. Just as in
the Talbot effect, the recoil an atom undergoes on scattering from a grating
must be taken into account. It turns out, however, that the Doppler
dephasing and rephasing encountered in analyzing the problem of classical
scattering by MS still can be given a classical interpretation when $L\sim
L_T$, provided that the angular divergence of the incident beam is
sufficiently large, $\theta _b\gg d/L_T=\lambda _{dB}/d=\theta _d$. In other
words, even though we must account for quantization of the atoms'
center-of-mass motion, effects related to Doppler dephasing (which are
automatically included in a quantized motion approach) are unchanged from
the classical case.

We have already alluded to this result in the Introduction. Recall that
matter-wave interference results from the overlap on the screen of atomic
wave functions associated with states having center-of-mass momenta $p$ and $%
p+n\hbar k.$ The relative dephasing between these states (\ref{intr7})
contains a Doppler part (\ref{intr7a}) and a quantum part (\ref{intr7b}).
Since the washing out and restoration of the macroscopic atomic gratings is
connected with an averaging over atomic velocities $v,$ one expects that the
Doppler part, proportional to $v,$ is responsible for the
dephasing-rephasing effect. This contribution is actually classical in
nature (i. e. it does not vanish in the limit $\hbar \rightarrow 0)$ and
enters the calculations whether or not the quantum contribution to the phase
has to be considered. As a consequence, the dephasing-rephasing process is
the same for the classical shadow effect and the quantum Talbot-Lau effect.
It turns out, however, that for separations of the MS equal to a rational
multiple of the Talbot length, the Talbot effect can actually result in a
decrease of the period of the atomic gratings from those periods which would
result from the classical shadow effect result. The decrease in period
occurs for MS consisting of open slits and opaque strips; it would not
occur, for example, in resonant standing wave fields.

Since the Doppler dephasing determines the position of the focal planes and
the period of the atomic grating, one can carry over the results Eqs. (\ref
{shdslt59}, \ref{shdslt11}, \ref{shdslt61}) obtained in Sec. IV for the
shadow effect. In this section, we are interested in the variation of the
atomic gratings in a given focal plane as a function of the separation of
the MS. In other words, we look for those separations $L$ for which the
Talbot effect significantly modifies the gratings that would have been
produced by the shadow effect alone. This is analogous to photon echo
studies of atomic relaxation in which the echo amplitude is monitored as a
function of the separation between the excitation pulses.

It should be noted that the Talbot-Lau effect has been studied using light
by Clauser and Reinsch $\left( 1992\right) $ for the parameters 
\begin{equation}
\frac{d_1}{d_2}=3;\,\,y_e=3L,  \label{tlbtvl66}
\end{equation}
corresponding to \{$m,\,\,n,\,\,j,\,\,\ell ,\bar{j},\bar{\ell}$\}=\{$%
3,\,\,1,\,\,3,\,\,1,1,1/3\}$ [recall that $\bar{j}=j/\mu ;\bar{\ell}=\ell
/\mu $, where $\mu $ is the largest common factor of $j(m-n)$ and $ml$] and
a grating period $d_g=d_1.$ The atomic Talbot-Lau effect was demonstrated by
Clauser and Li $\left( 1994\right) $ using K atoms for the parameters 
\begin{equation}
\frac{d_1}{d_2}=2,\,\,y_e=2L,  \label{tlbtvl67}
\end{equation}
$\left\{ m,\,\,n,\,\,j,\,\,\ell ,\bar{j},\bar{\ell}\right\} =\left\{
2,\,\,1,\,\,2,\,\,1,1,1/2\right\} $, $d_g=d_1$. A theoretical study of the
atomic Talbot-Lau effect was also carried out for the parameters (\ref
{tlbtvl67}) by Carnal et al. $\left( 1995\right) $. The conclusions as to
the period and location of the atomic gratings follows from purely classical
considerations in this case; there is no need to invoke arguments related to
the wave nature of matter (Clauser and Reinsch, 1992; Clauser and Li, 1994;
Carnal et al., 1995).

\subsection{Grating formation}

The geometry is the same as that considered for the classical shadow effect,
except that $L$ is no longer restricted to be less than $L_T$. Again, it is
convenient to work in the atomic rest frame defined by $t=y/u$. As discussed
above, it is necessary to quantize the atomic motion only in the $x$%
-direction. The atoms undergo scattering at the MS at times $T_s=L_s/u\
(T_1=0,\ T_2=T=L/u).$ For thin gratings, the atomic wave function $\psi
(x,t) $ undergoes jumps at the MS given by 
\begin{equation}
\psi (x,T_s^{+})=\eta _s(x)\psi (x,T_s^{-}),  \label{tlbtvl1}
\end{equation}
where $\psi (x,T_s^{\pm })$ is the wave function on either side of grating $%
s $, $\eta _s(x)$ is the amplitude transmission function of grating $s$, and 
\begin{equation}
\chi _s(x)=\left| \eta _s(x)\right| ^2  \label{tlbtvl2}
\end{equation}
is the transmission function of grating $s$ [for MS consisting of a series
of slits, $\eta _s(x)=\chi _s(x)]$.

To characterize the atomic beam using a quantized center-of-mass
description, one can use the Wigner distribution function defined by 
\begin{equation}
f(x,p,t)=\int \frac{d\hat{x}}{2\pi \hbar }\exp (-ip\hat{x}/\hbar )\psi (x+%
\frac{\hat{x}}{2},t)\psi ^{\ast }(x-\frac{\hat{x}}{2},t).  \label{tlbtvl3}
\end{equation}
For scattering at a MS, one finds
\end{multicols}
\begin{equation}
f(x,p,T_{s}^{+})=\int \frac{d\hat{x}dp^{\prime }}{2\pi \hbar }\exp \left[
-i\left( p-p^{\prime }\right) \hat{x}/\hbar \right] \eta _{s}\left( x+\frac{%
\hat{x}}{2}\right) \eta _{s}^{\ast }\left( x-\frac{\hat{x}}{2}\right)
f\left( x,p^{\prime },T_{s}^{-}\right) .  \label{tlbtvl4}
\end{equation}
When $\eta _{s}(x)$ is a periodic function of $x$, one can write Eq. (\ref
{tlbtvl4}) as 
\begin{equation}
f(x,p,T_{s}^{+})=\sum_{n_{s},n_{s}^{^{\prime }}}\exp (im_{s}k_{s}x)\eta
_{n_{s}}^{(s)}\left[ \eta _{n_{s}^{^{\prime }}}^{(s)}\right] ^{\ast }f\left[
x,p-\frac{\hbar k}{2}(n_{s}+n_{s}^{^{\prime }}),T_{s}^{-}\right] ,
\label{tlbtvl5}
\end{equation}
where 
\begin{equation}
m_{s}=n_{s}-n_{s}^{^{\prime }},  \eqnum{116a}  \label{tlbtvl5a}
\end{equation}
and

\begin{equation}
\eta _{n}^{(s)}=\int_{0}^{d_{s}}\frac{dx}{d_{s}}e^{-ink_{s}x}\eta _{s}(x) 
\eqnum{116b}  \label{tlbtvl5b}
\end{equation}
is a Fourier component of $\eta _{s}(x),$ having period $d_{s}$ and wave
number$\ k_{s}=2\pi /d_{s}$. For times other than $T_{s}$, the Wigner
distribution function evolves freely as 
\begin{equation}
f(x,p,t)=f(x-v(t-T_{s}),p,T_{s}^{+}),  \label{tlbtvl6}
\end{equation}
where $v=p/M$.

Applying Eqs. (\ref{tlbtvl5}, \ref{tlbtvl6}) one obtains the atomic
distribution function for times $t>T$ ($y>L$) to be 
\begin{eqnarray}
f(x,p,t) &=&\sum_{n_{i},\ n_{i}^{\prime }}\eta _{n_{1}}^{(1)}\left[ \eta
_{n_{1}^{\prime }}^{(1)}\right] ^{\ast }\eta _{n_{2}}^{(2)}\left[ \eta
_{n_{2}^{\prime }}^{(2)}\right] ^{\ast }\exp \left\{ im_{1}k_{1}\left[
x-v(t-T)-\left( v-\frac{\hbar k_{2}}{2M}\left( n_{2}+n_{2}^{\prime }\right)
\right) T\right] \right.  \nonumber \\
&&+\left. im_{2}k_{2}\left[ x-v\left( t-T\right) \right] \right\} f\left\{ x-%
\left[ v-\frac{\hbar k_{2}}{2M}\left( n_{2}+n_{2}^{\prime }\right) \right]
T-v(t-T),\right.  \nonumber \\
&&\left. p-\frac{\hbar }{2}\left[ k_{1}\left( n_{1}+n_{1}^{\prime }\right)
+k_{2}\left( n_{2}+n_{2}^{\prime }\right) \right] \right\} ,  \label{tlbtvl7}
\end{eqnarray}
where $f(x,p)$ is the Wigner distribution function of the incoming atomic
beam. The atomic spatial distribution is given by 
\begin{equation}
f\left( x,t\right) =\int dpf\left( x,p,t\right)  \label{tlbtvl8}
\end{equation}
which can be obtained from Eq. (\ref{tlbtvl7}) as 
\begin{eqnarray}
f(x,t) &=&\sum_{n_{i},\ n_{i}^{\prime }}\eta _{n_{1}}^{(1)}\left[ \eta
_{n_{1}^{\prime }}^{(1)}\right] ^{\ast }\eta _{n_{2}}^{(2)}\left[ \eta
_{n_{2}^{\prime }}^{(2)}\right] ^{\ast }\int dpf\left\{ x-\left[ v+\frac{%
\hbar }{2M}\left( k_{1}\left( n_{1}+n_{1}^{\prime }\right) +k_{2}\left(
n_{2}+n_{2}^{\prime }\right) \right) \right] \left( t-T\right) \right. 
\nonumber  \label{tlbtvl} \\
&&\left. -\left[ v+\frac{\hbar k_{1}}{2M}\left( n_{1}+n_{1}^{\prime }\right) %
\right] T,p\right\} \exp \left\{ i\left( m_{1}k_{1}+m_{2}k_{2}\right) \left[
x-\frac{\hbar k_{2}}{2M}\left( n_{2}+n_{2}^{\prime }\right) \right] \left(
t-T\right) \right.  \nonumber \\
&&\left. -i\left[ v+\frac{\hbar k_{1}}{2M}\left( n_{1}+n_{1}^{\prime
}\right) \right] \left[ m_{1}k_{1}t+m_{2}k_{2}\left( t-T\right) \right]
\right\} .  \label{tlbtvl9}
\end{eqnarray}
In this expression terms having $\left( m_{1}k_{1}+m_{2}k_{2}\right) \neq 0$
contribute to the atomic gratings. Owing to the assumption of an incident
beam having large angular divergence $\theta _{b}\sim v/u\gg d/L$, the
Doppler phases associated with these terms oscillate rapidly as a function
of $p$, except in the echo focal planes. As a consequence, the positions and
periods of the atomic gratings are the same as those in the classical shadow
effect [see Eqs. (\ref{shdslt59}, \ref{shdslt11}, \ref{shdslt61})]. In the
remainder of this section, we calculate the atomic density in the focal
planes $y_{e}$ or, equivalently, at times $t_{e}=y_{e}/u$ given in Eqs. (\ref
{shdslt59}, \ref{shdslt11}).

It is possible to simplify Eq. (\ref{tlbtvl9}) if we assume that the angular
divergence $\theta _{b}$ of the incident beam is less than $\theta _{D}=D/L$%
, such that a freely propagating beam would undergo negligible diffraction
over a distance of order $L$. For $p\sim Mu\theta _{b}\ll MD/T$, one can
neglect the dependence on $n_{i}$ and $n_{i}^{^{\prime }}$ of the
distribution function appearing in Eq. (\ref{tlbtvl9}). Then the sum over $%
n_{i}$ can be carried out using the formula 
\begin{eqnarray}
&&\sum_{n}e^{-in\alpha }\eta _{n}^{(s)}\left[ \eta _{n-\nu }^{(s)}\right]
^{\ast }  \nonumber \\
&=&e^{-i\nu \alpha /2}\left[ \eta _{s}\left( x-\frac{\alpha }{2k_{s}}\right)
\eta _{s}^{\ast }\left( x+\frac{\alpha }{2k_{s}}\right) \right] _{\nu },
\label{tlbtvl14}
\end{eqnarray}
where 
\begin{equation}
\left[ F(x)\right] _{\nu }=\int_{0}^{d_{s}}\frac{dx}{d_{s}}e^{-i\nu
k_{s}x}F(x)  \label{tlbtvl15}
\end{equation}
is a Fourier component of the function $F\left( x\right) $. As a result one
finds that the atomic density in the echo focal planes is given by 
\begin{equation}
f(x,t_{e})=f(x)\sum_{q}e^{iqk_{g}x}\chi _{-\bar{j}(m-n)q}^{(1)}\left[ \eta
_{2}\left( x-qd_{2}\frac{\phi _{T}(m,n)}{2\pi }\right) \eta _{2}^{\ast
}\left( x+qd_{2}\frac{\phi _{T}(m,n)}{2\pi }\right) \right] _{m\bar{\ell}q},
\label{tlbtvl16}
\end{equation}
\begin{multicols}{2}
where $\chi _{s}^{(1)}$ is a Fourier component of the transmission function $%
\chi _{1}\left( x\right) ,$%
\begin{equation}
\phi _{T}(m,n)=\frac{\bar{j}^{2}(m-n)}{\bar{\ell}}\omega _{k_{1}}T
\label{tlbtvl18}
\end{equation}
is a Talbot phase associated with a specific focal plane, and 
\begin{equation}
f\left( x\right) =\int dpf\left( x,p\right)   \label{tlbtvl25}
\end{equation}
is the initial spatial distribution in the atomic beam. Since the beam
diameter is much larger than the period of the gratings, 
\begin{equation}
D\gg d_{g}  \label{tlbtvl27}
\end{equation}
one can neglect the variation of $f\left( x\right) $ and set $f\left(
x\right) =1$ in Eq. (\ref{tlbtvl16})$.$

The distribution function (\ref{tlbtvl16}) is identical with the shadow
effect result (\ref{shdslt13}), except for the presence of the Talbot
phases. The main features of the dependence of the atomic density on the
Talbot phase in the Talbot-Lau effect are the same as those for the Talbot
effect considered in Sec. III. The density (\ref{tlbtvl16}) is an
oscillating function of $\phi _T(m,n)$ having period $2\pi $. If $\phi
_T(m,n)$ is increased by $2\pi $, or, equivalently, if the separation
between the MS is increased by $L_T(m,n)$, the density distribution in the
corresponding echo plane is unchanged. The Talbot distance associated with a
given focal plane is defined here as 
\begin{equation}
L_T(m,n)\equiv \frac{2d_1^2}{\lambda _{dB}}\frac{\bar{\ell}}{\bar{j}^2\left(
m-n\right) }.  \label{tlbtvl61}
\end{equation}
In terms of $L_T(m,n)$, the Talbot phase (\ref{tlbtvl18}) is equal to 
\begin{equation}
\phi _T(m,n)=2\pi [L/L_T(m,n)].  \label{tlbtvl100}
\end{equation}
In our notation, the Talbot phase is a function of $L$, while the Talbot
distance is independent of $L.$ Note that, as defined by Eq.(\ref{tlbtvl61}%
), there is a different Talbot length associated with the signal for
different focal planes, $y_e=(m/n)L$. We wish to examine the signal in a
given focal plane as a function of the separation $L$ of the MS or,
equivalently, as a function of $\phi _T(m,n)$.When $L=L_T(m,n)$ [$\phi
_T(m,n)=2\pi $] the atomic density (\ref{tlbtvl16}) is the same as that of
the shadow effect (\ref{shdslt13}).

Using arguments similar to those leading to Eqs. (\ref{talbot38}), one can
prove that, for pure amplitude modulation of the wave functions, i. e. for
real amplitude transmission functions $\eta _j(x)=\eta _j^{*}(x),$ the
dependence of the particles' distribution on the Talbot phase is symmetric
with respect to the point $\phi _T(m,n)=\pi $ [$L=L_T(m,n)/2$], 
\begin{equation}
\left. f\left( x,t_e\right) \right| _{\phi _T(m,n)}=\left. f\left(
x,t_e\right) \right| _{2\pi -\phi _T(m,n)},  \label{tlbtvl62}
\end{equation}
\begin{equation}
\left. f\left( x,t_e\right) \right| _L=\left. f\left( x,t_e\right) \right|
_{L_T-L}.  \label{tlbtvl62a}
\end{equation}

The question arises as to what values of $\phi _T(m,n)$ lead to especially
interesting results, i. e. atomic gratings that differ significantly from
the gratings that would be produced by the shadow effect. We have found that
the atomic gratings are significantly modified by the Talbot effect when the
Talbot phase is a rational multiple of $2\pi ,$%
\begin{equation}
\phi _T(m,n)=2\pi \frac{m_T}{n_T},  \label{tlbtvl19}
\end{equation}
where $m_T$ and $n_T$ are positive integers having no common factors. We
proceed to analyze the atomic density function in the focal planes for
separations of the MS corresponding to Eq. (\ref{tlbtvl19}), that is, for $%
L=L_T(m,n)\frac{m_T}{n_T}.$

For Talbot phases given by Eq. (\ref{tlbtvl19}), the sum in Eq. (\ref
{tlbtvl16}) can be divided into $n_T-1$ independent sums having 
\begin{equation}
q=n_Tq^{\prime }+r  \label{tlbtvl20}
\end{equation}
where $0\leq r\leq n_T.$ For Talbot phases given by Eq. (\ref{tlbtvl19}),
any dependence on $q^{\prime }$ disappears in the last factor of Eq. (\ref
{tlbtvl16}), allowing one to rewrite Eq. (\ref{tlbtvl16}) as 
\end{multicols}
\begin{eqnarray}
f(x,t_e) &=&\sum_{r=0}^{n_T-1}\sum_{q^{\prime }}\int_0^{d_1}\int_0^{d_2}%
\frac{dx_1dx_2}{d_1d_2}\exp \left\{ i\left( q^{\prime }n_T+r\right) \left[
k_gx+\bar{j}\left( m-n\right) k_1x_1-m\bar{\ell}k_2x_2\right] \right\} 
\nonumber \\
&&\times \chi _1\left( x_1\right) \eta _2\left( x_2-r\frac{m_T}{n_T}%
d_2\right) \eta _2^{*}\left( x_2+r\frac{m_T}{n_T}d_2\right) ,
\label{tlbtvl21}
\end{eqnarray}
to sum over $q^{\prime }$ using Eq. (\ref{shdslt16}), and to integrate over $%
x_1.$ The calculations are similar to those used to obtain Eq. (\ref
{shdslt42}) from Eq. (\ref{shdslt38}), and one can obtain
\begin{eqnarray}
f\left( x,t_e\right) &=&\frac 1{\bar{j}n_T(m-n)}\sum_{r=0}^{n_T-1}\,\,%
\sum_{s=0}^{\bar{j}(m-n)n_T-1}\int_0^{d_2}\frac{dx_2}{d_2}\exp \left\{ \frac{%
2\pi ir}{n_T}\left[ s+\left[ n_T\left( m\bar{\ell}\frac{x_2}{d_2}-\frac x{d_g%
}\right) \right] _I\right] \right\}  \nonumber  \label{tlbtvl} \\
&&\times \eta _2\left( x_2-r\frac{m_T}{n_T}d_2\right) \eta _2^{*}\left( x_2+r%
\frac{m_T}{n_T}d_2\right) \chi _1\left\{ \frac{d_1}{\bar{j}(m-n)n_T}\left[
s+\left\{ n_T\left( m\bar{\ell}\frac{x_2}{d_2}-\frac x{d_g}\right) \right\}
_F\right] \right\} .  \label{tlbtvl26}
\end{eqnarray}
\begin{multicols}{2}

\subsection{Higher-order gratings using the Talbot-Lau technique}

Equation (\ref{tlbtvl26}) is the basic result of this section. It gives the
atomic density function in the focal plane for separation of the MS that
corresponds to Talbot phases which are rational multiples of 2$\pi $. For
specified transmission functions, it can be evaluated numerically in focal
planes defined by $t_e=y_e/u=(m/n)T=(m/n)L/u$ for arbitrary $(j/\ell
)=d_1/d_2$ (recall that $\bar{j}=j/\mu ;\bar{\ell}=\ell /\mu $, where $\mu $
is the largest common factor of $j(m-n)$ and $\ell m$), $m_T$ and $n_T$ [$%
L=L_T(m,n)\frac{m_T}{n_T}$]. In this subsection, we are interested primarily
in showing that, owing to the Talbot effect, periodic atomic density
gratings can be produced whose periods $d_T^{}$ are smaller than the
corresponding periods $d_g$ which would have been produced by the shadow
effect.

The first thing to note is that the function $\chi _1$ in Eq. (\ref{tlbtvl26}%
), considered as a function of $x$, is periodic with period 
\begin{equation}
d_T=d_g/n_T,  \label{tlbtvl28}
\end{equation}
$n_T-$times smaller than the period $d_g$ of the shadow-effect grating$.$
Unfortunately, this does not guarantee that $f\left( x,t_e\right) $ is
periodic with period $d_T$, owing to the exponential term in Eq. (\ref
{tlbtvl26}). Under the transformation $x\rightarrow x+d_T$, the exponential
term is multiplied by the phase-factor 
\begin{equation}
\exp \left( 2\pi ir/n_T\right)  \label{tlbtvl30p}
\end{equation}
which is a function of the summation index $r.$ If the summation over $r$ in
Eq. (\ref{tlbtvl26}) somehow was restricted to $r=0$, the atomic grating
would have period $d_T$. Restricting the summation to $r=0$ can be
accomplished by choosing the amplitude transmission function such that the
product $\eta _2\left( x_2-\frac r{n_T}d_2\right) \eta _2^{*}\left( x_2+%
\frac r{n_T}d_2\right) $ is nonvanishing only for $r=0$. To simplify the
discussion, we have taken $m_T=1$ [$L=L_T(m,n)/n_T$].

\begin{figure}
\begin{minipage}{0.99\linewidth}
\begin{center}
\epsfxsize=.95\linewidth \epsfysize=1.05\epsfxsize \epsfbox{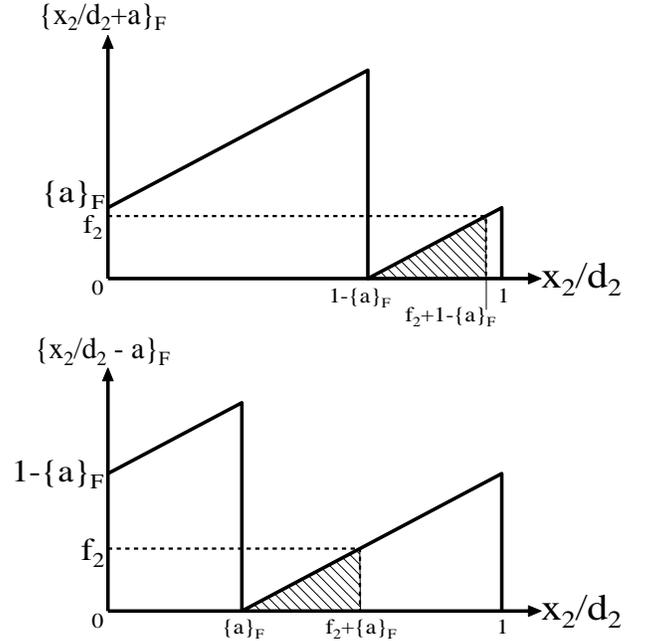}
\end{center}
\end{minipage}
\begin{minipage}{0.99\linewidth} \caption{Plots of the functions 
$\left\{ \frac{x_{2}}{d_{2}}\pm a\right\} .$
In the shaded areas only transmissions $\protect\phi _{2}(x_{2}\pm
ad_{2})\neq 0.$ When slits relative width $f_{2}<\min \left( \left\{
a\right\} ,1-\left\{ a\right\} \right) $ there are no shaded areas in the
vicinity of the point $x_{2}=0.$ If also $f_{2}<\left| 1-2\left\{ a\right\}
\right| ,$ then shaded areas have no common points and product $\protect\phi %
_{2}(x_{2}+ad_{2})\protect\phi _{2}(x_{2}-ad_{2})=0$ for any $x_{2}$ 
\label{tlbtvlf1}}
\end{minipage}
\end{figure}

For MS consisting of slits and opaque strips, both the amplitude
transmission functions $\eta _{j}(x)$ and transmission functions $\chi
_{j}(x)$ are equal to the Heaviside step-function 
\begin{equation}
\eta _{j}(x)=\chi _{j}(x)=\theta \left( f_{j}-\left\{ \frac{x}{d_{j}}%
\right\} _{F}\right) ,  \label{tlbtvl29}
\end{equation}
where $f_{j}$ is the duty cycle of MS $j$. The functions $\left\{ \frac{x_{2}%
}{d_{2}}\pm \frac{r}{n_{T}}\right\} _{F}$ shown in Fig. \ref{tlbtvlf1}
represent the profile of the second MS displaced by $\pm \frac{r}{n_{T}}%
d_{2} $. 
In the range $0\leq x_{2}<d_{2}$, the product $\eta _{2}\left( x_{2}-\frac{r%
}{n_{T}}d_{2}\right) \eta _{2}^{\ast }\left( x_{2}+\frac{r}{n_{T}}%
d_{2}\right) =$ $\theta \left( f_{2}-\left\{ \frac{x_{2}}{d_{2}}+\frac{r}{%
n_{T}}\right\} _{F}\right) $ $\theta \left( f_{2}-\left\{ \frac{x_{2}}{d_{2}}-%
\frac{r}{n_{T}}\right\} _{F}\right) $, which represents a product of
profiles of the second MS displaced by $\pm \frac{r}{n_{T}}d_{2}$, vanishes
for $r\neq 0$ provided that $f_{2}$ is sufficiently small and provided that $%
\frac{r}{n_{T}}$ $\neq \frac{1}{2}$ (if $\frac{r}{n_{T}}$ $=\frac{1}{2}$,
the gratings are displaced by $\frac{d_{2}}{2}$ and overlap for any $f_{2}$).

If 
\begin{equation}
f_2\leq \min \left( \frac r{n_T},1-\frac r{n_T}\right) ,  \label{tlbtvl30}
\end{equation}
the only regions where $\eta _2(x_2\pm \frac r{n_T}d_2)$ does not vanish are 
\begin{equation}
\frac{x_2}{d_2}\in \left[ 1-\frac r{n_T},\,\,f_2+1-\frac r{n_T}\right] \,\,%
\text{and}\,\,\frac{x_2}{d_2}\in \left[ \frac r{n_T},f_2+\frac r{n_T}\right]
,  \label{tlbtvl32}
\end{equation}
respectively. These two intervals have no common regions if $f_2+1-\frac r{%
n_T}\leq \frac r{n_T}$ or $f_2+\frac r{n_T}\leq 1-\frac r{n_T},$ i. e. if 
\begin{equation}
f_2\leq \left| 1-2\frac r{n_T}\right| .  \label{tlbtvl33}
\end{equation}
Inequality (\ref{tlbtvl33}) must hold for all $r\neq 0$ to guarantee that
the atomic grating has period $d_T=d_g/n_T$. Clearly, inequality (\ref
{tlbtvl33}) does not hold for $r=\frac{n_T}2$ when $n_T$ is even. While this
does not preclude the possibility of higher order gratings for $n_T$ even,
it does suggest that we consider the cases of even and odd $n_T$ separately.

\subsubsection{$n_T$ odd}

$\ $In this case, we write 
\begin{equation}
n_{T}=2n^{\prime }+1,  \label{tlbtvl36}
\end{equation}
where $n^{\prime }$ is a positive integer or zero. For the summation range $%
0\leq r\leq n_{T}-1$ in Eq. (\ref{tlbtvl26}), the minimum value of the rhs
of both inequalities (\ref{tlbtvl30}) and (\ref{tlbtvl33}) is $1/n_{T}$,
which occurs for $r=1$ or $r=2n^{\prime }$ in (\ref{tlbtvl30}) and $%
r=n^{\prime }$ or $r=n^{\prime }+1$ in (\ref{tlbtvl33}). Thus, provided that 
\begin{equation}
f_{2}\leq \frac{1}{2n^{\prime }+1}  \label{tlbtvl38}
\end{equation}
one can produce atomic gratings having period $d_{T}=d_{g}/(2n^{\prime }+1)$
in the focal plane $y=\frac{m}{n}L$ for separations between the MS equal to $%
L=\frac{L_{T}}{\left( 2n^{\prime }+1\right) }$ or, equivalently, for a
Talbot phase (\ref{tlbtvl18}) equal to $\frac{2\pi }{2n^{\prime }+1}$. Under
these conditions one omits terms having $r\neq 0$ in Eq. (\ref{tlbtvl26})
and finds 
\begin{equation}
f(x,t_{e})=\frac{f_{2}}{\bar{j}(m-n)n_{T}}\sum_{s=0}^{\left[ \beta ^{\prime }%
\right] _{I}}h_{s}\left( w\right) ,  \label{tlbtvl41}
\end{equation}
\begin{equation}
h_{s}\left( w\right) =\int_{0}^{1}dz\theta \left( \beta ^{\prime }-\left(
s+\left\{ \alpha ^{\prime }z-w\right\} _{F}\right) \right) ,  \eqnum{143a}
\label{tlbtvl41a}
\end{equation}
\begin{eqnarray}
\alpha ^{\prime } &=&n_{T}\alpha =m\bar{\ell}\left( 2n^{\prime }+1\right)
f_{2},  \nonumber \\
\,\beta ^{\prime } &=&n_{T}\beta =\bar{j}\left( m-n\right) \left( 2n^{\prime
}+1\right) f_{1},  \eqnum{143b}  \label{tlbtvl41b}
\end{eqnarray}
where $\alpha $ and $\beta $ are given by (\ref{shdslt23b}), and
dimensionless variables 
\begin{equation}
w=\frac{x}{d_{T}},\,\,z=\frac{x_{2}}{d_{2}f_{2}}  \label{tlbtvl42}
\end{equation}
have been introduced. Note that the ratio 
\begin{equation}
\alpha /\beta =\alpha ^{\prime }/\beta ^{\prime }=\frac{m}{m-n}\frac{f_{2}}{%
f_{1}}\frac{\ell }{j}=\frac{m}{m-n}\frac{f_{2}d_{2}}{f_{1}d_{1}}
\label{tlbtvl100a}
\end{equation}
depends on the focal plane and ratio of slit widths. In a manner similar to
arriving at Eq. (\ref{shdslt31}), one can obtain 
\begin{eqnarray}
f(x,t) &=&\left[ \alpha ^{\prime }[\beta ^{\prime }]_{I}+\left\{ \beta
^{\prime }\right\} _{F}[\alpha ^{\prime }]_{I}+S\left( w\right) \right] 
\nonumber \\
&&\times \left[ m(m-n)\bar{\ell}\bar{j}\left( 2n^{\prime }+1\right) ^{2}%
\right] ^{-1}  \label{tlbtvl43}
\end{eqnarray}
where $S\left( w\right) $ is given by Eqs. (\ref{shdslt32}-\ref{shdslt32c})
with the replacements $\alpha \rightarrow \alpha ^{\prime },\,\,\beta
\rightarrow \beta ^{\prime }$.
\begin{figure}
\begin{minipage}{0.99\linewidth}
\begin{center}
\epsfxsize=.95\linewidth \epsfysize=\epsfxsize \epsfbox{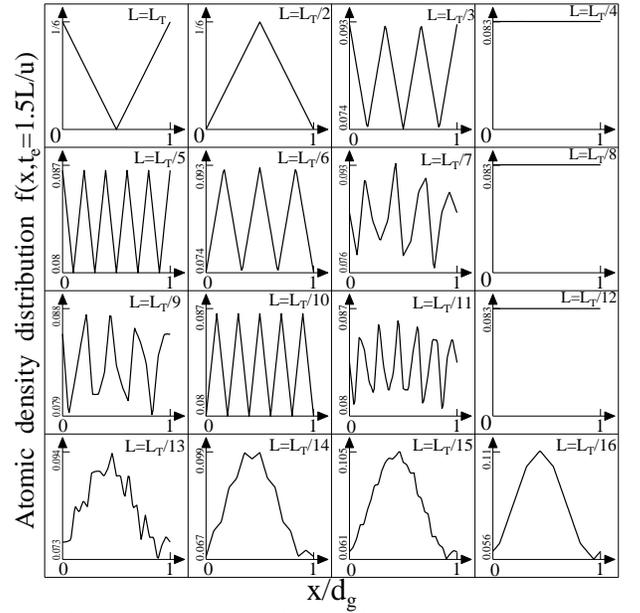}
\end{center}
\end{minipage}
\begin{minipage}{0.99\linewidth} \caption{Atomic spatial distribution 
created by two microfabricated
structures, having the same spacings $d_{1}=d_{2},$ relative widths $f_{1}=%
\frac{1}{2},$ $f_{2}=\frac{1}{6}$ and separated on the distance $L.$ At the
focal plane $y=\frac{3}{2}L$ $\left( m=3,\,\,n=2\right) ,$ where owing to
the shadow effect one expects grating with period $d_{g}=\frac{d_{1}}{2},$
the wave matter interference leads to the higher order gratings if one
chooses distance between fields coinciding with the integer fraction of the
Talbot distance $\left( L=\frac{L_{T}}{n_{T}}=\frac{4d_{1}^{2}}{n_{T}\protect%
\lambda _{dB}},\,\,n_{T}=1,\,\,\ldots \,\,16\right) ,$ which corresponds to
the Talbot phase $\protect\phi _{T}=\frac{2\protect\pi }{n_{T}}.$ For
sufficiently small relative width of the second microfabricated structure, $%
f_{2}\leq \frac{1}{n_{T}}$ $\left( f_{2}\leq \frac{2}{n_{T}}\right) ,$ one
observes gratings having spacing $d_{T}=\frac{d_{g}}{n_{T}}=\frac{d_{1}}{%
2n_{T}}$ $\left( d_{T}=\frac{2d_{g}}{n_{T}}=\frac{d_{1}}{n_{T}}\right) $ for
distance between fields equal to the odd $\left( \text{even}\right) $
fraction part of the Talbot distance, i. e. for odd $\left( \text{even}%
\right) $ $n_{T}.$ Exceptions here are the cases when $n_{T}$ is multiple $4$
$\left( L=\frac{L_{T}}{4},\,\,L=\frac{L_{T}}{8},\,\,L=\frac{L_{T}}{12}%
\right) ,$ where for $f_{2}\leq \frac{2}{n_{T}}$ wave-matter interference
leads to the entire washing out of the gratings. 
\label{tlbtvlf2}}
\end{minipage}
\end{figure}

The amplitude of the grating (\ref{tlbtvl43}) is maximum when $%
m=n+1,\,\,j=\ell =1,$ and 
\begin{equation}
\left\{ \alpha ^{\prime }\right\} _{F}=\left\{ \beta ^{\prime }\right\} _{F}=%
\frac{1}{2},  \label{tlbtvl44}
\end{equation}
for which 
\begin{equation}
f_{1}=\frac{2q_{1}+1}{2\left( 2n^{\prime }+1\right) },\,\,f_{2}=\frac{%
2q_{2}+1}{2\left( n+1\right) \left( 2n^{\prime }+1\right) },
\label{tlbtvl45}
\end{equation}
where $0\leq q_{1}\leq 2n^{\prime },\,\,0\leq q_{2}\leq n$ are integers$.$
Under these conditions one finds 
\begin{equation}
f(x,t_{e})=\frac{q_{1}q_{2}+\frac{1}{2}\left( q_{1}+q_{2}\right) +\left|
\left\{ \frac{xn\left( 2n^{\prime }+1\right) }{d_{1}}\right\} _{F}-\frac{1}{2%
}\right| }{\left( n+1\right) \left( 2n^{\prime }+1\right) ^{2}}.
\label{tlbtvl46}
\end{equation}
This grating has amplitude 
\begin{equation}
A=1/\left[ 2\left( n+1\right) \left( 2n^{\prime }+1\right) ^{2}\right] 
\label{tlbtvl47}
\end{equation}
and a background term whose amplitude is ($2q_{1}q_{2}+q_{1}+q_{2})$ times
larger than $A$. Talbot-Lau gratings for different values of the Talbot
phase are shown in Fig. \ref{tlbtvlf2}. 

\subsubsection{$n_T$ even}

The atomic density patterns in Fig. \ref{tlbtvlf2} have been drawn for both
even and odd values of $n_T.$ From this figure one sees that qualitatively
new features appear for even values of $n_T$. When the Talbot phase $\phi
_T(m,n)=\pi ,\,\,\pi /3,\,\,\pi /5\,\,(n_T=2,\,\,6,\,\,10)$ gratings having
period $d_g,\,\,d_g/3,\,\,d_g/5,\,\,$or 
\begin{equation}
d_T^{\prime }=2d_g/n_T=d_g/(n_T/2)  \label{tlbtvl51}
\end{equation}
are focused. When the Talbot phase $\phi _T(m,n)=\pi /2,\,\,\pi /4,\,\,\pi
/6\,\,(n_T=4,\,\,8,\,\,12)$, the gratings are washed out entirely. To
explain these results one needs to return to the general expression (\ref
{tlbtvl26}). If 
\begin{equation}
n_T=2n^{\prime },  \label{tlbtvl48}
\end{equation}
where $n^{\prime }$ is a positive integer, one divides the sum over $r$ in
Eq. (\ref{tlbtvl26}) into two parts having 
\begin{equation}
r=n^{\prime }q+r^{\prime },  \label{tlbtvl50}
\end{equation}
with $q=0$ or $1$ with $r^{\prime }$ restricted to the range $0\leq
r^{\prime }\leq n^{\prime }-1.$ In the second part $(q=1)$ one shifts the
integration variable from $x_2$ to $\left( x_2+\frac 12d_2\right) ,$ which
leads to the same factors $\eta _2,\,\,\eta _2^{*},\,\,\chi _1$ in both the $%
q=0$ and $q=1$ terms. In this manner, one arrives at the expression
\end{multicols}
\begin{eqnarray}
f\left( x,t_{e}\right)  &=&\frac{1}{2\bar{j}n^{\prime }(m-n)}\sum_{r^{\prime
}=0}^{n^{\prime }-1}\,\,\sum_{s=0}^{2\bar{j}(m-n)n^{\prime
}-1}\int_{0}^{d_{2}}\frac{dx_{2}}{d_{2}}\left\{ 1+\left( -1\right) ^{\left[
\left( r^{\prime }+n^{\prime }\right) m_{T}m\bar{\ell}+s+\left[ 2n^{\prime
}\left( m\bar{\ell}(\frac{x_{2}}{d_{2}}-\frac{x}{d_{g}}\right) \right] _{I}%
\right] }\right\}   \nonumber  \label{tlbtvl} \\
&&\times \exp \left\{ \frac{\pi ir^{\prime }}{n^{\prime }}\left[ s+\left[
2n^{\prime }\left( m\bar{\ell}\frac{x_{2}}{d_{2}}-\frac{x}{dg}\right) \right]
_{I}\right] \right\} \eta _{2}\left( x_{2}-r^{\prime }\frac{m_{T}}{%
2n^{\prime }}d_{2}\right)   \nonumber \\
&&\times \eta _{2}^{\ast }\left( x_{2}+r^{\prime }\frac{m_{T}}{2n^{\prime }}%
d_{2}\right) \chi _{1}\left\{ \frac{d_{1}}{\bar{j}(m-n)n_{T}}\left[
s+\left\{ 2n^{\prime }\left( m\bar{\ell}\frac{x_{2}}{d_{2}}-\frac{x}{dg}%
\right) \right\} _{F}\right] \right\} .  \label{tlbtvl52}
\end{eqnarray}
\begin{multicols}{2}
The transmission function $\chi _{1}$ still has period $d_{g}/n_{T}=$ $%
d_{g}/(2n^{\prime })$, but the first factor in the integrand has twice this
period, $d_{T}=2d_{g}/n_{T}=$ $d_{g}/n^{\prime }$. As in the case of odd $%
n_{T}$, one must choose $f_{2}$ sufficiently small to eliminate all but the $%
r^{\prime }=0$ terms in the sum to ensure that the atomic grating has period 
$d_{T}$. Inequalities (\ref{tlbtvl30}) and (\ref{tlbtvl33}) are satisfied if 
\begin{equation}
f_{2}\leq \frac{1}{2n^{\prime }},  \label{tlbtvl53}
\end{equation}
which is a sufficient condition for neglect of terms with $r^{\prime }\neq 0$
in Eq. (\ref{tlbtvl52})$.$ As a result one arrives at 
\begin{equation}
f(x,t_{e})=\frac{f_{2}}{2\bar{j}n^{\prime }(m-n)}\sum_{s=0}^{\left[ \beta
^{\prime }\right] _{I}}h_{s}\left( w\right) ,  \label{tlbtvl54}
\end{equation}
\begin{eqnarray}
h_{s}\left( w\right)  &=&\int_{0}^{1}dz\left( 1+\left( -1\right) ^{n^{\prime
}m_{T}m\bar{\ell}+s+\left[ \alpha ^{\prime }z-w\right] _{I}}\right)  
\nonumber \\
&&\times \theta \left\{ \beta ^{\prime }-\left[ s+\left\{ \alpha ^{\prime
}z-w\right\} _{F}\right] \right\} ,  \eqnum{156a}  \label{tlbtvl54a}
\end{eqnarray}
\begin{eqnarray}
\alpha ^{\prime } &=&n_{T}\alpha =2n^{\prime }m\bar{\ell}f_{2}; 
\eqnum{156b}  \label{tlbtvl54b} \\
\beta ^{\prime } &=&n_{T}\beta =2n^{\prime }\bar{j}\left( m-n\right) f_{1}; 
\eqnum{156c}  \label{tlbtvl54c} \\
\,w &=&\frac{x}{d_{T}}.  \eqnum{156d}  \label{tlbtvl54d}
\end{eqnarray}

This expression can be evaluated in the same manner used to evaluate Eq. (%
\ref{tlbtvl41}), but the evaluation is more complicated owing to two
factors: (i) contributions $h_s\left( w\right) $ with $s<\left[ \beta
^{\prime }\right] _I$ are not independent of $w,$ and (ii) for $s=\left[
\beta ^{\prime }\right] _I$ one has to consider separately contributions
from odd and even $\left[ \alpha ^{\prime }z-w\right] _I.$ The situation
simplifies for integer $\beta ^{\prime },$ when the $\theta $-factor equals $%
1$ for $s<\beta ^{\prime }$ and $0$ for $s=\beta ^{\prime }$, independent of
the values of $z,\,\,w,\,\,$and $\alpha ^{\prime }.$ This is the only limit
considered in the subsection. For integer $\beta ^{\prime }$, one can carry
out the summation over $s$ in Eq. (\ref{tlbtvl54})$.$

For even $\beta ^{\prime },$ when $\sum_{s=0}^{\beta ^{\prime }-1}\left(
-1\right) ^s=0,$ the gratings are washed out and 
\begin{equation}
f\left( x,t\right) =f_1f_2.  \label{tlbtvl55}
\end{equation}
This result is consistent with the vanishing of the atomic gratings in Fig. 
\ref{tlbtvlf2} for Talbot phases equal to $\pi /2,$ $\pi /4$ and $\pi /6$,
corresponding to values of $\beta ^{\prime }$ equal to $2,\,\,4,\,\,6.$

When $\beta ^{\prime }$ is odd one finds 
\begin{equation}
f\left( x,t\right) =f_{1}f_{2}\left( 1+\frac{\left( -1\right) ^{n^{\prime
}m_{T}m\bar{\ell}}}{\alpha ^{\prime }\beta ^{\prime }}h^{\prime }(w)\right) ,
\label{tlbtvl56}
\end{equation}
\begin{equation}
h^{\prime }\left( w\right) =\alpha ^{\prime }\left\{ 2\int_{0}^{1}dz\theta 
\left[ \frac{1}{2}-\left\{ \frac{\alpha ^{\prime }z-w}{2}\right\} _{F}\right]
-1\right\} ,  \eqnum{158a}  \label{tlbtvl56a}
\end{equation}
where the equality 
\begin{equation}
\left( -1\right) ^{\left[ x\right] _{I}}\equiv 2\theta \left[ \frac{1}{2}%
-\left\{ \frac{x}{2}\right\} _{F}\right] -1  \label{tlbtvl57}
\end{equation}
has been used. Equation (\ref{tlbtvl56a}) can be reduced to Eq.(\ref
{shdslt25}) with the replacements $\beta ,\ \alpha $ $,$ $w\rightarrow $ $%
\frac{1}{2},\,\,\frac{\alpha ^{\prime }}{2}$ $,$ $\frac{w}{2}.$ Using these
values in Eqs. (\ref{shdslt32a},\ref{shdslt32b}), one obtains
\end{multicols}

\begin{equation}
\begin{array}{cc}
h^{\prime }\left( w\right) = & 2\left\{ 
\begin{array}{l}
\left\{ 
\begin{array}{ll}
\left\{ \frac{\alpha ^{\prime }}{2}\right\} _{F}-w, & 0<w<2\left\{ \frac{%
\alpha ^{\prime }}{2}\right\} _{F}, \\ 
-\left\{ \frac{\alpha ^{\prime }}{2}\right\} _{F}, & 2\left\{ \frac{\alpha
^{\prime }}{2}\right\} _{F}<w<1, \\ 
w-1-\left\{ \frac{\alpha ^{\prime }}{2}\right\} _{F}, & 1<w<1+2\left\{ \frac{%
\alpha ^{\prime }}{2}\right\} _{F}, \\ 
\left\{ \frac{\alpha ^{\prime }}{2}\right\} _{F}, & 1+2\left\{ \frac{\alpha
^{\prime }}{2}\right\} _{F}<w<2
\end{array}
\right. ,\,\,\text{for }\left\{ \frac{\alpha ^{\prime }}{2}\right\} _{F}<%
\frac{1}{2} \\ 
\,\,\, \\ 
\left\{ 
\begin{array}{ll}
1-\left\{ \frac{\alpha ^{\prime }}{2}\right\} _{F}, & 0<w<2\left\{ \frac{%
\alpha ^{\prime }}{2}\right\} _{F}-1, \\ 
\left\{ \frac{\alpha ^{\prime }}{2}\right\} _{F}-w, & 2\left\{ \frac{\alpha
^{\prime }}{2}\right\} _{F}-1<w<1, \\ 
\left\{ \frac{\alpha ^{\prime }}{2}\right\} _{F}-1, & 1<w<2\left\{ \frac{%
\alpha ^{\prime }}{2}\right\} _{F}, \\ 
w-1-\left\{ \frac{\alpha ^{\prime }}{2}\right\} _{F}, & 1+2\left\{ \frac{%
\alpha ^{\prime }}{2}\right\} _{F}<w<2
\end{array}
\right. ,\,\,\text{for }\left\{ \frac{\alpha ^{\prime }}{2}\right\} _{F}>%
\frac{1}{2}
\end{array}
\right. .
\end{array}
\label{tlbtvl58}
\end{equation}
\begin{multicols}{2}
\begin{figure}
\begin{minipage}{0.99\linewidth}
\begin{center}
\epsfxsize=.95\linewidth \epsfysize=.89\epsfxsize \epsfbox{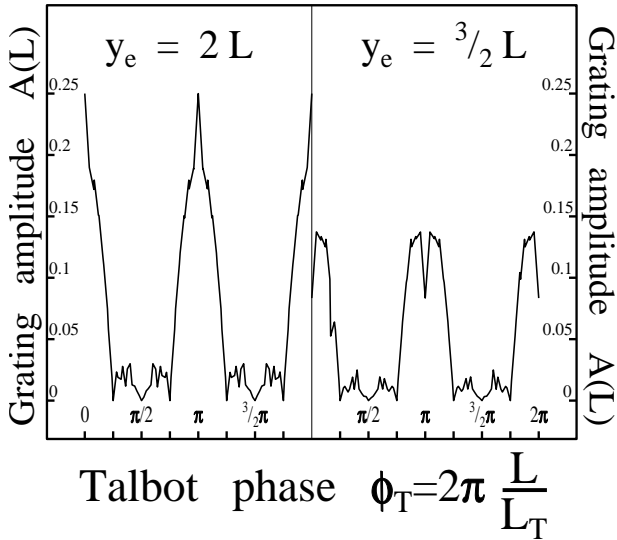}
\end{center}
\end{minipage}
\begin{minipage}{0.99\linewidth} \caption{Gratings' amplitudes 
$A\left( L\right) $ at the focal planes $y=2L$
and $y=\frac{3}{2}L$ as a function of the distance $L$ between
microfabricated structures. Two separated microfabricated structures have
the same period $\left( d_{1}=d_{2}\right) $ and relative width $f_{1}=\frac{%
1}{2},\,\,f_{2}=\frac{1}{4}.$ 
\label{tlbtvlf3}}
\end{minipage}
\end{figure}
This expression describes an atomic grating having period $\Delta w=2$ or $%
\Delta x=\frac{d_g}{n^{\prime }}$, which is $n^{\prime }$ times narrower
than that caused by the shadow effect. For the parameters chosen in Fig. \ref
{tlbtvlf2}, values of the Talbot phase are equal to $\phi _T(m,n)=\pi
/n^{\prime }$ ($n^{\prime }=1,\,3\,\,$or $5)$, $\alpha ^{\prime }=\beta
^{\prime }=n^{\prime }$, and the atomic density is given by 
\begin{equation}
f\left( x,t_e\right) =\frac 1{12}\left\{ 1-\frac 2{n^{\prime 2}}\left[
\left| 2\left\{ \frac{nn^{\prime }x}{d_1}\right\} _F-1\right| -\frac 12%
\right] \right\} .  \label{tlbtvl59}
\end{equation}

Since the atomic density is a periodic function of the distance between the
MS having period $L_{T}$, the dependence of the grating amplitude $A$ at a
given focal plane must also be a periodic function of $L$ having period $%
L_{T}$ for a fixed value of the ratio $y_{e}/L$. One period of $A(L)$ is
shown in Fig. \ref{tlbtvlf3} at the focal planes $y_{e}=2L,\,\,3L/2$. This
dependence is plotted for values of $L$ equal to rational multiples of the
Talbot length $L=\frac{m_{T}}{n_{T}}L_{T}$. One can not expect the
dependence of $A\left( L\right) $ to be smooth, because the transmission
function (\ref{tlbtvl29}) is discontinuous; even small changes in the ratio $%
\frac{m_{T}}{n_{T}}$ can lead to dramatic changes in the atomic density (\ref
{tlbtvl26}). 

\subsection{Comparison of the Talbot and Talbot-Lau effects}

Qualitatively, the transition from the shadow effect to the Talbot-Lau
effect for a beam having a large angular divergence that is scattered by two
MS parallels the transition from spatial modulation to Talbot self-imaging
of a collimated beam that is scattered by a single MS. Similarities and
differences of these transitions, which occur when the characteristic length
scale in the problem changes from $L\ll L_T$ to $L\sim L_T,$ can be
summarized as follows:
\end{multicols}
\newpage 
$---------------------------------------------------------$

\vspace{.2in}

\parbox{3.3in}{\begin{center}
Transition from shadow to Talbot-like profile\\
(collimated beam)
\end{center}}\hspace{.3in}%
\parbox{3.3in}{\begin{center}
Transition from shadow effect to Talbot-Lau\\
 effect (divergent beam)
\end{center}}

\vspace{.3in}

\parbox[t]{3.3in}{Atomic density is a periodic function of the distance $L$ between the MS and the screen
having period $L_T=\frac{2d^2}{\lambda _{dB}}.$}\hspace{.3in}%
\parbox[t]{3.3in}{Atomic density at a given focal plane $y=\frac mnL$ is a periodic function of the distance $L$
between the two separated MS$,$ having period $L_T=\frac{2d_1^2}{\lambda
_{dB}}\frac{\bar{\ell}}{\bar{j}^2\left( m-n\right) }$.}

\vspace{.2in}

\parbox[t]{3.3in}{Higher order gratings (with respect to the MS-grating) can be obtained at distances
$L=\frac{L_T}{n_T};$ if, for example, $n_T$ is odd, an atomic grating having period $\frac d{n_T}$ is produced if
the MS's duty cycle $f<\frac 1{n_T}.$}\hspace{.3in}%
\parbox[t]{3.3in}{Higher order gratings (with respect to those focused in the shadow-effect-regime) can be
obtained at distances $L=\frac{L_T}{n_T};$ if, for example, $n_T$ is odd, an atomic grating having
period $\frac{d_1}{\bar jnn_T}$ is produced if the second MS's duty cycle $f_2<\frac 1{n_T}.$}

\vspace{.2in}

\parbox[t]{3.3in}{The atomic grating's profile is the same as MS's profile, no
compression occurs.}\hspace{.3in}%
\parbox[t]{3.3in}{The atomic grating's profile is the corresponding shadow effect grating's
 profile compressed by a factor $n_T$.}
$---------------------------------------------------------$
\begin{multicols}{2}

\subsection{Additional examples including a quantum Talbot-Lau effect}

To make some connection with previous work, we analyze the atomic density
for the parameters of Eq. (\ref{tlbtvl67}), corresponding to the Talbot-Lau
effect studied theoretically by Carnal et al. $\left( 1995\right) $ and
realized experimentally by Clauser and Li $\left( 1994\right) $. The
appropriate parameters are $\left\{ m,\,\,n,\,\,j,\,\,\ell ,\bar{j},\bar{\ell%
}\right\} =$ $\left\{ 2,\,\,1,\,2,\,\,1,1,1/2\right\} $, 
\begin{equation}
d_g=d_1,\text{ }L_T(2,1)=\frac{d_1^2}{\lambda _{dB}},d_{1\ }=2d_2
\label{tlbtvl69}
\end{equation}
When the distance between the MS is $L=\frac{L_T(2,1)}2$, corresponding to $%
\phi _T=\pi $ [the case analyzed by Carnal et al. $\left( 1995\right) $],
one has $n_T=2n^{\prime }=L_T(2,1)/L=2$, and the corresponding values of $%
\alpha ^{\prime }$ and $\beta ^{\prime }$ obtained from Eqs. (\ref{tlbtvl54}%
) are 
\begin{equation}
\alpha ^{\prime }=2f_2,\,\,\beta ^{\prime }=2f_1.  \label{tlbtvl70}
\end{equation}
As in (Carnal et al., $1995)$, we choose $f_2=\frac 12$ and $f_1=\frac 12$
or $f_1=\frac 14.$ In order to apply the results of Sec. V.B.2, it is
necessary that $f_2\leq 1/n_T=1/2$; clearly, this requirement is met.

When $f_1=\frac 12$, the parameter $\beta ^{\prime }$ is an integer $\left(
\beta ^{\prime }=1\right) $ and one can use Eqs. (\ref{tlbtvl56}, \ref
{tlbtvl58}) to obtain the atomic density 
\begin{equation}
f\left( x,t_e\right) =\frac 12\left( 1-\left| 2\left\{ \frac x{d_1}\right\}
_F-1\right| \right) ,  \label{tlbtvl71}
\end{equation}
coinciding with the profile obtained by Carnal et al. $\left( 1995\right) $.
Since in this case both parameters $\alpha ^{\prime }$ and $\beta ^{\prime }$
are integers, the shadow effect does not lead to the any atomic grating (see
below). The grating (\ref{tlbtvl71}) arises entirely as a result of
matter-wave interference.

When $f_1=\frac 14,$ the parameter $\beta ^{\prime }=\frac 12,$ and one has
to return to Eq. (\ref{tlbtvl54}), in which only the $s=0$ term in the sum
contributes. Carrying out the integration in Eq. (\ref{tlbtvl54a}), one
finds 
\begin{equation}
f\left( x,t_e\right) =\left\{ 
\begin{array}{cc}
0, & for\,\,\left\{ \frac x{d_1}\right\} _F\leq \frac 14 \\ 
\left\{ \frac x{d_1}\right\} _F-\frac 14, & for\,\,\frac 14\leq \left\{ 
\frac x{d_1}\right\} _F\leq \frac 12 \\ 
\frac 14, & for\,\,\frac 12\leq \left\{ \frac x{d_1}\right\} _F\leq \frac 34
\\ 
1-\left\{ \frac x{d_1}\right\} _F, & for\,\,\frac 34\leq \left\{ \frac x{d_1}%
\right\} _F\leq 1
\end{array}
\right. ,  \label{tlbtvl72}
\end{equation}
coinciding with the distribution calculated by Carnal et al. $\left(
1995\right) $. To compare this profile with that caused by the shadow
effect, one finds from Eqs. (\ref{shdslt31}, \ref{shdslt32c}), that the
shadow effect distribution function is given by $\left. f\left( x,t_e\right)
\right| _{shadow}=f(x-\frac{d_1}2,t_e)$. Thus, owing to matter-wave
interference, the atomic grating (\ref{tlbtvl72}) is shifted by a
half-period from the grating that would have been produced by the shadow
effect alone.

\begin{figure}
\begin{minipage}{0.99\linewidth}
\begin{center}
\epsfxsize=.95\linewidth \epsfysize=\epsfxsize \epsfbox{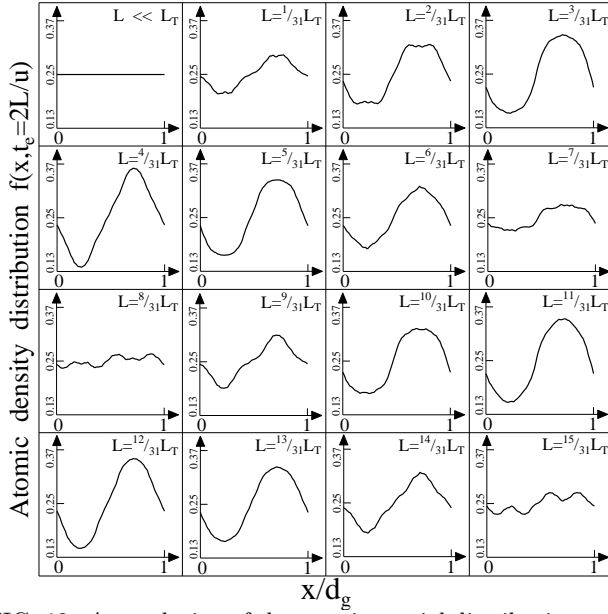}
\end{center}
\end{minipage}
\begin{minipage}{0.99\linewidth} \caption{An evolution of the atomic 
spatial distribution at the echo-point $%
t=2T$ $\left( \text{focal plane }y=2L\right) ,$ induced by two separated
microfabricated structures having the same spacings and 50\% relative widths 
$\left( f_{1}=f_{2}=\frac{1}{2}\right) $, with an increase of the distance $%
L $ between structures on the scale of the Talbot distance $L_{T}.$ Chosen
case allows exclude any influence of the classical shadow effect and
interpret the atomic grating as an exact consequence of the wave-matter
interference. The atomic grating as a function of $L$ is monitored for $L<%
\frac{1}{2}L_{T}$ with step $\frac{1}{31}L_{T}$ starting from the point
where one has no influence of the atom interference. Gratings at $\frac{1}{2}%
L_{T}<L\leq L_{T}$ can be obtained using the relation (\ref{tlbtvl62a}). 
\label{tlbtvlf4}}
\end{minipage}
\end{figure}

In general, the atomic gratings produced when a beam scatters from two,
separated MS cannot be attributed entirely to quantum effects since the
classical shadow effect contributes to grating formation. If the parameters
are chosen in such a way, however, that the classical shadow effect
vanishes, then any atomic gratings that are formed can be attributed solely
to quantum matter-wave interference. We have already alluded to this result
above. Returning to Eqs. (\ref{shdslt32}, \ref{shdslt31}), one finds that
the shadow effect grating $S\left( w\right) $ disappears if the parameters $%
\alpha =m\bar{\ell}f_{2}$ or $\beta =\bar{j}(m-n)f_{1}$ are integers. One
can guarantee that $\alpha $ is integral by choosing 
\begin{equation}
f_{1}=f_{2}=\frac{1}{2},\,\,n=1,\,\,m=2,\,\,j=1,\,\,\ell =1,
\label{tlbtvl60}
\end{equation}
which corresponds to $y_{e}=2L$, $d_{1\ }=d_{2}$, $\bar{j}=\bar{\ell}%
=1,\,\,\alpha =1$, $\beta =\frac{1}{2}$ and a Talbot phase $\phi _{T}=2\pi 
\frac{L}{L_{T}(2,1)}$. The atomic density in this focal plane as a function
of Talbot phase is shown in Fig. \ref{tlbtvlf4}. 

\section{Talbot and Talbot--Lau effects in a thermal atomic beam.}

Up to this point, all effects related to a distribution of longitudinal
velocities $u$ in the atomic beam have been neglected. Averaging over $u$ is
not important for the shadow effect since the focal planes are located at $%
y_e=(m/n)L$, independent of $u$. In both the Talbot and Talbot-Lau effects,
the Talbot phase depends on the Talbot length $L_T=2d^2/\lambda _{dB}$,
which, in turn, is proportional to $u$ owing to the presence of the De
Broglie wavelength. To achieve the maximum contrast in the Talbot and
Talbot-Lau effects, it is necessary to longitudinally cool the atomic beam
(Clauser and Li, 1994). The results of Secs. IV and VI must be averaged over 
$u$ once changes in the Talbot phase originating from the distribution of
longitudinal velocities becomes of order unity. For a thermal beam having a
Maxwellian distribution over longitudinal velocities, the averaging can be
carried out using the function tabulated by Kruse and Ramsey $\left(
1951\right) $. For other distributions numerical integration is needed. Such
calculations are not included in this contribution.
\begin{figure}
\begin{minipage}{0.99\linewidth}
\begin{center}
\epsfxsize=.95\linewidth \epsfysize=.89\epsfxsize \epsfbox{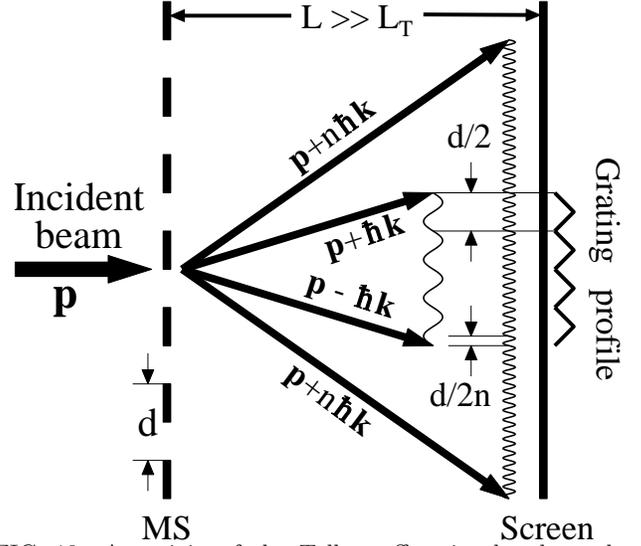}
\end{center}
\end{minipage}
\begin{minipage}{0.99\linewidth} \caption{An origin of the Talbot effect 
in the thermal beam at an
asymptotic distance $L>L_{T}.$ When incident beam of atoms having momenta $%
{\bf p}$ splits in a set of scattered beams having momenta ${\bf p\pm }%
n\hbar {\bf k.}$ The wave functions of the states associated with momenta $%
{\bf p+}\hbar {\bf k}$ and ${\bf p-}\hbar {\bf k,\,\,p+}2\hbar {\bf k}$ and $%
{\bf p-}2\hbar {\bf k,\,\,\ldots \,\,p+}n\hbar {\bf k}$ and ${\bf p-}n\hbar 
{\bf k}$ acquire the same Talbot phase and therefore no dephasing between
these states occurs, independently on the distance $L$ and atomic velocity $%
u.$ Interference of these states remains at the asymptotic distance, leads
to the 2nd, 4th $\ldots $ harmonics in the atomic distribution which form
the $2-$order atomic grating on the screen. 
\label{thrmbef3}}
\end{minipage}
\end{figure}

Instead, we examine the role of the longitudinal velocity distribution when
the width $\bar{u}$ of the longitudinal velocity distribution is of order of
the average velocity, 
\begin{equation}
\bar{u}\sim u  \label{thrmbeam1}
\end{equation}
for distances 
\begin{equation}
(D/d)L_{T}\gg y\gg L_{T}  \label{thrmbeam2}
\end{equation}
in the Talbot effect and separations $L$ between the MS 
\begin{equation}
(D/d)L_{T}\gg L\gg L_{T}  \eqnum{168a}  \label{thrmbeam2a}
\end{equation}
in the Talbot-Lau effect. We want to examine whether or not it is possible
under these conditions to obtain atomic gratings having periods smaller than
the MS producing the scattering.

To understand how the gratings can survive the average over longitudinal
velocities $u$, one should note that it is the atomic density (\ref{talbot9}%
) and not the wave function amplitude that is averaged. The phase factors in
the atomic density depending on the Talbot phase can be unity for specific
combinations of the spatial harmonics in the atomic wave functions. The way
in which this can be achieved is illustrated in Fig. \ref{thrmbef3}. When
one combines on the screen components of the scattered atomic states
associated with momenta $\hbar k$ and $-\hbar k,\,\,2\hbar k$ and $-2\hbar
k,\,\,\ldots ,$ etc., the amplitudes of the combining states acquire the
same Talbot phases since the energy of the scattered atoms does not depend
on the direction of scattering. The interference from these pairs of states
leads to a superposition of atomic gratings having period $d/2$, $d/4$,
etc.; the overall period of the grating is $d/2$. Gratings originated from
Talbot-Lau effect can survive in a similar manner. 

\subsection{ Atomic density profile for a thermal beam.}

\subsubsection{Talbot effect}

Consider first the Talbot effect, i. e., the atomic grating produced when an
atomic beam having negligible angular divergence, but a finite spread of
longitudinal velocities, is scattered by a MS having period $d$. For a given 
$u$, the atomic density in the plane $y=ut$ is given by [see Eqs. (\ref
{talbot8}, \ref{talbot9})] 
\begin{equation}
f_{u}(x,t)=\sum_{n,n^{\prime }}\eta _{n}\eta _{n^{\prime }}^{\ast }\exp 
\left[ i\left( n-n^{\prime }\right) kx-i\left( n^{2}-n^{\prime \,2}\right)
\phi _{t}(u)\right] ,  \label{thrmbeam3}
\end{equation}
where the Talbot phase $\phi _{t}(u)=\omega _{k}t=\omega _{k}y/u$, as given
by Eq. (\ref{talbot36}), is a function of $u$ for fixed $y$. Recall that $%
\eta _{n}$ is a Fourier component of the amplitude transmission function $%
\eta (x)$. Since the Talbot phase is much greater than unity in the
asymptotic region (\ref{thrmbeam2}), the distribution (\ref{thrmbeam3})
oscillates rapidly as a function of $u.$ After averaging over $u,$ a
non-zero result arises from only those terms having 
\begin{equation}
n^{\prime }=\pm n.  \label{thrmbeam4}
\end{equation}
There is a constant background term $\bar{f}$ ( for a MS consisting of an
array of slits, $\bar{f}$ is equal to the duty cycle $f$ of the MS)
corresponding to contributions with $n^{\prime }=n$ and an interference term 
$\tilde{f}(x,t)$ corresponding to contributions from $n^{\prime }=-n$.
Neglecting all other terms, one finds 
\begin{equation}
f(x,t)=\bar{f}+\tilde{f}(x,t),  \label{thrmbeam5}
\end{equation}
where 
\begin{equation}
\bar{f}=\sum_{n}\left| \eta _{n}\right| ^{2}=\int_{0}^{d}\frac{dx^{\prime }}{%
d}\chi (x^{\prime }),  \eqnum{171a}  \label{thrmbeam5a}
\end{equation}
and 
\begin{eqnarray}
\tilde{f}(x,t) &=&\int_{0}^{d}\frac{dx^{\prime }}{d}\int_{0}^{d}\frac{%
dx^{\prime \prime }}{d}\eta \left( x^{\prime }\right) \eta ^{\ast }\left(
x^{\prime \prime }\right)  \nonumber \\
&&\times \sum_{n\neq 0}\exp \left[ ink\left( 2x-x^{\prime }-x^{\prime \prime
}\right) \right] .  \eqnum{171b}  \label{thrmbeam5b}
\end{eqnarray}
The atomic density profile has a period given by 
\begin{equation}
d_{g}=\frac{d}{2}.  \label{thrmbeam27}
\end{equation}
Note that the density profile is independent of $t$ for the times $t\gg
L_{T}/u$ under consideration.

To evaluate the atomic distribution (\ref{thrmbeam5}), it is convenient to
introduce new variables 
\begin{equation}
\bar{x}=\frac 12\left( x^{\prime }+x^{\prime \prime }\right) ,\,\,\hat{x}%
=x^{\prime }-x^{\prime \prime }.  \label{thrmbeam6}
\end{equation}
After adding and subtracting a term having $n=0$ in Eq. (\ref{thrmbeam5b})
one can carry out the summation to obtain $\frac d2\sum_s\delta \left( \bar{x%
}-x-\frac{sd}2\right) $, making use of Eq. (\ref{talbot17}). Switching
integration variables from $(x^{\prime },x^{\prime \prime })$ to $(\bar{x}$,$%
\,\hat{x})$ one sees that the $\delta -$functions having $s=0$ and $1$ are
the only ones that contribute to the sum, and it follows from Eq. (\ref
{thrmbeam5b}) that 
\begin{equation}
f(x,t)=\bar{f}-\left| \int_0^d\frac{dx^{\prime }}d\eta \left( x^{\prime
}\right) \right| ^2+\frac 12\left[ F\left( x\right) +F\left( x+\frac d2%
\right) \right] ,  \label{thrmbeam8}
\end{equation}
where 
\begin{equation}
F(\bar{x})=\int_{\left| \hat{x}\right| <2\min \left( \bar{x},d-\bar{x}%
\right) }\frac{d\hat{x}}d\eta \left( \bar{x}+\frac{\hat{x}}2\right) \eta
^{*}\left( \bar{x}-\frac{\hat{x}}2\right) .  \label{thrmbeam7}
\end{equation}
For a transmission function corresponding to a periodic array of slits
having duty cycle $f$, one finds $\bar{f}=\int_0^d\frac{dx^{\prime }}d\chi
\left( x^{\prime }\right) =f,$ and 
\begin{equation}
F\left( x\right) =\left\{ 
\begin{array}{c}
2\left( f-2\left| \left\{ \frac{x}{d}\right\} _{F}-\frac{f}{2}\right|
\right) ,\text{ for }\left\{ \frac{x}{d}\right\} _{F}<f \\ 
0,\text{ for }\left\{ \frac{x}{d}\right\} _{F}>f
\end{array}
\right. .  
\label{thrmbeam9}
\end{equation}
When $f<\frac 12,$ the two $F-$functions in Eq. (\ref{thrmbeam8}) do not
overlap with one another, and the atomic density is given by 
\begin{equation}
f\left( x,t\right) =f(1-f)+2\left\{ 
\begin{array}{ll}
w, & \text{for }0<w<\frac f2 \\ 
f-w, & \text{for }\frac f2<w<f \\ 
0, & \text{for }f<w<\frac 12
\end{array}
\right. ,  \label{thrmbeam10}
\end{equation}
where $w=\frac 12\left\{ \frac{2x}d\right\} _F.$ This function has period $%
\frac d2$. For $f>\frac 12$ one arrives at the distribution 
\begin{equation}
f\left( x,t\right) =f(1-f)+2\left\{ 
\begin{array}{ll}
f-\frac 12, & \text{for }0<w<f-\frac 12 \\ 
w, & \text{for }f-\frac 12<w<\frac f2 \\ 
f-w, & \text{for }\frac f2<w<\frac 12
\end{array}
\right. ,  \label{thrmbeam11}
\end{equation}
which also has period $\frac d2$. The amplitude of the atomic grating (\ref
{thrmbeam10}, \ref{thrmbeam11}) is given by 
\begin{equation}
A=\min \left[ f,\left( 1-f\right) \right] .  \label{thrmbeam12}
\end{equation}

\begin{figure}
\begin{minipage}{0.99\linewidth}
\begin{center}
\epsfxsize=.95\linewidth \epsfysize=1.54\epsfxsize \epsfbox{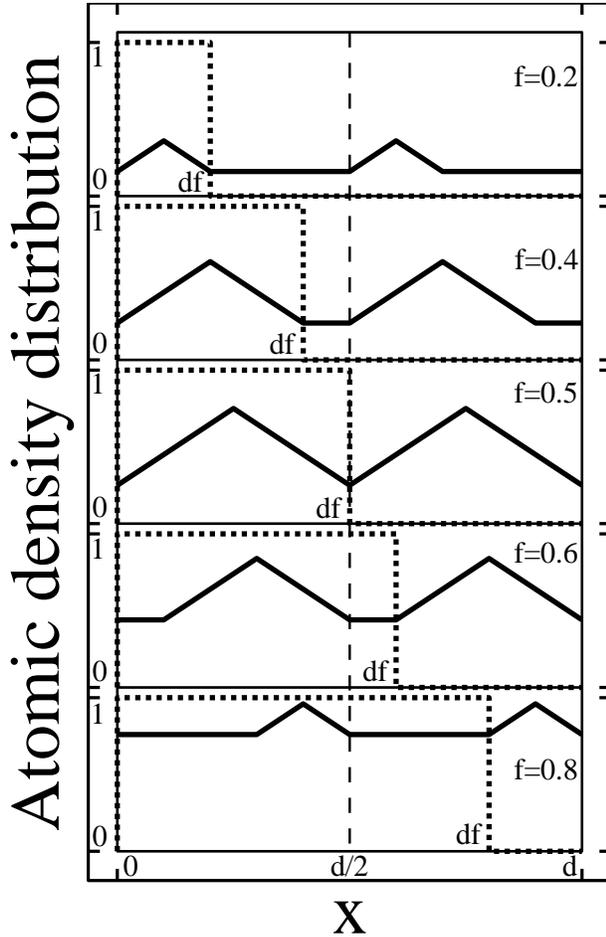}
\end{center}
\end{minipage}
\begin{minipage}{0.99\linewidth} \caption{Talbot effect in the thermal 
beam. The initial atomic distribution
(dashed lines), created just after passing through the microfabricated
structure with the slits' relative width $f$ and spasing $d,$ transforms
into the second order garting $\left( \text{having spacing }\frac{d}{2}%
\right) $ on the distance much larger than the Talbot distance owing to the
wave-matter interference. 
\label{thrmbef1}}
\end{minipage}
\end{figure}

The manner in which the atomic density profile changes as $y$ varies from $%
y\ll L_{T}$ to $y\gg L_{T}$ is shown in Fig. \ref{thrmbef1}. 

\subsubsection{Talbot-Lau effect}

To evaluate the Talbot-Lau density profile in the asymptotic limit (\ref
{thrmbeam2}), one must return to Eq. (\ref{tlbtvl16}) and average it over
longitudinal velocities. Using the Fourier expansion of the amplitude
transmission functions in Eq. (\ref{tlbtvl16}) and setting the smooth
envelope function $f\left( x\right) $ equal to unity, one finds 
\begin{eqnarray}
f(x,t_{e}) &=&\sum_{q,n_{2}}\exp \left( iq\left[ k_{g}x-\phi
_{T}(m,n;u)\left( 2n_{2}-m\bar{\ell}q\right) \right] \right)   \nonumber \\
&&\times \chi _{-\bar{j}(m-n)q}^{(1)}\eta _{n_{2}}^{(2)}\left[ \eta _{n_{2}-m%
\bar{\ell}q}^{(2)}\right] ^{\ast },  \label{thrmbeam13}
\end{eqnarray}
where $\phi _{T}(m,n;u)$ is given by (\ref{tlbtvl18})$.$

On averaging over $u$ for $\phi _{T}(m,n;u)\gg 1$, one finds that only those
terms in the sum having $q=0$ or 
\begin{equation}
n_{2}=m\bar{\ell}\frac{q}{2},\,\,q\neq 0  \label{thrmbeam14}
\end{equation}
contribute to the density. The $q=0$ term is a background, given by $\bar{f}%
_{1}\bar{f}_{2}$ [$\bar{f}_{j}$ is defined in terms of $\chi _{j}$ in the
same way that $\bar{f}$ is defined in terms of $\chi $ in Eq. (\ref
{thrmbeam5a})]$,$ while the terms satisfying Eq. (\ref{thrmbeam14}) lead to
the atomic grating. Explicitly, one finds 
\begin{equation}
f\left( x,t_{e}\right) =\bar{f}_{1}\left[ \bar{f}_{2}-\left| \int \frac{%
dx_{2}}{d_{2}}\eta _{2}\left( x_{2}\right) \right| ^{2}\right] +\tilde{f}%
\left( x,t_{e}\right) ,  \label{thrmbeam15}
\end{equation}
\begin{equation}
\tilde{f}\left( x,t_{e}\right) =\sum_{q}\exp \left( iqk_{g}x\right) \chi _{-%
\bar{j}\left( m-n\right) q}^{(1)}F_{m\bar{\ell}q}^{\left( 2\right) }, 
\eqnum{182a}  \label{thrmbeam15a}
\end{equation}
where the summation over $q$ includes $q=0$ and other values of $q$ leading
to integral $n_{2}$ in Eq. (\ref{thrmbeam14}). The function $F_{\nu
}^{\left( 2\right) }$ is a Fourier component of the function $F_{2}(x)$
defined in terms of $\eta _{2}\left( x\right) $ in the same way that $F$ is
defined in terms of $\eta $ in Eq. (\ref{thrmbeam7})$.$ The density is
independent of $t_{e}$ for spatial separations $L\gg L_{T}$ of the MS.

When $m\bar{\ell}$ is even, all $q$ are allowed according to Eq. (\ref
{thrmbeam14}). When $m\bar{\ell}$ is odd, only even values $q$ contribute,
which means that the grating (\ref{thrmbeam15a}) has a period equal to $%
d_{g}/2.$ Equation (\ref{thrmbeam15a}) has the same structure as Eq. (\ref
{shdslt13}). Repeating the calculations leading to Eq. (\ref{shdslt42}) one
arrives at the formula 
\begin{eqnarray}
&&\left. \tilde{f}(x,t_{e})=\frac{1}{\bar{j}(m-n)}\sum_{s=0}^{\bar{j}%
(m-n)-1}\int_{0}^{d_{2}}\frac{dx_{2}}{d_{2}}F_{2}\left( x_{2}\right) \right. 
\nonumber \\
&&\left. \times \chi _{1}\left\{ \frac{d_{1}}{\bar{j}(m-n)}\left[ s+\left\{
\xi \left( m\bar{\ell}\frac{x_{2}}{d_{2}}-\frac{x}{d_{g}}\right) \right\}
_{F}\right] \right\} \right. ,  \label{thrmbeam16}
\end{eqnarray}
where 
\begin{equation}
\xi =\left\{ 
\begin{array}{c}
1 \\ 
2
\end{array}
\right. ,\text{ for }m\bar{\ell}\text{ }
\begin{array}{c}
\text{even} \\ 
\text{odd}
\end{array}
.  \label{thrmbeam17}
\end{equation}

Consider now the case of MS having transmission functions (\ref{shdslt21p}),
when the function $F_{2}\left( x\right) $ is given by the rhs of Eq. (\ref
{thrmbeam9}) with $d$ and $f$ replaced by $d_{2}$ and $f_{2}.$ Using
dimensionless variables 
\begin{equation}
w=\xi x/d_{g},\,\,z=x_{2}/f_{2}d_{2},  \label{thrmbeam18}
\end{equation}
one arrives at equations that are the analogues of Eqs. (\ref{shdslt23}),
namely 
\begin{equation}
f\left( x,t_{e}\right) =f_{1}f_{2}\left( 1-f_{2}\right) +\tilde{f}\left(
x,t_{e}\right) ,  \label{thrmbeam19}
\end{equation}
\begin{equation}
\tilde{f}\left( x,t_{e}\right) =\frac{f_{2}^{2}}{\bar{j}\left( m-n\right) }%
\sum_{s=0}^{\left[ \beta \right] _{I}}h_{s}\left( w\right) ,  \eqnum{186a}
\label{thrmbeam19a}
\end{equation}
\begin{equation}
h_{s}\left( w\right) =\int_{0}^{1}dz\tilde{F}(z)\theta \left( \beta -\left(
s+\left\{ \alpha z-w\right\} _{F}\right) \right) ,  \eqnum{186b}
\label{thrmbeam19b}
\end{equation}
\begin{equation}
\tilde{F}\left( z\right) =4\left\{ 
\begin{array}{c}
z,\text{ for }z<\frac{1}{2} \\ 
\frac{1}{2}-z,\text{ for }z>\frac{1}{2}
\end{array}
\right. ,  \eqnum{186c}
\label{thrmbeam19c}
\end{equation}
\begin{equation}
\alpha =\xi m\bar{\ell}f_{2},\,\,\beta =\bar{j}\left( m-n\right) f_{1}. 
\eqnum{186d}  \label{thrmbeam19d}
\end{equation}

Omitting further calculations which are essentially the same as those used
to derive Eqs. (\ref{shdslt32}, \ref{shdslt31}), one finds 
\begin{equation}
\tilde{f}\left( x,t_{e}\right) =\frac{f_{2}^{2}}{\bar{j}\left( m-n\right) }%
\left( \left[ \beta \right] _{I}+\sum_{s=1}^{\left[ \alpha \right]
_{I}}F\left( a_{s},b_{s}\right) +f^{\prime }\left( w\right) \right) ,
\label{thrmbeam20}
\end{equation}
where 
\begin{equation}
F\left( a,b\right) \equiv \int_{a}^{b}dz\tilde{F}\left( z\right)
\label{thrmbeam22}
\end{equation}
is given by 
\begin{equation}
F\left( a,b\right) =\left\{ 
\begin{array}{ll}
2\left( b^{2}-a^{2}\right) & \max \left( a,b\right) \leq \frac{1}{2} \\ 
4b-2\left( a^{2}+b^{2}\right) -1 & a\leq \frac{1}{2}\leq b \\ 
2\left( b-a\right) \left( 2-b-a\right) & \min \left( a,b\right) \geq \frac{1%
}{2}
\end{array}
\right. ,  \label{thrmbeam23}
\end{equation}
the quantities $a_{s}$ and $b_{s}$ are given in the Appendix by Eqs. (\ref
{shdslt26}, \ref{shdslt27}), and
\end{multicols}

\begin{equation}
f^{\prime }\left( w\right) =\left\{ 
\begin{array}{ll}
F\left( a_{\left[ \alpha \right] _{I}+1},1\right) , & \text{for }0\leq w\leq
1-\left\{ \beta \right\} _{F} \\ 
F\left( a_{\left[ \alpha \right] _{I}+1},1\right) +F\left( 0,b_{0}\right) ,
& \text{for }1-\left\{ \beta \right\} _{F}\leq w\leq \left\{ \alpha \right\}
_{F} \\ 
F\left( 0,b_{0}\right) , & \text{for }\left\{ \alpha \right\} _{F}\leq w\leq
1+\left\{ \alpha \right\} _{F}-\left\{ \beta \right\} _{F} \\ 
F\left( 0,b_{0}\right) +F\left( b_{\left[ \alpha \right] _{I}},1\right) , & 
\text{for }1+\left\{ \alpha \right\} _{F}-\left\{ \beta \right\} _{F}\leq
w\leq 1
\end{array}
\right\} ,\text{ if }\left\{ \beta \right\} _{F}\geq \max \left( \left\{
\alpha \right\} _{F},1-\left\{ \alpha \right\} _{F}\right) ,
\label{thrmbeam21}
\end{equation}
\begin{equation}
f^{\prime }\left( w\right) =\left\{ 
\begin{array}{ll}
F\left( a_{\left[ \alpha \right] _{I}+1},1\right) , & \text{for }0\leq w\leq
\left\{ \alpha \right\} _{F} \\ 
0, & \text{for }\left\{ \alpha \right\} _{F}\leq w\leq 1-\left\{ \beta
\right\} _{F} \\ 
F\left( 0,b_{0}\right) , & \text{for }1-\left\{ \beta \right\} _{F}\leq
w\leq 1+\left\{ \alpha \right\} _{F}-\left\{ \beta \right\} _{F} \\ 
F\left( 0,b_{0}\right) +F\left( b_{\left[ \alpha \right] _{I}},1\right) , & 
\text{for }1+\left\{ \alpha \right\} _{F}-\left\{ \beta \right\} _{F}\leq
w\leq 1
\end{array}
\right\} ,\text{ if }\left\{ \alpha \right\} _{F}\leq \left\{ \beta \right\}
_{F}\leq 1-\left\{ \alpha \right\} _{F},  \eqnum{190a}  \label{thrmbeam21a}
\end{equation}
\begin{equation}
f^{\prime }\left( w\right) =\left\{ 
\begin{array}{ll}
F\left( a_{\left[ \alpha \right] _{I}+1},b_{\left[ \alpha \right]
_{I}+1}\right) , & \text{for }0\leq w\leq \left\{ \alpha \right\}
_{F}-\left\{ \beta \right\} _{F} \\ 
F\left( a_{\left[ \alpha \right] _{I}+1},1\right) , & \text{for }\left\{
\alpha \right\} _{F}-\left\{ \beta \right\} _{F}\leq w\leq 1-\left\{ \beta
\right\} _{F} \\ 
F\left( 0,b_{0}\right) +F\left( a_{\left[ \alpha \right] _{I}+1},1\right) ,
& \text{for }1-\left\{ \beta \right\} _{F}\leq w\leq \left\{ \alpha \right\}
_{F} \\ 
F\left( 0,b_{0}\right) , & \text{for }\left\{ \alpha \right\} _{F}\leq w\leq
1
\end{array}
\right\} ,\text{ if }1-\left\{ \alpha \right\} _{F}\leq \left\{ \beta
\right\} _{F}\leq \left\{ \alpha \right\} _{F},  \eqnum{190b}
\label{thrmbeam21b}
\end{equation}
\begin{equation}
f^{\prime }\left( w\right) =\left\{ 
\begin{array}{ll}
F\left( a_{\left[ \alpha \right] _{I}+1},b_{\left[ \alpha \right]
_{I}+1}\right) , & \text{for }0\leq w\leq \left\{ \alpha \right\}
_{F}-\left\{ \beta \right\} _{F} \\ 
F\left( a_{\left[ \alpha \right] _{I}+1},1\right) , & \text{for }\left\{
\alpha \right\} _{F}-\left\{ \beta \right\} _{F}\leq w\leq \left\{ \alpha
\right\} _{F} \\ 
0, & \text{for }\left\{ \alpha \right\} _{F}\leq w\leq 1-\left\{ \beta
\right\} _{F} \\ 
F\left( 0,b_{0}\right) , & \text{for }1-\left\{ \beta \right\} _{F}\leq
w\leq 1
\end{array}
\right\} ,\text{ if }\left\{ \beta \right\} _{F}\leq \min \left( \left\{
\alpha \right\} _{F},1-\left\{ \alpha \right\} _{F}\right) .  \eqnum{190c}
\label{thrmbeam21c}
\end{equation}
\begin{multicols}{2}
In principle one can use Eqs. (\ref{thrmbeam15}, \ref{thrmbeam20}, \ref
{thrmbeam21}) to derive a general analytical expression for the atomic
density distribution, but, given the large number of cases, such an
expression is of limited use. Instead, one can write a computer code based
on Eqs. (\ref{thrmbeam15}, \ref{thrmbeam20}, \ref{thrmbeam21}) to obtain the
atomic density profile. Using this code, we varied the duty cycles of the MS
to optimize the atomic grating amplitude in the focal planes $y=\frac{n+1}nL$
for $n=1-4$ and equal periods of the MS, $j=\ell =1.$ Calculations show that
one has to choose 
\begin{equation}
f_1=f_2=\frac 12  \label{thrmbeam24}
\end{equation}
in all cases except $n=3,$ where the optimal duty cycles are given by 
\begin{equation}
f_1=\frac 12,\,\,f_2=\frac 14.  \label{thrmbeam24a}
\end{equation}

For these optimal values of the duty cycles, it is a simple matter to obtain
analytical expressions for the atomic density profile in a given focal
plane. For example, at the echo plane $y=2L$ ($n=1,\,\,m=2$) the parameters $%
\alpha $ and $\beta $ are equal $1$ and $\frac{1}{2}$,$\ $respectively, and
the atomic density is given by 
\begin{eqnarray}
f\left( x,t_{e}\right)  &=&\frac{1}{8}+\frac{1}{4}\left\{ F\left(
a_{1},b_{1}\right) \right.   \nonumber \\
&&+\left. \theta \left( w-\frac{1}{2}\right) \left[ F\left( 0,b_{0}\right)
+F\left( b_{1},1\right) \right] \right\} ,  \label{thrmbeam25}
\end{eqnarray}
where $w=\left\{ \frac{x}{d_{1}}\right\} _{F},\,\,b_{0,1}=w\mp \frac{1}{2}%
,\,\,a_{1}=w.$ Using Eq. (\ref{thrmbeam23}) one arrives at the atomic
grating profile 
\begin{equation}
f\left( x,t_{e}\right) =\frac{1}{4}\left\{ 
\begin{array}{c}
1+2w(1-2w),\text{ for }w<\frac{1}{2} \\ 
3-2w(3-2w),\text{ for }w>\frac{1}{2}
\end{array}
\right. .  \label{thrmbeam26}
\end{equation}
\begin{figure}
\begin{minipage}{0.99\linewidth}
\begin{center}
\epsfxsize=.95\linewidth \epsfysize=1.12\epsfxsize \epsfbox{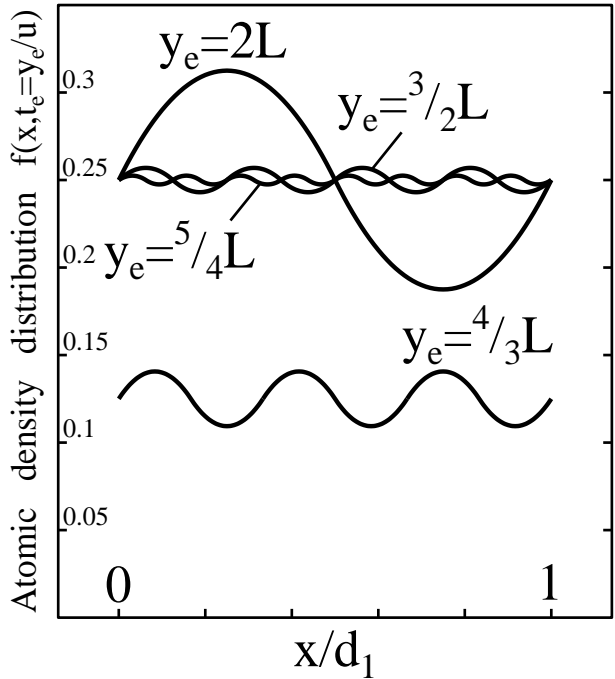}
\end{center}
\end{minipage}
\begin{minipage}{0.99\linewidth} \caption{Talbot-Lau effect in the thermal 
beam. The atomic density
distribution at the different focal planes $y$ plotted for the optimum
relative widths of the separated microfabricated structutres ($f_{1}=\frac{1%
}{2}$ and $f_{2}=\frac{1}{2}$ in all cases exept that for focal plane $y=%
\frac{4}{3}L,$ where $f_{2}=\frac{1}{4}).$ The distance between fields is
assumed to be much larger than the Talbot distance. In spite on this, one
can observe higher-order spatial gratings. Moreover, one finds from Eqs. (%
\ref{shdslt31}, \ref{shdslt32}) that for optimum widths there are no
gratings caused by classical shadow effect and conclude that using
microfabricated structures one observes gratings originated entirely from
the wave-matter interference. 
\label{thrmbef2}}
\end{minipage}
\end{figure}
Similar calculations leads to the atomic grating profiles 
\begin{equation}
f\left( x,t_{e}\right) =\frac{1}{36}\left\{ 
\begin{array}{c}
9+2w(1-2w),\text{ for }w<\frac{1}{2} \\ 
11-2w(3-2w),\text{ for }w>\frac{1}{2}
\end{array}
\right.   \eqnum{194a}  \label{thrmbeam26a}
\end{equation}
at the focal plane $y=\frac{3}{2}L,$ where $w=\left\{ \frac{4x}{d_{1}}%
\right\} _{F};$%
\begin{equation}
f\left( x,t_{e}\right) =\frac{1}{8}\left\{ 
\begin{array}{c}
1+w(1-2w),\text{ for }w<\frac{1}{2} \\ 
2-w(3-2w),\text{ for }w>\frac{1}{2}
\end{array}
\right.   \eqnum{194b}  \label{thrmbeam26b}
\end{equation}
at the focal plane $y=\frac{4}{3}L,$ where $w=\left\{ \frac{3x}{d_{1}}%
\right\} _{F};$ and 
\begin{equation}
f\left( x,t_{e}\right) =\frac{1}{100}\left\{ 
\begin{array}{c}
25+2w(1-2w),\text{ for }w<\frac{1}{2} \\ 
27-2w(3-2w),\text{ for }w>\frac{1}{2}
\end{array}
\right.   \eqnum{194c}  \label{thrmbeam26c}
\end{equation}
at the focal plane $y=\frac{5}{4}L,$ where $w=\left\{ \frac{8x}{d_{1}}%
\right\} _{F}.$ These atomic density profiles are shown in Fig. \ref
{thrmbef2}. Since the optimal duty cycles (\ref{thrmbeam24}, \ref
{thrmbeam24a}) correspond to the limit where the shadow effect vanishes, the
density profiles (\ref{thrmbeam26}) cannot be vestiges of the shadow effect.
They must originate from matter-wave interference. One can compare the
density profile (\ref{thrmbeam26a}), valid for distances $L$ between the MS
larger than the Talbot distance, with that of Fig. \ref{tlbtvlf2} for $L\sim
L_{T}$ (Talbot-Lau effect). 

\section{Conclusion}

Atom interferometry is an emerging field of atomic, molecular and optical
physics. In this review, we have focused on the scattering of atoms by one
or more microfabricated structures (MS). We have seen that the scattering
can be described in purely classical terms for characteristic length scales $%
L\ll L_T$, where $L_T$ is the so-called Talbot length. For $L\gtrsim L_T$, a
classical description of the atomic center-of-mass motion is no longer
adequate. Our approach has relied on an interpretation of the phenomena in
terms of the recoil an atom acquires when it is scattered from a MS. With
this approach, we could make a connection with the theory of coherent
transients, for which a rich literature has already been developed. We have
considered both collimated beams (Talbot effect) and beams having large
angular divergence (shadow effect, Talbot-Lau effect) and have allowed for a
broad distribution of longitudinal velocities in the atomic beam (Talbot and
Talbot-Lau effects in a thermal beam). The next step would be to extend our
considerations to regimes corresponding to Bragg scattering and Fraunhofer
diffraction, allowing for an analysis of atom interferometers which split
atomic wave functions into nonoverlapping paths.

Scattering of atoms by MS shares both similarities and differences with
scattering of atoms by standing-wave optical fields (SW). Similarities
include a periodical recovery of the atomic interference pattern at
multiples of the Talbot distance (\ref{talbot11}) [or (\ref{tlbtvl61}) for
the Talbot-Lau effect], a compression of the atomic gratings with respect to
the periods of the MS or SW, spatial separation of the higher order atomic
gratings in different focal planes, and splitting of the incident beam into
an infinite set of scattered beams having momenta ($p\pm n\hbar k$). The
differences are due in large part to the nature of the scattering. The MS
produce a piecewise continuous atomic density profile while the SW produce a
smooth atomic density profile. As a result, the decrease in period relative
to that of the classical shadow effect observed in the Talbot-Lau effect
using MS does not occur for scattering by standing-wave fields. Moreover,
the possibility of observing a Talbot-Lau effect caused entirely by
matter-wave interference (see Fig. \ref{tlbtvlf4}) does not occur for the
smooth amplitude modulation by SW (Dubetsky and Berman, 1994). In the case
of scattering by MS, the fact that the shadow effect does not give rise to
atomic gratings in the focal plane $y=2L$ for MS having duty cycles $f_i=%
\frac 12,$ is directly related to the stepwise amplitude modulation of the
atomic beam produced by the MS, as is evident from Fig. \ref{sltshf14}.

For scattering by MS, the qualitative nature of the atomic density profile
depends on the properties of the incident atomic beam. When one observes the
Talbot effect using a monovelocity beam, the atomic gratings are
discontinuous functions (see Fig. \ref{talbotf1}), but when one averages
these gratings over longitudinal velocities, the atomic density is
transformed into a piecewise continuous profile (compare with Fig. \ref
{thrmbef1}). Similarly, the shadow effect and Talbot-Lau effect atomic
gratings (which are averaged over transverse velocities) are piecewise
continuous, but they are transformed into profiles which are discontinuous
only in the second derivative when averaged over longitudinal velocities
(compare Figs. \ref{tlbtvlf2} and \ref{thrmbef2}). These examples show that
averaging over transverse or longitudinal velocities tends to smooth out the
atomic density profiles.

It is clear that many of the situations analyzed in this chapter have direct
applications to atom lithography. We can expect that future developments in
this emerging field will incorporate many of the basic ideas which have been
encountered in our discussion.

\acknowledgments

We are pleased to acknowledge helpful discussions with J. L. Cohen and Yu.
V. Rozhdestvensky. This material is based upon work supported by the U. S.
Army Research office under grant number DAAH04-93-G-0503, by the National
Science Foundation under grant PHY-9414020 and by the National Science
Foundation through the Center for Ultrafast Optical Science under STC\ PHY
8920108.

\appendix
\section*{}
%

\begin{figure}
\begin{minipage}{0.99\linewidth}
\begin{center}
\epsfxsize=.95\linewidth \epsfysize=\epsfxsize \epsfbox{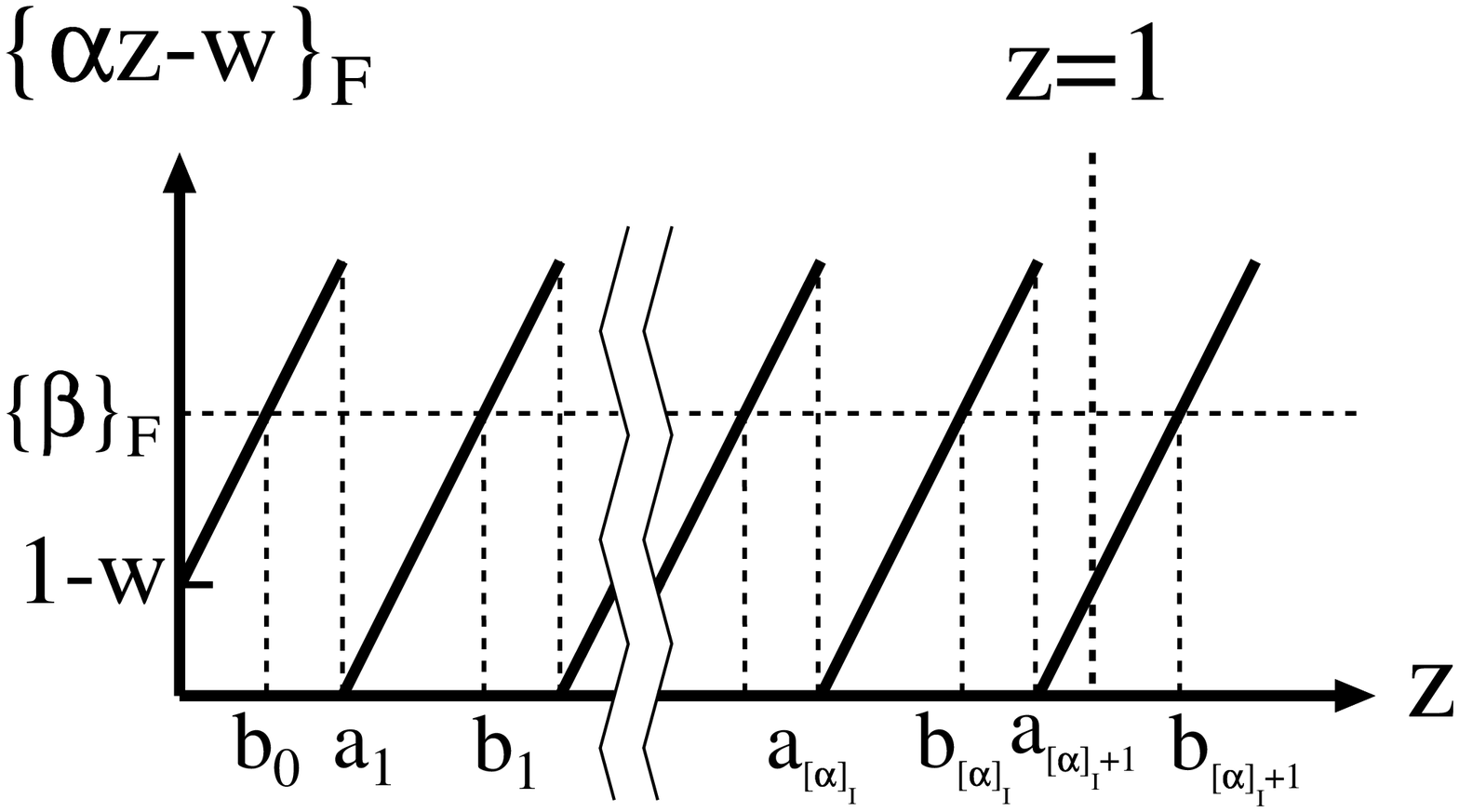}
\end{center}
\end{minipage}
\begin{minipage}{0.99\linewidth} \caption{Function $\left\{ \protect\alpha 
z-w\right\} $ undergoes jumps at
the points $a_{r}=\left( r-1+w\right) /\protect\alpha $ and is equal to $%
\left\{ \protect\beta \right\} $ at the points $b_{r}=\left( r-1+w+\left\{ 
\protect\beta \right\} \right) /\protect\alpha $. For $\left[ \protect\alpha %
\right] >1$ and $1\leq r\leq \left[ \protect\alpha \right] $, $a_{r}<1,$ i.
e. contribution to Eq. (\ref{shdslt25}) from interim $\left[ a_{1},a_{[%
\protect\alpha ]}\right] $ is $\left( \left[ \protect\alpha \right]
-1\right) \left( b_{r}-a_{r}\right) $; when $w>1-\left\{ \protect\beta
\right\} $ one gets term $b_{0}$ from $[0,a_{1}]$; term from $z\in \left[ a_{%
\left[ \protect\alpha \right] },b_{[\protect\alpha ]}\right] $ is equal to $%
\left\{ \protect\beta \right\} /\protect\alpha $ or $1-a_{\left[ \protect%
\alpha \right] }$ for $b_{[\alpha ]}<1$ or $b_{[\alpha ]}>1$; in the same manner
one calculates term from interim $\left[ a_{\left[ \protect\alpha \right]
+1},b_{[\protect\alpha ]+1}\right] $ in different possible cases. 
\label{sltshf11}}
\end{minipage}
\end{figure}

A method for calculating the integral 
\begin{equation}
h_{\left[ \beta \right] _{I}}(w)=\int_{0}^{1}dz\theta \left[ \left\{ \beta
\right\} _{F}-\left\{ \alpha z-w\right\} _{F}\right] ,  \label{shdslt25p}
\end{equation}
with $0\leq w\leq 1$ is presented here. The function $\theta $ is the
Heaviside step function and $[a]_{I}$ and $\{a\}_{F}$ refer to the integral
and fractional parts of $a$. Depending on the values of the parameters $%
\alpha ,$ $\beta $ and $w$ the integrand in this equation can jump from $0$
to $1$ several times inside the range $z\in \left[ 0,1\right] .$ The value
of the integral is the total length of the intervals in this range where 
\begin{equation}
\zeta \left( z\right) \equiv \left\{ \alpha z-w\right\} _{F}\leq \left\{
\beta \right\} _{F}.  \label{ap1}
\end{equation}
To determine this length one needs to find the values of $z$ for which the
function $\left\{ \alpha z-w\right\} _{F}$ equals $0$ and where it equals $%
\left\{ \beta \right\} _{F}.$This function is shown in Fig. \ref{sltshf11}. 
When $\alpha z-w=r-1,$ where $r$ is an integer, $\zeta \left( z\right) =0,$
i. e. zeros of $\zeta \left( z\right) $ are given by 
\begin{equation}
z=a_{r}=\left( r-1+w\right) /\alpha .  \label{shdslt26}
\end{equation}
For $z>$ $a_{r}$ the function $\zeta \left( z\right) $ evolves as 
\begin{equation}
\zeta \left( z\right) =\alpha \left( z-a_{r}\right)  \label{ap2}
\end{equation}
and equals $\left\{ \beta \right\} _{F}$ at the point

\begin{equation}
z=b_r=a_r+\frac{\left\{ \beta \right\} _F}\alpha =(r-1+w+\left\{ \beta
\right\} _F)/\alpha .  \label{shdslt27}
\end{equation}
Since contributions to the integral (\ref{shdslt25p}) vanish unless $a_r\leq
1$ and $b_r\geq 0$ and since $0\leq w\leq 1$, it follows from Eqs. (\ref
{shdslt26}, \ref{shdslt27}) that $0\leq r\leq (1+[\alpha ]_I)$. For the time
being, we assume that $\alpha >1$.

All intervals $\left[ a_r,b_r\right] $ totally or partially within the range 
$\left[ 0,1\right] $ contribute to the integral (\ref{shdslt25p}). Let us
denote the contribution from the range $\left[ a_r,b_r\right] $ by $A_r$ and
the total value of the integral by 
\begin{equation}
h_{\left[ \beta \right] _I}(w)=\sum_{r=0}^{1+[\alpha ]_I}A_r.  \label{ap50}
\end{equation}
When $a_r>0$ and $b_r<1$ the interval $\left[ a_r,b_r\right] $ lies entirely
to the range $\left[ 0,1\right] $ and 
\begin{equation}
A_r=b_r-a_r=\frac{\left\{ \beta \right\} _F}\alpha ;\hspace{0.2in}a_r>0\text{
and }b_r<1  \label{ap3}
\end{equation}
When $1>a_r>0$ and $b_r\geq 1$, the maximum value of $z$ contributing to the
integral (\ref{shdslt25p}) is $z=1$ and

\begin{equation}
A_r=1-a_r;\hspace{0.2in}1>a_r>0\text{ and }b_r\geq 1  \label{ap4}
\end{equation}
Similarly, for $a_r<0$ and $0<b_r<1$, 
\begin{equation}
A_r=b_r;\hspace{0.2in}a_r<0\text{ and }0<b_r<1  \eqnum{A8a}  \label{ap4aa}
\end{equation}
and for $b_r\leq 0$, 
\begin{equation}
A_r=0;\hspace{0.2in}\text{ }b_r\leq 0  \eqnum{A8b}  \label{ap4bb}
\end{equation}
For given $r$, $\alpha $, and $\beta $, the values of $a_r$ and $b_r$ can
depend on $w$, giving rise to a dependence of $h_{\left[ \beta \right] _I}\ $%
on $w$. Note that ($b_r-a_r)\leq 1$, which follows from Eqs. (\ref{shdslt27}%
) and the assumption that $\alpha >1$.

We first consider the range $1\leq r\leq \left[ \alpha \right] _I-1$, for
which $b_r<1$ and $a_r>0$ for any $w\in \left[ 0,1\right] $. It then follows
from Eq. (\ref{ap3}) that the total contribution $A^{\prime }$ to the
integral (\ref{shdslt25p}) from the region $\left[ a_1,b_{\left[ \alpha %
\right] _I-1}\right] $ is given by 
\begin{equation}
A^{\prime }=\sum_{r=1}^{[\alpha ]_I-1}A_r=\left\{ \beta \right\} _F(\left[
\alpha \right] _I-1)/\alpha .  \label{ap51}
\end{equation}
This contribution is independent of $w$ and represents a constant background
term.

Since $r\leq [\alpha ]_I+1$, the only remaining contributions to the
integral can come from $A_0,$ $A_{\left[ \alpha \right] _I}$ and $A_{\left[
\alpha \right] _I+1}$. These terms depend on $w$ and represent the atomic
gratings. Let us first consider $A_0$. If $r=0$, $b_0=(-1+w+\{\beta
\}_F)/\alpha <1$ and $a_0=(-1+w)/\alpha <0$. It then follows from Eqs. (\ref
{ap4}, \ref{ap4aa}, \ref{ap4bb}) that 
\begin{equation}
A_0=\left\{ 
\begin{array}{cc}
(w+\left\{ \beta \right\} _F-1)/\alpha , & \text{for\thinspace }\,w\geq
(1-\{\beta \}_F)\equiv \bar{w} \\ 
0 & \text{for}\,\,w\leq \bar{w}
\end{array}
\right. .  \label{ap6}
\end{equation}

We now turn our attention to $A_{[\alpha ]_I}$ and $A_{\left[ \alpha \right]
_I+1}$. It follows from Eq. (\ref{shdslt26}) that $a_{[\alpha ]_I}\in [0,1]$%
. Only the points $b_{\left[ \alpha \right] _I},\,\,a_{\left[ \alpha \right]
_I+1},\,\,b_{\left[ \alpha \right] _I+1}$ can lie to the right of the range $%
\left[ 0,1\right] $, which occurs when 
\begin{eqnarray}
w &\geq &(\{\alpha \}_F+1-\left\{ \beta \right\} _F)\equiv w_1,  \nonumber
\label{ap53} \\
w &\geq &\left\{ \alpha \right\} _F\equiv w_2,  \label{ap53} \\
w &\geq &(\{\alpha \}_F-\left\{ \beta \right\} _F)\equiv w_3,  \nonumber
\end{eqnarray}
respectively. From Eqs. (\ref{ap4}, \ref{ap4aa}, \ref{ap4bb}), one finds
that $A_{[\alpha ]_I}$, and $A_{\left[ \alpha \right] _I+1}$ are given by 
\begin{equation}
A_{[\alpha ]}=\left\{ 
\begin{array}{cc}
b_{[\alpha ]_{I}}-a_{[\alpha ]_{I}}=\{\beta \}_{F}/\alpha  & w<w_{1} \\ 
1-a_{[\alpha ]_{I}}=\frac{\{\alpha \}_{F}-w+1}{\alpha } & w>w_{1}
\end{array}
\right. ,  \label{ap54}
\end{equation}
\begin{equation}
A_{[\alpha ]_I+1}=\left\{ 
\begin{array}{cc}
0 & w>w_2 \\ 
1-a_{[\alpha ]_I+1}=\frac{\{\alpha \}_F-w}\alpha & w_3<w<w_2 \\ 
b_{[\alpha ]_I+1}-a_{[\alpha ]_I+1}=\{\beta \}_F/\alpha & w<w_3
\end{array}
\right.  \eqnum{A12a}  \label{ap54a}
\end{equation}
In order to sum $A_0$, $A_{[\alpha ]_I}$, and $A_{\left[ \alpha \right]
_I+1} $ it is convenient to separate regions of $\alpha $ and $\beta $
according to the relative values of $w_1,\,\,w_2,\,\,w_3\,\,$and $\bar{w}$.
Since 
\begin{equation}
w_3\leq w_2\leq w_1,  \label{shdslt46}
\end{equation}
\begin{equation}
w_3\leq \bar{w}\leq w_1,  \eqnum{A13a}  \label{shdslt46a}
\end{equation}
and $w_2=\left\{ \alpha \right\} _F\geq 0$ one can distinguish four cases 
\begin{equation}
w_3\leq 0\leq \bar{w}\leq w_2\text{;}\hspace{0.15in}\left[ \left\{ \beta
\right\} _F\geq \max \left( \left\{ \alpha \right\} _F,1-\left\{ \alpha
\right\} _F\right) \right] ,  \label{shdslt29}
\end{equation}
\begin{equation}
w_3\leq 0\leq w_2\leq \bar{w}\text{;}\hspace{0.15in}\left[ \left\{ \alpha
\right\} _F\leq \left\{ \beta \right\} _F\leq 1-\left\{ \alpha \right\} _F%
\right] ,  \eqnum{A14a}  \label{shdslt29a}
\end{equation}
\begin{equation}
0\leq w_3\leq \bar{w}\leq w_2\text{;}\hspace{0.15in}\left[ 1-\left\{ \alpha
\right\} _F\leq \left\{ \beta \right\} _F\leq \left\{ \alpha \right\} _F%
\right] ,  \eqnum{A14b}  \label{shdslt29b}
\end{equation}
\begin{equation}
0\leq w_3\leq w_2\leq \bar{w}\text{ ;}\hspace{0.15in}\left[ \left\{ \beta
\right\} _F\leq \min \left( \left\{ \alpha \right\} _F,1-\left\{ \alpha
\right\} _F\right) \right] .  \eqnum{A14c}  \label{shdslt29c}
\end{equation}
Consider, for example, the case (\ref{shdslt29}) for the range of $w$ given
by 
\begin{equation}
0\leq w\leq \bar{w}.  \label{shdslt45}
\end{equation}
For integer $\beta $ this corresponds to the entire range of allowed $w$, $%
0\leq w\leq 1$. From Eqs. (\ref{ap6}, \ref{ap54}, \ref{ap54a}) one finds $%
A_0=0$, $A_{\left[ \alpha \right] _I}=\left\{ \beta \right\} _F/\alpha $,
and $A_{\left[ \alpha \right] _I+1}=\left( \left\{ \alpha \right\}
_F-w\right) /\alpha $. As a result, one finds that $A\left( w\right) =A_0+A_{%
\left[ \alpha \right] _I}+A_{\left[ \alpha \right] _I+1}$ is given by 
\begin{equation}
A(w)=\left( \left\{ \beta \right\} _F+\left\{ \alpha \right\} _F-w\right)
/\alpha .  \label{shdslt30}
\end{equation}
By combining Eqs. (\ref{ap51}, \ref{shdslt30}), one obtains 
\begin{equation}
h_{\left[ \beta \right] _I}\left( w\right) =\left( \left\{ \beta \right\} _F%
\left[ \alpha \right] _I+\left\{ \alpha \right\} _F-w\right) /\alpha .
\label{ap7}
\end{equation}
Even though Eq. (\ref{ap7}) have been derived for $\alpha \geq 1$, one can
verify that it holds for arbitrary $\alpha .$

Other values of $w$ and other cases (\ref{shdslt29a}-\ref{shdslt29c}) can be
considered in the same manner. As a result, one arrives at Eqs. (\ref
{shdslt32p}, \ref{shdslt32}) of the text.

\begin{center}
\bf{REFERENCES}\\
\end{center}%
%

\begin{description}
\item  Altschuler, S. and Frantz, L. M. $\left( 1973\right) $ US Patent No.
3,761,721.

\item  Baklanov, Ye. V., Dubetsky, B., Chebotayev, V. P. $\left( 1976\right) 
$ Appl. Phys. {\bf 9,} 171-173.

\item  Barger, R. L., Bergquist, J. C., English, T. C., Glaze, D. J. $\left(
1979\right) $ Appl. Phys. Lett. {\bf 34} 850-852.

\item  Batelaan, H., Bernet, S., Oberthaler, K., Rasel, E. M., Schmiedmayer,
J., Zeilinger, A. $\left( 1996\right) ,$ in ''{\it Atom Interferometry,'' }%
P. R. Berman (ed.) (Academic, Chestnut Hill, 1997).

\item  Bord\'{e}, Ch. J. $\left( 1989\right) $ Phys. Lett. A {\bf 140} 10-12.

\item  Carnal, O., Mlynek., J. (1991) Phys. Rev. Lett. {\bf 66, }2689-2692.

\item  Carnal, O., Turchette, Q. A., Kimble, H. J. $\left( 1995\right) $
Phys. Rev. A {\bf 51 }3079-3087.

\item  Chapman, M. S., Ekstrom, C. R., Hammond, T. D., Schmiedmayer J.,
Tannian, B. E., Wehinger, S., Pritchard, D. E. $\left( 1995\right) $ Phys.
Rev. A {\bf 51,} R14-17.

\item  Chebotayev, V. P. $\left( 1978\right) $ Appl. Phys. {\bf 15}, 219-222.

\item  Chebotayev, V. P., Dyuba, N. M., Skvortsov, M. N., Vasilenko, L. S. $%
\left( 1978a\right) $ Appl. Phys. {\bf 15,} 319-322.

\item  Chebotayev, V. P., Dubetsky, B., Kazantsev, A. P., Yakovlev, V. P. $%
\left( 1985\right) $ J. Opt.Soc. Am. B {\bf 2,} 1791-1798.

\item  Chebotayev, V. P. $\left( 1986\right) $ Simposium hold in the memory
of Yu. B. Rumer, Novosibirsk (unpublished).

\item  Clauser, J. F., Reinsch, M. W. $\left( 1992\right) $ Apll. Phys. B 
{\bf 54} 380-395.

\item  Clauser, J. F., Li, S. $\left( 1994\right) $ Phys. Rev. A {\bf 49,}
R2213-2216.

\item  Dubetsky, B. (1976) Kvantovaya Elektronika {\bf 3}, 1258-1265 [Sov.
J. Quantum Electron. {\bf 6}, 682-686 (1976)]

\item  Dubetsky, B., Semibalamut, V. M. $\left( 1978\right) .$ In {\it Sixth
International conference on atomic physics. Abstaracts, }edited by E.
Anderson, E. Kraulinya, R. Peterkop $\left( \text{Riga, USSR}\right) ,$ p.
21.

\item  Dubetsky, B.,Semibalamut, V. M. (1982) Kvantovaya Elektronika {\bf 9}%
, 1688-1691 [Sov. J. Quantum Electron. {\bf 12,} 1081-1083 (1982)].

\item  Dubetsky, B., Chebotayev, V. P., Kazantsev, A. P., Yakovlev, V. P.
(1984) Pis'ma Zh. Eksp. Teor. {\bf 39}, 531-533 (1984) [JETP Lett. {\bf 39},
649 (1984)].

\item  B. Dubetsky, P. R. Berman. $\left( 1994\right) $ Phys. Rev. A {\bf 50,%
} 4057-4068.

\item  Ekstrom, C. R., Schmiedmayer, J., Chapman, M. S., Hammond, T. D.,
Pritchard, D. E. $\left( 1995\right) $ Phys Rev. A {\bf 51,} 3883-3888.

\item  Friedberg, R., Hartmann, S. R. $\left( 1993\right) $ Phys. Rev. A 
{\bf 48, }1446-1472.

\item  Friedberg, R., Hartmann, S. R. $\left( 1993a\right) $ Laser Physics 
{\bf 3}, 526-534.

\item  Giltner, D. M., McGowan, R. W., Lee, S. A.$\left( 1995\right) $ Phys.
Rev. Lett. {\bf 75, }2638-2641.

\item  Hall, J. L., Borde, C. J., Uehara, K. $\left( 1976\right) $ Phys.
Rev. Lett. {\bf 37}, 1339-1342.

\item  Janicke, U., Wilkens, M. $\left( 1994\right) $ J. Phys. II France 
{\bf 4,} 1975-1980.

\item  Kapitza, P. L., Dirac, P. A. M. (1933) Proc. Camb. Phil. Soc. {\bf 29}%
, 297-300.

\item  Kasevich, M., Chu, S. $\left( 1991\right) $ Phys. Rev. Lett. {\bf 67,}
181-184.

\item  Kazantsev, A. P., Surdutovich, G. I., Yakovlev, V. P. (1980) Pis'ma
Zh. Eksp. Teor. Fiz. {\bf 31}, 542-545 [ JETP Lett. {\bf 31,} 509-512 $%
\left( 1980\right) $].

\item  Keith, D. W., Ekstrom, C. R., Turchette, Q. A., D. E. Pritchard, D.
E. $\left( 1991\right) $ Phys. Rev. Lett. {\bf 66,} 2693-2696.

\item  Kol'chenko, A. P., Rautian S. G., Sokolovski\"{i}, R. I. $\left(
1968\right) .$ Zh. Eksp. Teor. Fiz. {\bf 55,} 1864-1873 [Sov. Phys. JETP 
{\bf 28, }986-990 (1969)].

\item  Kruse, U. E., Ramsey, N. F. $\left( 1951\right) $ J. Math. Phys., 
{\bf 30,} 40-43.

\item  LeGou\"{e}t, J.-L., Berman, P. R. $\left( 1979\right) $ Phys. Rev. A 
{\bf 20,} 1105-1115.

\item  Martin, P. J., Oldaker, B. J., Miklich, A. H., Pritchard, D. E. $%
\left( 1988\right) $ Phys. Rev. Lett. {\bf 60,} 515-518.

\item  Mossberg, T, Kachru, R., Whittaker, E., Hartmann, S. R. (1979) Phys.
Rev. Lett. {\bf 43,} 851-855.

\item  Moskovitz, P. E., Gould, P.L., Atlas, S. R., Pritchard, D. E. $\left(
1983\right) $ Phys. Rev. Lett. {\bf 51,} 370-373.

\item  M\"{u}ller, J. H., Bettermann, D., Rieger, V., Ruschewitz, F.,
Sengstock, K., Sterr, U., Christ, M., Schiffer, M., Scholz, A., Ertmer, W. $%
\left( 1995\right) $ In {\it Atomic Physics 14,} Editors Wineland D. J.,
Wieman, C. E., Smith, S. J., AIP Conference Proceedings (AIP Press, New
York) {\bf 323}, 240-257.

\item  Patorski, K. $\left( 1989\right) $ In E. Wolf, Progress in optics
XXVII, Elsevier Science Publishers B. V., pp 1-108.

\item  Rasel, E. M., K. Oberthaler, K., Batelaan, H., Schmiedmayer, J.,
Zeilinger, A. $\left( 1995\right) $ Phys. Rev. Lett. {\bf 75,} 2633-2637.

\item  Riehle, F., Kisters, Th., Witte, A., and Helmcke, J. $\left(
1991\right) $ Phys. Rev. Lett. {\bf 67,} 177-180.

\item  Shimizu, F., Shimizu, K., Takuma, H. $\left( 1992\right) $ Phys. Rev.
A {\bf 46,} R17-20.

\item  Timp, G., Behringer, R. E., Tennant, D. M., Cunningham, J. E.,
Prentiss, Berggren, M. K. $\left( 1992\right) $ Phys. Rev. Lett. {\bf 69, }%
1636-1639.

\item  Turchette, Q. A., Pritchard, D. E., Keith, D. W. $\left( 1992\right) $
J. Opt. Soc. Am. A {\bf 9} 1601-1606.

\item  Winthrop, J. T., Worthington, C. R. $\left( 1965\right) $ J. Opt.
Soc. Am. {\bf 55, }373-381.
\end{description}

\end{multicols}


\begin{references}
\bibitem{1}  The scattering of atoms by standing-wave fields, rather than
MS, is a bit more subtle. For resonant standing wave fields, which can act
as amplitude gratings, the situation is unchanged. On the other hand,
off-resonant fields act as phase gratings for the atoms; as such, they
produce no effect on classically moving particles. Strictly speaking,
therefore, one must quantize the center-of-mass motion to calculate the
scattering of the atoms by the fields. Nevertheless, in a manner analogous
to the normal photon echo, it is possible to assign phases to the atoms
while they are freely evolving between the MS and the screen and still
consider the motion as classical in these regions.
\end{references}
\end{document}